\newcommand{\secondLanguage}{icelandic} 	
\documentclass[final,english]{volcanica-template}
\DeclareUnicodeCharacter{2212}{\ensuremath{-}}
\nonstopmode									% can toggle to see compilation warnings
\usepackage{graphicx}
\usepackage{subcaption}
\usepackage{multirow}
\usepackage{placeins}
\usepackage{booktabs} % optional, if you later want cleaner tables
\FloatBarrier
\newcommand{\ArticleType}{Research article}			
\newcommand{\Title}{Extension of openCOSMO-RS Into a Full Open-Source Equation of State: Implementation, Parameterization, and Benchmarking} 	% Manuscript title goes here.
\newcommand{\shortTitle}{Extension of openCOSMO-RS Into a Full Open-Source Equation of State} 		        % Short title for header goes here, less than 60 characters.
\newcommand{\Author}{Markgraf et al.,}    	% First author goes here.

\author[{{\affiliation{1}}}] 				% affiliation number
{\orcidaffil{0009-0008-8037-7671}~			% orcid number
Jan Markgraf} 					        	% first author

\author[{{\affiliation{1}}}] 				% affiliation number
{\orcidaffil{0000-0002-3580-3386}~			% orcid number
D. G. Lisboa Girardi 				        % second author
}

\author[{{\affiliation{1}}}] 				% affiliation number
{\orcidaffil{0000-0003-4503-4039}~			% orcid number
Irina Smirnova}						        % fourth author

\author[{{\affiliation{1}}}
\Email{simon.mueller@tuhh.de}] 				% affiliation number
{\orcidaffil{0000-0003-1684-6994}~			% orcid number
Simon Müller} 					        	% corresponding author

\affil[{{\affiliation{1}}}]{					% affiliation #1
Institute of Thermal Separation Processes, Hamburg University of Technology, Hamburg, Germany
}
\begin{document}

%------------------------------------------------------------
%  Abstract(s) and Keywords
%------------------------------------------------------------	
% Replace dummy text (\protect{\lipsum[45]}) with the abstract in the first argument for \FrontMatter. 
% If there is a second-language abstract, it goes between the square brackets []. 
% Make sure you have included the correct language in line 1, above.
% If no second abstract, ensure there is no space between the square brackets [].
\FrontMatter{
The COSMO-SAC-Phi model developed by Soares et al. extends the COSMO-SAC activity-coefficient framework into a full equation of state by explicitly accounting for pressure effects. In this approach, pure substances and mixtures are represented as pseudo-mixtures consisting of the actual number of moles and an additional pseudo-component that describes free volume, or holes.
In this work, we implement this extension within the openCOSMO-RS framework and evaluate it using a large and diverse set of molecules and binary systems. The resulting equation of state includes an extensive open-source parameter set with around 1800 pure-component entries, made freely available to the academic community.
The four pure-component parameters were fitted to vapor-pressure and liquid-molar volume data for each substance. Model performance was assessed against two benchmark equation-of-state databases, one for pure compounds and one for binary mixtures, without introducing any binary interaction parameters.
The resulting openCOSMO-RS-Phi model reproduces the accuracy of the original COSMO-SAC-Phi formulation while providing a fully open-source and accessible implementation for the scientific community. Beyond its immediate utility, it also establishes a foundation for future development of predictive EoS for electrolyte solutions.
}
[]% 
{					
{Open-source}{Parametrization}{ORCA}{Equation of State}	% add up to six keywords in curly braces {}
}

%------------------------------------------------------------
%  Maintext
%------------------------------------------------------------	
\hypertarget{introduction}{
\section{Introduction}\label{introduction}}

Thermodynamic models are essential for the design, development, and control of chemical processes, and the selection of an appropriate model is critical for these applications. They describe the interactions between temperature, pressure, volume, and material compositions, and are crucial to calculate thermodynamic properties and determine phase equilibrium \cite{lin_thermodynamic_2006}. Commonly used are equations of state (EoS), which can be applied to high pressure conditions, and activity coefficient models, also known as excess Gibbs energy ( $G^E$ ) models, which perform well for liquid mixtures and are pressure independent \cite{privat_state_2023}.

While all thermodynamic models possess some predictive capability, Jirasek et al. \cite{jirasek_making_2022} distinguish between two types of prediction. The first refers to frameworks that generalize across different conditions, such as temperature, pressure, and concentration, for a fixed set of components. The second refers to models that generalize over different systems. In the present work, the term 'predictive models' refers to the second category proposed by Jirasek et al., which predict properties and phase behavior of systems without requiring prior data for the specific system under consideration. This ability is particularly valuable in situations where experimental data are unavailable, scarce, difficult to obtain, or expensive.

Among the early and most widely used predictive thermodynamic models are those in which EoS are combined with $G^E$ models. This approach merges the advantages of both EoS and $G^E$ approaches, expanding their range of application \cite{kontogeorgis_thirty_2012, lopez-echeverry_peng-robinson_2017}. This was first achieved following the publication of the Huron-Vidal mixing rule \cite{huron_new_1979}, which enabled the incorporation of a predictive activity coefficient model into an equation of state through a mixing rule. Since then, several mixing rules have been developed, such as MHV1, MHV2 \cite{michelsen_modified_1990}, WS \cite{wong_theoretically_1992}, UMR \cite{voutsas_universal_2004}, UHVM \cite{privat_let_2025}, along with combinations of cubic EoS models like PR \cite{peng_new_1976} and SRK \cite{soave_equilibrium_1972} with group contribution activity coefficients such as UNIFAC \cite{fredenslund_groupcontribution_1975}, modified-UNIFAC \cite{gmehling_modified_1993}, LIFAC \cite{yan_prediction_1999}, and ASOG \cite{kojima_prediction_1979}. This development has resulted in several useful models, including PSRK \cite{holderbaum_psrk_1991}, LCVM \cite{boukouvalas_prediction_1994}, VTPR \cite{ahlers_development_2001, ahlers_development_2002, ahlers_development_2002-1, wang_development_2003, ahlers_development_2004}, UMR-PRU \cite{voutsas_thermodynamic_2006}, and others \cite{dahl_highpressure_1990, tochigi_prediction_1995, li_prediction_2001}. Recently, similar approaches have been proposed for an EoS/ $G^E$ using PC-SAFT \cite{gross_perturbed-chain_2001, gross_application_2002} \cite{walker_toward_2022} or CPA \cite{kontogeorgis_equation_1996} \cite{tasios_development_2023} as equation of state.

Other recent advancements in predictive EoS focus on the development of methods for predicting pure component parameters, binary interaction parameters, and utilizing QSPR methods for activity coefficient model \cite{panayiotou_partial_2015, mastrogeorgopoulos_toward_2017}. Pure component parameter predictions are primarily carried out in SAFT EoS \cite{chapman_saft_1989, chapman_new_1990} and its variants, using either group contribution methods, as comprehensively reviewed by Shaahmadi and colleagues \cite{shaahmadi_group-contribution_2023}, or machine learning algorithms \cite{zhu_generating_2020, matsukawa_estimation_2021, abdallah_el_hadj_ai-pcsaft_2022, habicht_predicting_2023, winter_understanding_2023, felton_ml-saft_2024}. For binary interaction parameters, predictive EoS have been developed for PR EoS using group contribution models, leading to models like PPR78 \cite{jaubert_vle_2004} and EPPR78 \cite{xu_e_2017, jaubert_impressive_2022}, as well as QSPR \cite{abudour_generalized_2014} and machine learning \cite{abudour_predicting_2017}. Similarly, both QSPR \cite{stavrou_estimation_2016} and machine learning \cite{abbasi_estimation_2020} approaches have been applied to PC-SAFT.

The conductor-like screening model for real solvents (COSMO-RS) \cite{klamt_conductor-like_1995, klamt_refinement_1998, klamt_cosmo-rs_2000} is a completely a priori predictive activity coefficient model. It treats a molecule as a collection of surface segments and calculates the molecular chemical potential by summing the chemical potentials of each segment. These segment potentials are determined by the interaction energies between segment screening charges, obtained through quantum chemical \textit{ab initio}  solvation calculations. In contrast to other established group contribution $G^E$ models, such as UNIFAC and modified-UNIFAC, COSMO-RS requires a much smaller set of parameters and can partially differentiate between isomers \cite{grensemann_performance_2005}. Several versions of the COSMO-RS model have emerged in the literature, such as COSMO-SAC \cite{lin_priori_2002}, COSMO-RS(Ol) \cite{grensemann_performance_2005}, COSMO-vac \cite{shimoyama_development_2009}, COSMO-SAC-dsp \cite{hsieh_considering_2014} and the closely related F-SAC \cite{soares_functional-segment_2013, soares_functional-segment_2013-1}.

Although COSMO-based models are effective, they share the same limitations as other $G^E$ models, as they do not account for pressure effects and are restricted to representing incompressible liquids. As a result, various approaches have been developed to integrate them with existing EoS. Early methods involved using an EoS/ $G^E$ framework with UNIQUAC as the activity coefficient model, with parameters adjusted to either COSMO-RS or COSMO-SAC \cite{constantinescu_vaporliquid_2005, shimoyama_prediction_2006}. Within the EoS/ $G^E$ framework, recent studies have explored coupling cubic EoS with COSMO-based models through both traditional and novel mixing rules \cite{lee_prediction_2007, shimoyama_prediction_2008, leonhard_comparison_2009, staudt_self-consistent_2012, chen_prediction_2018, wang_prediction_2018, mambo-lomba_predictions_2021, wang_vapor-liquid_2023, paes_prediction_2023, paes_predicting_2024, prinos_phase_2025}, and PC-SAFT with COSMO-RS has been implemented as well \cite{song_prediction_2025}. Other significant contributions include NRCOSMO \cite{panayiotou_solvation_2010, panayiotou_toward_2011, panayiotou_partial_2015}, based on the Non-Random Hydrogen-Bonding EoS; $\sigma$-MTC \cite{costa_equation_2016, costa_equation_2016-1}, derived from the Mattedi-Tavares-Castier EoS; COSMO-SAC-Phi \cite{de_p_soares_pairwise_2019}, a new equation of state inspired by lattice-fluid theory; and the recent notes on the essentials for the formulation of a COSMO-RS-based equation of state published by Panayiotou and colleagues \cite{panayiotou_cosmo-rs_2023}.

In this work, an open source implementation of the original COSMO-SAC-Phi is proposed as an extension of the openCOSMO-RS \cite{gerlach_open_2022} activity coefficient model. COSMO-SAC-Phi is a unique and novel approach that, unlike most of the literature, expands the COSMO-SAC model into a full equation of state rather than coupling it with or modifying existing EoS. Consequently, the resulting EoS does not rely on binary interaction parameters to describe the mixtures, but uses the underlying COSMO-RS theory. Although an improved version of COSMO-SAC-Phi exists \cite{zini_improved_2021}, featuring an additional parameter to increase precision and thermodynamic consistency, it was not employed in this study. Another related model is F-SAC-Phi \cite{baladao_functional-segment_2019} which uses the F-SAC model instead of the COSMO-SAC model as backend. To prevent any ambiguity when referring to the original or to the open-source model, the proposed method is designated as openCOSMO-RS-Phi. This work serves as a foundation for the further development of the first a priori predictive equation of state based on COSMO-RS theory for mixtures.

\hypertarget{methodology}{%
\section{Methodology}\label{methodology}}

This work follows a three-step approach. First, COSMO-SAC-Phi is implemented in the open-source form of openCOSMO-RS-Phi. To this end, the theoretical framework is revisited in this section, and openCOSMO-RS-Phi is formulated as a Helmholtz-explicit equation of state. For the assessment of model performance, pure-component parameters are fitted, and the resulting model is benchmarked against both a pure-component database and a binary-mixture database. The corresponding methodological and theoretical foundations are presented in the second part of this section.

\subsection{Mathematical Derivation}\label{Mathematical Derivation}

openCOSMO-RS-Phi, as an open-source implementation of COSMO-SAC-Phi, is a perturbation model designed to predict the behavior of fluids. It employs an attraction term developed by integrating COSMO-RS with concepts from lattice-fluid theory. This concept introduces holes that occupy all available free volume and account for pressure and volume effects. Therefore, all systems are considered to be pseudo-mixtures from the moles of real mixture or pure substance \(\mathbf{n}\ = [n_1, n_2,..., n_i, ..., n_N]\), and the moles of holes \(n_h\), with molar vector \(\mathbf{\tilde n} = [n_1, n_2,..., n_i, ..., n_N, n_h]\). This leads to the following expression for the total system volume \(V\), which, given the total moles of the real fluid, enables the calculation of the moles of holes:
\begin{equation}
  V = \sum_i{n_ib_i + n_hb_h}
\label{eq:01}\end{equation}
\begin{equation}
  n_h = \frac{V - \sum_i{n_ib_i}}{b_h}
\label{eq:02}\end{equation}
where \(b_i\) is the molar cavity of species \(i\) and \(b_h\) is the molar cavity of a hole. The values from \(b_i\) and \(b_h\) are the first among the four existing parameters that need to be adjusted.

Although the concept of vacancy/holes within COSMO-SAC was initially introduced with the COSMO-vac model \cite{shimoyama_development_2009}, a decade prior to the development of COSMO-SAC-Phi, it is applied only to computing the activity coefficients of solutes diluted in a supercritical phase. Rather than describing the vacancy as an additional pseudo-substance, COSMO-vac calculates a vacancy fraction \(\alpha_{vac}\) that modifies the COSMOSPACE equation described at \autoref{eq:24}. Another similar approach is the use of a vacuum pseudo-liquid within the COSMOtherm software \cite{cosmotherm2025}. In the latter case, the vacuum molecule is used only to approximate the gas phase at a liquid-gas interface. Similarly to COSMO-SAC-Phi, this pseudo-substance is characterized by the presence of a single surface segment with a fixed area, and the possession of volume.
%\begin{equation}
%      \Gamma_I^{vac} = \frac{1}{\sum_J{X_J(1-\alpha_{vac})\Gamma_J^{vac}\tau_{IJ}}}
%\label{eq:03}\end{equation}

It is possible to derive the relevant residual properties from the residual Helmholtz energy \(A^r(T,V,\mathbf{n})\). The residual Helmholtz energy \(A^r(T,V,\mathbf{n})\) is defined as the difference between the Helmholtz energy of the real phase and the Helmholtz energy of an ideal gas mixture having the same temperature \(T\), the same volume \(V\), and the same composition as the real mixture. The openCOSMO-RS-Phi equation for \(A^r\) can be divided into a repulsive contribution, \(A_R^r\), and an attractive contribution, \(A_A^r\):
\begin{equation}
A^r(T,V,\mathbf{n})=A_R^r(T,V,\mathbf{n})+A_A^r(T,V,\mathbf{n})
    \label{eq:35}
\end{equation}

The repulsive interactions are represented using the classical Van der Waals approach. This simplified approximation yields \cite{prausnitz_molecular_1999}:
\begin{equation}
  P_R = \frac{nRT}{V-\sum_i{n_ib_i}}
\label{eq:07}\end{equation}

It is important to note that the repulsive pressure \(P_R\) according to \autoref{eq:07} is actually the sum of the repulsive contribution to the residual pressure \(P_R^r\) and the pressure of an ideal gas mixture \(P^{IGM}\): \(P_R=P_R^r+P^{IGM}\). \(A_R^r\) can then be calculated from classical thermodynamics as follows:
\begin{align}
    A_R^\text{r}(T,V,\mathbf{n}) &= - \int_\infty^V \frac{nRT}{V-\sum_i n_ib_i} - \frac{nRT}{V} \mathrm{d}V
    \label{eq:36}\\
    &= -nRT \ln \left(\frac{V-\sum_i n_ib_i}{V} \right) \label{eq:37}
\end{align}

The attractive contribution, \(A_A^r\), can be obtained from Euler's theorem, provided that the Helmholtz energy is a homogeneous function to the first degree in \(V\) and \(n\) which is correct within the frame of openCOSMO-RS-Phi:
\begin{equation}
    A_A^r(T,V,\mathbf{n}) = \left(\frac{\partial A_A^r}{\partial V} \right)_{T,\mathbf{n}} V + \sum_{i=1}^N n_i \left( \frac{\partial A_A^r}{\partial n_i}\right)_{T,V,n_j} \label{eq:38} 
\end{equation}

An important detail when converting the partial derivative of the attractive contribution to the Helmholtz energy of the real fluid \(A_A^r\) into the partial derivative of the Helmholtz energy of the pseudo-mixture \(\tilde A_A^r\) in regards to the moles of holes, is that free volume changes are governed completely in the pseudo-mixture by the variation of the amount of holes, while the overall volume of the pseudo-mixture remains constant. Therefore:
\begin{equation}
  \left(\frac{\partial A^r_A}{\partial V}\right)_{T,\mathbf{n}} = 
  \left(\frac{\partial \tilde A^r_A}{\partial n_h}\right)_{T,V,\mathbf{n}} 
  \left(\frac{\partial n_h}{\partial V}\right)_{T,\mathbf{n}} 
\label{eq:10}\end{equation}

The attractive residual chemical potential of the holes in the pseudo-mixture is defined:
\begin{equation}
  \tilde{\mu}^r_{Ah} = 
  \left(\frac{\partial \tilde A^r_A}{\partial n_h}\right)_{T,V,\mathbf{n}}
\label{eq:11}\end{equation}

Moreover, differentiating \autoref{eq:02} in regards to the volume while \(\mathbf{n}\) is constant yields the final term of \autoref{eq:10}:
\begin{equation}
  \left(\frac{\partial n_h}{\partial V}\right)_{T,\mathbf{n}} =
  \frac{1}{b_h}
\label{eq:12}\end{equation}

Thus, and considering for perturbation terms, pressure due to attractive forces is derived from the fundamental thermodynamic relations:
\begin{equation}
  \left(\frac{\partial A^r_A}{\partial V}\right)_{T,\mathbf{n}} = 
  - P_A = \frac{\tilde{\mu}^r_{Ah}}{b_h}
\label{eq:09}\end{equation}

In \autoref{eq:38}, the partial derivative of the attractive Helmholtz contribution with respect to the amount of species \(i\) must be reformulated for the pseudo-mixture. In this framework, a change in \(n_i\) affects the attractive contribution both directly through species \(i\) and indirectly through the number of holes. Since the total volume of the pseudo-mixture is kept constant, any change in the number of holes is solely induced by the change in \(n_i\):
\begin{equation}
  \left(\frac{\partial A^r_A}{\partial n_i}\right)_{T, V, n_j} = 
  \left(\frac{\partial \tilde A^r_A}{\partial n_i}\right)_{T, V, n_j, n_h} + 
  \left(\frac{\partial \tilde A^r_A}{\partial n_h}\right)_{T, V, \mathbf{n}}
  \left(\frac{\partial n_h}{\partial n_i}\right)_{T, V, n_j}
\label{eq:14}\end{equation}

The attractive residual chemical potential of species \(i\) in the pseudo-mixture is defined:
\begin{equation}
  \tilde{\mu}^r_{Ai} = 
  \left(\frac{\partial \tilde A^r_A}{\partial n_i}\right)_{T,V,n_j, n_h}
\label{eq:15}\end{equation}

Although the hole molar cavity \(b_h\) has thus far been treated as constant, it is represented by a sum of the contribution of the hole molar volume of each molecule \(b_{h,i}\) within the system through the following mixing rule:
\begin{equation}
  b_h = \sum_i{\frac{n_i}{n}b_{h,i}}
\label{eq:17}\end{equation}

Taking this into consideration, the final term of \autoref{eq:14} can be obtained by differentiating \autoref{eq:02} in regards to the \(n_i\) while \(V\) and \(n_j\) are constant:
\begin{equation}
  \left(\frac{\partial n_h}{\partial n_i}\right)_{T,V,n_j} =
  -\left[\frac{b_i}{b_h}+\frac{n_h}{n}\left(\frac{b_{h,i}}{b_h}-1\right)\right]
\label{eq:18}\end{equation}

Therefore, the chemical potential of the components in the real mixture are calculated from the chemical potentials of the pseudo-mixture as follows:
\begin{equation}
  \left(\frac{\partial A^r_A}{\partial n_i}\right)_{T, V, n_j} = 
  \tilde{\mu}^r_{Ai}-\tilde{\mu}^r_{Ah}
  \left[\frac{b_i}{b_h}+\frac{n_h}{n}\left(\frac{b_{h,i}}{b_h}-1\right)\right]
\label{eq:19}\end{equation}

If \(b_{h,i}\) have the same value throughout all the components of the mixture, this is to say \(b_{h,i} = b_h\), \autoref{eq:19} can be simplified to:
\begin{equation}
  \left(\frac{\partial A^r_A}{\partial n_i}\right)_{T, V, n_j} = 
  \tilde{\mu}^r_{Ai}-\tilde{\mu}^r_{Ah}\frac{b_i}{b_h}
\label{eq:20}\end{equation}

For both pure components and mixtures, \autoref{eq:38} becomes:
\begin{equation}
    A_A^r(T,V,\mathbf{n}) = \sum_i n_i \left(\tilde{\mu}_i^r - \frac{b_i}{b_h}\tilde{\mu}_h^r \right) - P_A V 
    \label{eq:39}
\end{equation}

Combining \autoref{eq:37} and \autoref{eq:39} yields the complete residual Helmholtz energy, making openCOSMO-RS-Phi a Helmholtz-explicit EoS. As a consequence of linear homogeneity of \(A^r\), one directly obtains a relation for the molar residual Helmholtz energy \(a^r\) as a function of temperature \(T\), molar volume \(v\) and mole fractions \(x_i\):
\begin{equation}
    a^r(T,v,\mathbf{x}) = \sum_i x_i \left(\tilde{\mu}_i^r - \frac{b_i}{b_h}\tilde{\mu}_h^r \right) - P_A v - RT \ln \left(\frac{v-\sum_i x_ib_i}{v} \right)
    \label{eq:40}
\end{equation}

At a defined temperature \(T\), the pressure \(P\) is described as the sum of the reference fluid repulsive term and the perturbation attractive term \cite{poling_properties_2001}:
\begin{equation}
  P = P_R + P_A
\label{eq:04}\end{equation}
where \(P_R\) is the repulsive contribution pressure, described in \autoref{eq:07}, and \(P_A\) is the attractive contribution pressure, shown in \autoref{eq:09}. Therefore:
\begin{equation}
  P = \frac{nRT}{V-\sum_i{n_ib_i}} -\frac{\tilde{\mu}^r_{Ah}}{b_h}
\label{eq:33}\end{equation}

At the same temperature, the fugacity coefficient \({\phi}_i\) of species \(i\) can be described by \cite{michelsen_thermodynamic_2007}:
\begin{equation}
  \ln{{\phi}_i} = 
  \frac{1}{RT}\left(\frac{\partial A^r}{\partial n_i}\right)_{T, V, n_j} - 
  \ln{Z}
\label{eq:05}\end{equation}
where \(A^r\) is the residual Helmholtz energy,  \(Z = PV/nRT\) is the compressibility factor, and \(n =\sum_i{n_i}\) is the molar amount of the real fluid.

Similarly to pressure, the partial derivative of residual Helmholtz energy in regards to the moles of species \(i\) can be described as the sum of a repulsive and an attractive contributions \cite{poling_properties_2001}:
\begin{equation}
  \left(\frac{\partial A^r}{\partial n_i}\right)_{T, V, n_j} = 
  \left(\frac{\partial A^r_R}{\partial n_i}\right)_{T, V, n_j} + 
  \left(\frac{\partial A^r_A}{\partial n_i}\right)_{T, V, n_j}
\label{eq:06}\end{equation}
where:
\begin{equation}
  \left(\frac{\partial A^r_R}{\partial n_i}\right)_{T, V, n_j} = RT \left[ 
  \ln{\frac{V}{V-\sum_i{n_ib_i}}} +
  \frac{nb_i}{V-\sum_i{n_ib_i}} \right]
\label{eq:08}\end{equation}

Using \autoref{eq:08} and \autoref{eq:19}, the fugacity coefficient of component \(i\) in the real mixture is obtained through the following expressions:
\begin{equation}
  \ln{{\phi}_i}(T,V,\mathbf{n}) =  \tilde{\mu}^r_{Ai}-\tilde{\mu}^r_{Ah}
  \left[\frac{b_i}{b_h}+\frac{n_h}{n}\left(\frac{b_{h,i}}{b_h}-1\right)\right] + \ln{\frac{V}{V-\sum_i{n_ib_i}}} +
  \frac{nb_i}{V-\sum_i{n_ib_i}}  - 
  \ln{\frac{PV}{nRT}}
\label{eq:34}\end{equation}

The expressions for \(P_A\), \(P_R\), and \(\ln \phi_i\) relevant to the application of the EoS have so far been written in terms of \(T\), \(V\), and \(n_j\). However, all of these expressions are homogeneous of the order zero with respect to the number of moles and may then also be expressed in terms of \(T\), the molar volume \(v\), and the mole fractions \(x_j=n_j/n\). 

Finally, to completely describe attractive terms, the residual chemical potential in the pseudo-mixture of species \(i\) and of the holes are similarly defined. To put it differently, the mathematical foundation starting in \autoref{eq:21} is applicable to both species \(i\) and holes. However, for the purpose of clarity, their derivation will be presented solely for species \(i\), with brief comments provided when the derivation for holes differ. Therefore, the attractive chemical potential of species \(i\) is related to the activity \(a_i\) by:
\begin{equation}
  \tilde{\mu}_{Ai} = \tilde{\mu}_{Ai}^* + RT \ln{a_i}
\label{eq:21}\end{equation}
where \(\tilde{\mu}_{Ai}^*\) is the attractive chemical potential of species \(i\) in a defined reference state. In the model, the reference state used is the ideal gas, thus:
\begin{equation}
  \tilde{\mu}^r_{Ai} = RT \ln{a_i}
\label{eq:22}\end{equation}

The activity of species \(i\) for COSMO-RS is defined as:
\begin{equation}
  \ln{a_i} = 
  \sum_I{\left[\frac{\mathcal{A}^i_I}{a_{eff}}\ln{\left(\frac{\Gamma_I}{\Gamma_I^*}\right)}\right]}
\label{eq:23}\end{equation}
where \(\mathcal{A}^i_I\) is the surface area of segment \(I\) of molecule \(i\), \(a_{eff}\) is the average molecular contact area \(\Gamma_I\) is the activity coefficient of segment \(I\), and \(\Gamma_I^*\) is the activity coefficient of segment \(I\) at the reference state. The area of one hole is kept equal to that of a 1 \text{\AA} radius sphere. Thus, \(\mathcal{A}^h_I = 12.57\) \AA$^2$.

Through the COSMOSPACE equation system \cite{klamt_cosmospace_2002}, the segment activity coefficient can be calculated:
\begin{equation}
  \Gamma_I=\frac{1}{\sum_J{X_J\Gamma_J\tau_{IJ}}}
\label{eq:24}\end{equation}
where \(X_J\) is the fraction of segment \(J\) that can be found within all segments, and \(\tau_{IJ}\) is the interaction parameter from segments \(I\) and \(J\). They are defined as:
\begin{equation}
  X_J = \frac{\sum_i{x_i\mathcal{A}^i_J}}{\sum_i{\sum_I{x_i\mathcal{A}^i_I}}}
\label{eq:25}\end{equation}
\begin{equation}
  \tau_{IJ}=\exp{\left(-\frac{G_{IJ}^{int}}{RT}\right)}
\label{eq:26}\end{equation}
 where \(\mathcal{A}^i_J\) is the surface area of segment \(J\) of molecule \(i\), \(x_i\) is the mole fraction of molecule \(i\), and \(G_{IJ}^{int}\) describes the interaction energy between segments \(J\) and \(I\). For openCOSMO-RS-Phi, the mole fraction of molecule \(i\) in the pseudo-mixture, \(\Tilde{x}_i = n_i/(n+n_h)\), is used in \autoref{eq:25}.

In this framework, where ideal gas is considered as the reference state, \(\Gamma_I^*\) is calculated at infinite molar volume. Hence, in ideal gas, the mole fractions of all species are considered to approach the limit zero, while the mole fraction of the holes is regarded to approach the limit of one:
\begin{equation}
  \Gamma_I^* = \Gamma_I^{IG} \implies \left( \forall i \in \{1, 2, \ldots, n\}, \; x_i = 0 \right)  \land \left( x_h = 1 \right) 
\label{eq:27}\end{equation}

Furthermore, the interaction energy between segments remains to be described. This step marks the primary distinctions between COSMO-SAC-Phi and openCOSMO-RS-Phi. Both models describe the interaction energy as a sum of the contribution from electrostatic misfit interactions \(E_{IJ}^{mf}\), the contribution from hydrogen bond interactions \(G_{IJ}^{hb}\), and the contribution from dispersion forces \(G_{IJ}^{Disp}\), although the formulation for these individual free energies differ:
\begin{equation}
  G_{IJ}^{int} = E_{IJ}^{mf} + G_{IJ}^{hb} + G_{IJ}^{Disp}
\label{eq:28}\end{equation}

Among the three contributions, both the dispersion forces energy and the electrostatic misfit interactions energy are represented identically as the original model. The first is described as a simple combining rule from two distinct segment dispersions from species \(i\), \(\delta_I^i\) and \(\delta_J^i\), which are temperature dependent:
\begin{equation}
  G_{IJ}^{Disp} =-\frac{1}{2}\sqrt{\delta_I^i\delta_J^i}
\label{eq:29}\end{equation}
\begin{equation}
  \delta_I^i = \delta^{i,0} \left[ 1-\text{exp} \left( -\frac{\delta^{i,T}}{T} \right) \right] 
\label{eq:30}\end{equation}
where \(\delta^{i,0}\) and \(\delta^{i,T}\) are parameters without assigned names, they are adjusted for each individual component, rather than to each segment. As dispersion is not considered for holes, \(\delta^{h}\) is equal to zero.

The misfit contribution is calculated from the charge densities of segments \(I\) and \(J\), denoted as \(\sigma_I\) and \(\sigma_J\):
\begin{equation}
  E_{IJ}^{mf} =
  \frac{a_{eff} \alpha }{2}  \left( \sigma_I + \sigma_J \right) ^2 
\label{eq:31}\end{equation}
with \(\alpha\) as the misfit prefactor.

With regard to the calculation of the hydrogen bond interaction energy, the approach adopted is analogous to that proposed by \textsc{Gerlach, et al.} \cite{gerlach_open_2022} in the derivation of openCOSMO-RS. However, it differs in that the hydrogen bond strength parameter \(c_{hb}\) is considered temperature-independent. This change was made to improve the numerical performance and bring the implementation closer to the original publication of the model.
\begin{equation}
  G_{IJ}^{hb} =
  \begin{cases}
  c_{hb} a_{eff}\left[ 
  \text{min}\left(0; \sigma_I+\sigma_{hb}\right) \cdot 
  \text{max}\left(0; \sigma_J+\sigma_{hb}\right) \right]
  & \text{if } \sigma_I < \sigma_J\\
  c_{hb} a_{eff}\left[ 
  \text{min}\left(0; \sigma_J+\sigma_{hb}\right) \cdot 
  \text{max}\left(0; \sigma_I+\sigma_{hb}\right) \right]
  & \text{if } \sigma_I \geq \sigma_J
  \end{cases}
\label{eq:32}\end{equation}
where \(c_{hb}\) is the hydrogen bond strength parameter, and \(\sigma_{hb}\) is hydrogen bond threshold parameter.

For a more thorough overview of the openCOSMO-RS model and its derivation, please refer to the original publication \cite{gerlach_open_2022}. Its open-source workflow is maintained in this work.

\subsection{Parametrization of the openCOSMO-RS model}
\autoref{tab:Constants} lists the optimized global parameters used in this study for the underlying openCOSMO-RS model. In contrast to previous parameterizations \cite{gerlach_open_2022, muller2025predicting}, \(c_\text{hb}^\text{T}\) was set to 0 and the basic misfit term was used instead of the improved one. 

\begin{table}[!ht]
    \centering
    \begin{tabular}{llr}
    \textbf{Parameter} & \textbf{Unit} & \textbf{Value} \\
    \hline
    \(a_\text{eff}\) & \(\text{AA}^2\) & 4.857 \\
    \(\alpha\) & \(\si{kJ \cdot \text{\AA}^2/(mol \cdot \text{e}^2)}\) & 8130 \\
    \(c_\text{hb}\) & \(\si{kJ \cdot \text{\AA}^2/(mol \cdot \text{e}^2)}\) & 42564 \\
    \(\sigma_\text{hb}\) & \(\si{e/\text{\AA}^2}\) & \(9.443 \times 10^{-3}\) \\
    \end{tabular}
    \caption{Universal constants of openCOSMO-RS.}
    \label{tab:Constants}
\end{table}

\subsection{Experimental Databases and Optimization for pure compound input}\label{Experimental Data and Optimization}
There are four parameters that need to be defined for each pure component: the molar cavity of the component \(b_i\), the hole molar volume of said component \(b_{h,i}\), and the two dispersion parameters \(\delta^{i,0}\) and \(\delta^{i,T}\). They are obtained from the minimization of the following objective function using experimental vapor pressure \(P_i^{sat}\) and molar volume \(v_i^l\):
\begin{equation}
    OF = \frac{1}{N_p} \sum_{i}{w_P\left(\frac{P_i^{sat} - P_i^{sat, calc}}{P_i^{sat}}\right)^2 + w_v\left(\frac{v_i^l - v_i^{l, calc}}{v_i^l}\right)^2}
    \label{eq:44}
\end{equation}
Here, \(N_p\) is the number of experimental points, calculated entries are indicated by the \(calc\) superscript, and \(w_P\) and \(w_v\) are the weighting factors for the saturated pressure and molar volume, respectively. 

The $\sigma$-profile (screened molecular surface) is generated by the quantum-chemical calculation in ORCA, in which the molecular cavity is discretized into individual surface segments. The total molecular surface area is calculated as the sum of the areas of these segments. These quantum-chemically calculated surface quantities are retained in openCOSMO-RS and are independent of the adjustable molar cavity parameter used in the equation of state. For more details the reader is referred to the original publication of the free and open source conformer pipeline \cite{muller2025predicting}.

Experimental data for the vapor pressure (\(P^{sat}\)) and molar volumes (\(v^{liq}\)) of pure components were collected from the DIPPR 801 database \cite{wilding_dippr_1998}. The guidelines proposed by \textsc{Piña-Martinez et al.} \cite{pina-martinez_use_2021} were applied over the database, limiting it to 1800 pure fluids. Furthermore, ten additional substances were removed from the analysis, either because they were mixtures (air), salts or organometallic (ammonium chloride, ammonium sulfide, iron pentacarbonyl, mercury dichloride, nickel carbonyl, vanadium oxytrichloride, vanadium tetrachloride), or lead to convergence issues (eicosamethylnonasiloxane, hexacosamethyldodesiloxane). Similarly to COSMO-SAC-Phi, the present study has focused exclusively on data constrained within a specified temperature range. In this instance, the analysis has been conducted within the temperature range of 50\% to 90\% of the critical temperature, with the data evaluated at 10 K intervals. Thus, for each component, vapor pressure and liquid molar volume are evaluated on the same temperature grid.

The performance of openCOSMO-RS-Phi for pure components was studied by calculating the mean absolute percentage error (MAPE) for each property \(x\), represented by \autoref{eq:44.2}. This method was previously used for COSMO-SAC-Phi, as well as in the aforementioned work by \textsc{Piña-Martinez et al.} \cite{pina-martinez_use_2021}.

\begin{equation}
    \text{MAPE}_x = \frac{100}{N_p} \sum_i {\left|\frac{x_i-x_i^{calc}}{x_i}\right|}
\label{eq:44.2}\end{equation}

\textsc{Piña-Martinez et al.} propose a classification of 'well-modeled' and 'badly-modeled' based on the distribution of the calculated data, analyzing over the complete dataset, as well over the self-associating (SA) and non-self-associating (NSA) fluids. In \autoref{eq:44.3}, a component \(i\) is considered 'well-modeled' if both conditions are satisfied; \(\sigma\) is the standard deviation and the NSA superscript indicates that only non-self-associating components are considered. Despite its limitations, which stem from its reliance on the distribution of the reference set rather than absolute accuracy and its omission of an evaluation of meaningful precision requirements, the analysis was conducted. To circumvent these concerns, the results are also categorized into four performance levels based on the MAPE for both vapor pressure and molar volume: 'excellent', 'good', 'satisfactory' and 'poor'. A model is classified as 'excellent' when the MAPE for both properties is $\le$1\%, 'good' when $\le$5\%, and 'satisfactory' when $\le$10\%. Cases in which the MAPE for either property exceeds 10\% are considered 'poor'. The coefficient of determination, \(R^2\), is also available.

\begin{equation}
    \begin{cases}
        \left( \text{MAPE}_{P^{sat}} \right)_i \le 
        \overline{\text{MAPE}}_{P_{sat}}^{\text{NSA}} + 
        \sigma_{P_{sat}}^{\text{NSA}} 
        \\[2mm]
        \left( \text{MAPE}_{v^{liq}} \right)_i \le 
        \overline{\text{MAPE}}_{v^{liq}}^{\text{NSA}} + 
        \sigma_{v^{liq}}^{\text{NSA}} 
    \end{cases}
\label{eq:44.3}\end{equation}

%Initially, the weighting factors were set to be the same as those used by \textsc{Soares et al.} \cite{de_p_soares_pairwise_2019}. However, analysis of a larger group of molecules revealed bias towards low MAPE values for vapor pressure, with high MAPE for molar volumes. An alternative set of weighting factors proposed in \textsc{Ramírez-Vélez et al.} \cite{ramirez-velez_assessing_2022} evaluation of the same 1800 pure fluids was examined. Nevertheless, given the significant impact of the weighting factors on the average MAPE of each property, and in the value of the objective funtion, a new approach is proposed in order to obtain less arbitrary weights. In this approach, the MAPEs of each property for 4 different EoS (tc-RK, tc-PR, PC-SAFT, and I-PC-SAFT)\cite{pina-martinez_use_2021}\cite{ramirez-velez_assessing_2022} using the same component data were used. his strategy maintains the bias towards saturated pressure present in the aforementioned works while ensuring that molar volume data are not underrepresented. The weighting factors are then defined at \autoref{eq:45} as a normalized inverse of the average MAPE of the respective property \(X\). Their values are 0.6717 for \(w_P\) and 0.3283 for \(w_v\).
Initially, the weighting factors were set to be the same as those used by \textsc{Soares et al.} \cite{de_p_soares_pairwise_2019}. However, analysis of a larger group of molecules revealed bias towards low MAPE values for vapor pressure, with high MAPE for molar volumes. An alternative set of weighting factors proposed in \textsc{Ramírez-Vélez et al.} \cite{ramirez-velez_assessing_2022} evaluation of the same 1800 pure fluids was examined. Nevertheless, given the significant impact of the weighting factors on the average MAPE of each property, and in the value of the objective function, a new approach is proposed in order to obtain less arbitrary weights. In this approach, the MAPEs of each property for 4 different EoS (tc-RK, tc-PR, PC-SAFT, and I-PC-SAFT)\cite{pina-martinez_use_2021}\cite{ramirez-velez_assessing_2022} using the same component data were used. This strategy maintains the bias towards saturated pressure present in the aforementioned works while ensuring that molar volume data are not underrepresented. The weighting factors are then defined at \autoref{eq:45} as a normalized inverse of the average MAPE of the respective property \(X\). Their values are roughly 0.67 for \(w_P\) and 0.33 for \(w_v\).% Their values are 0.6717 for \(w_P\) and 0.3283 for \(w_v\).

\begin{equation}
    w_X = \frac{\frac{1}{\overline{\text{MAPE}}_X}} {\frac{1}{\overline{\text{MAPE}}_{P^{sat}}} + \frac{1}{\overline{\text{MAPE}}_{v^{liq}}}}
\label{eq:45}\end{equation}

The identification of the optimal set of parameters was achieved through the implementation of two minimization methods, executed in a three-step process. Initially, the Nelder-Mead method\cite{nelder_simplex_1965} was implemented, with the parameters constrained to values that were more restrictive and closer to the results presented by \textsc{Soares et al}. \cite{de_p_soares_pairwise_2019}. The initial values for both \(b_i\) and \(b_h\) were respectively obtained through linear and quadratic regression between the parameter values and the COSMO volume, using the data from the COSMO-SAC-Phi publication. The starting values for both \(\delta^{i,T}\) and \(\delta^{i,0}\) were obtained by calculating the mean value of each parameter from the same publication. The best parameter set obtained from the initial step was used as starting values for the subsequent step, which incorporated Nelder-Mead minimization once more, albeit without bounds. This was done intentionally to circumvent the occurrence of local minima in proximity to the boundaries, thereby establishing a suitable initial point for an broader parameter evaluation. The final minimization step was executed via differential evolution \cite{storn_differential_1997} with a range of bounds for each parameter that was wider than those from the first step. For each substance, the parameter set with the smallest value from the objective function of the three described steps was selected. The bounds for each adjusted parameter in the first and third steps are shown in \autoref{tab:01}, with the COSMO volume being represented as \(v_{COSMO}\).

\begin{table}[]
    \centering
    \begin{tabular}{c|c|c|c|c}
        \hline
         \textbf{Adjusted} & \multicolumn{2}{c|}{\textbf{Nelder-Mead}} & \multicolumn{2}{c}{\textbf{Differential Evolution}} \\
        \cline{2-5}
           \textbf{Parameters} & \textbf{min} & \textbf{max} & \textbf{min} & \textbf{max} \\
        \hline
        \(b_i\) & 0.5 \(v_{COSMO}\) & 2 \(v_{COSMO}\) & 0.5 \(v_{COSMO}\) & 2 \(v_{COSMO}\) \\
        \(b_{h,i}\) & 2 & 0.1 \(v_{COSMO}\) + 20 & 8 & 0.1 \(v_{COSMO}\) + 20 \\
        \(\delta^{i,0}\) & 0.05 & 6 & 0.05 & 0.005 \(v_{COSMO}\) + 8 \\
        \(\delta^{i,T}\) & 0.05 & 30 & 0.05 & 30 \\
        \hline
    \end{tabular}
    \caption{openCOSMO-RS-Phi adjusted parameters boundaries for minimization methods.}
    \label{tab:01}
\end{table}

\subsection{Benchmarking database from Jaubert et al. \cite{jaubert_benchmark_2020}}
The assessment of openCOSMO-RS-Phi for binary mixtures is based on a high-quality reference database proposed by \textsc{Jaubert} and coworkers \cite{jaubert_benchmark_2020}. To cover a broad range of system types, the authors selected 107 nonelectrolytic pure compounds, forming a total of 200 binary mixtures. Its classification by association behavior and standardized evaluation procedure allow for a rigorous, systematic comparison of thermodynamic models over a broad range of mixtures, ensuring that predictive performance is assessed on physically meaningful and practically relevant systems.
Pure components are classified according to their association behavior (where "association" refers exclusively to hydrogen bonding) as follows:
\begin{itemize}
    \item Non-Associating (NA): neither a hydrogen atom capable of hydrogen bonding nor a lone pair available.
    \item Hydrogen-Donor (HD): a hydrogen atom capable of hydrogen bonding availablr but no lone pair of electrons.
    \item Hydrogen-Acceptor (HA): lone pair of electrons available but no hydrogen atom capable of hydrogen bonding.
    \item Self-Associating (SA): hydrogen atom capable of hydrogen bonding and a lone pair of electrons available.
\end{itemize}
Based on this classification, binary systems are divided into nine groups, each assigned a binary association code (BAC), as summarized in Table S1.

The dataset of the 200 binary systems contains low- and high-pressure phase-equilibrium data (VLE, LLE, VLLE, azeotropic, and critical points) as well as energetic data, namely enthalpy (\(h^\text{M}\)) and heat-capacity (\(c_\text{P}^\text{M}\)) changes upon mixing.

To quantify a model's accuracy against this database, the original publication also introduced a standard evaluation procedure. Subsequently, two modifications to this grading procedure were proposed by \textsc{Piña-Martinez et al.} \cite{pina2021optimal}. The first modification concerns the grading of three-phase-line data. The second introduces a Success Ratio (SR) into the grading scheme in order to account for out-of-model (OM) data points. OM points are points for which a MAPE cannot be evaluated because of an incorrect qualitative prediction.

For the exact methodology, the reader is referred to the original publications \cite{jaubert_benchmark_2020, pina2021optimal} or the Supporting Information. In brief:
\begin{enumerate}
    \item For each BAC, the MAPE is determined for each property type (10 at the most).
    \item The MAPE is then converted into a mark ranging from 0 and 20 for each BAC, with higher values indicating better model performance.
    \item After calculating the scores for each BAC and each property, three averaging steps are performed to obtain the final model score. First, a mark is assigned to each BAC by averaging the up to ten property scores for that BAC. By averaging the scores of some BACs, a mark is then calculated for each of in total four categories: non-associating (N-A: \(\text{BAC}_1\)-\(\text{BAC}_4\)), self-associating (S-A: \(\text{BAC}_5\)), cross-associating (C-A: \(\text{BAC}_6\)), and combined self- and cross-associating systems (S-A + C-A: \(\text{BAC}_7\)-\(\text{BAC}_9\)). Finally, the overall mark of the model is obtained as the average of the four category marks.
\end{enumerate}

The final mark of a thermodynamic model quantifies its predictive capability and is already available for several EoS, including the translated-consistent Peng-Robinson equation of state combined with different activity-coefficient models through advanced mixing rules \cite{pina-martinez_use_2021, paes2026using}, as well as PC-SAFT with and without binary interaction parameters (BIPs) \cite{nikolaidis2021assessment, nikolaidis2023effect}. 

In the present work, openCOSMO-RS-Phi is evaluated using zero BIPs and the modified grading procedure proposed by \textsc{Piña-Martinez et al.}; the model is assessed based on the resulting marks. It should be emphasized that, since no BIPs were regressed, the results are true predictions.

\subsection{Thermodynamic Calculations}\label{Thermodynamic Calculations}
This section briefly explains how the 10 properties of the \textsc{Jaubert} database are calculated.

\paragraph{Two-phase data}
VLE calculations are done by first generating the complete VLE region of the binary phase diagram using either an isothermal or isobaric, fast-converging bubble point algorithm consisting of two iterative loops \cite{elliott_introductory_2012}. This approach offers two main advantages. First, so-called \textit{out of model} points of type 1 according to \textsc{Jaubert} et al. \cite{jaubert_benchmark_2020}, i.e. a two-phase system is experimentally observed but model predicts one single phase, can be readily identified. Second, the procedure provides good initial estimates for the \(K\) values, which can subsequently be used in a \(PT\)-flash calculation to determine the vapor- and liquid-phase compositions at a specified experimental pressure and temperature. \\
A necessary condition for the occurrence of LLE is the existence of a local maximum in the plots of the activity of component 1, \(a_1\), and the activity of component 2, \(a_2\), versus mole fraction. Therefore, the \(a_i(x_1)\) curves were first calculated using openCOSMO-RS-Phi. If both curves exhibit a local maximum, an LLE \(PT\)-flash is carried out using the initialization scheme proposed by \textsc{Ohanomah} and \textsc{Thompson} \cite{ohanamah_computation_1984}. 

\paragraph{Three-phase line}
When only the temperature is specified,  the three-phase pressure and composition is calculated using the horizontal line method described by \textsc{Elliott} and \textsc{Lira} \cite{elliott_introductory_2012}. In this approach, a conventional bubble-point pressure algorithm for VLE is first applied. Odd loops and self-intersection of the dew curve that might then be observed are characteristic of a VLLE. A horizontal line through the intersection point of the dew curve then gives the three-phase line data. The complete binary phase diagram is obtained by overlapping the VLE diagram and the LLE diagram. If the method outlined above does not work, this overlap also provides an alternative means of determining the three-phase line, as the VLLE occurs at that pressure where the LLE curve intersects the bubble-point curve.

\paragraph{Critical and azeotropic points}
The calculation of the critical and azeotropic points is performed by systematically tracing the coexistence curves in the isothermal phase diagram until the difference in the mole fraction of component 1 between the two coexisting phases is smaller than 0.00015.

\paragraph{Mixing functions} 
As detailed in \cite{privat2012discussion} and \cite{qian2013enthalpy}, the mixing function of any intensive thermodynamic property is equal to the difference between the actual property \(d\) of the mixture (for \(P, T, \mathbf{z}\)) and the mole-fraction-weighted sum of the properties of the pure components considered in their stable state at equal \(P\) and \(T\). It is therefore necessary to determine the aggregation state of the mixture and the stable state of the pure components. To determine the phase state of the mixture, the bubble- and dew-point pressures at the given mixture composition \(\mathbf{z}\) and temperature \(T\) are calculated. If \(P_\text{dew} \leq P \leq P_\text{bubble}\), a subsequent flash calculation is performed to obtain the compositions of the two phases. If \( P \geq P_{\text{bubble}} \), the mixture exists entirely as a compressed (subcooled) liquid. In this liquid case, an additional check for liquid-liquid equilibrium (LLE) is performed.
If \( P \leq P_{\text{dew}} \), the mixture exists entirely as superheated vapor. \\
The identification of the stable state for the pure substance at the same \( T \), \( P \) is carried out in a similar way by comparing the pressure \( P \) with the vapor pressure of the pure component at temperature \( T \). \\
In the most general (i.e. multi-phase fluid) case, the mixing property \(d^\text{M}\) (here \(d \in \{h, c_\text{P} \}\)) at fixed temperature, pressure, and composition is obtained from a linear combination of the individual phases' mixing functions using the molar proportions \(\theta_k\) of phase \(k\) as weighting factors:
\begin{equation}
    d^\text{M} = \sum_k \theta_k \cdot d_k^\text{M}(T,P,\mathbf{x}^{(k)})
    \label{eq:46}
\end{equation}
In \autoref{eq:46}, \(\mathbf{x}^{(k)}\) denotes the composition vector of phase \(k\). \\
According to \textsc{Qian} et al. \cite{qian2013enthalpy}, a mixing function at constant pressure is connected to the residual properties as follows:
\begin{equation}
    d^\text{M}(T,P,\mathbf{z})=d^\text{r}(T,v,\mathbf{z})-\sum_i^N z_i \cdot d_{\text{pure }i}^\text{r,stable state}(T,v_{\text{pure }i})
    \label{eq:47}
\end{equation}
where \(v_{\text{pure }i}\) is the molar volume of pure compound \(i\) at the same temperature \(T\) and the same pressure \(P\) as the mixture. \\
Residual molar enthalpy, \(h^\text{R}\), can be calculated by using the following thermodynamic relation:
\begin{equation}
    h^\text{r}(T,v,\mathbf{z})=a^\text{r}(T,v,\mathbf{z}) - T \left(\frac{\partial a^\text{r}}{\partial T} \right)_{v,\mathbf{z}} + P(T,v,\mathbf{z})v-RT
    \label{eq:48}
\end{equation}
The residual heat capacity at constant pressure, \(c_\text{P}^\text{r}\), is given by the following expression:
\begin{equation}
    c_\text{P}^\text{r}(T,P,\mathbf{z}) =  - T\left(\frac{\partial^2 a^\text{r}}{\partial T^2} \right)_{v,\mathbf{z}} - R -T \left( \frac{\left[\left( \frac{\partial P}{\partial T}\right)_{v,\mathbf{z}} \right]^2}{\left(\frac{\partial P}{\partial v} \right)_{T,\mathbf{z}}} \right)
    \label{eq:49}
\end{equation}
As can be seen from \autoref{eq:47} and \autoref{eq:48}, \(h^\text{r}\) and \(c_\text{P}^\text{r}\) are derivative properties and the derivates with respect to temperature present in these equations include terms accounting the temperature dependence of the segment activity coefficients \(\Gamma\). It follows that both the first and second partial derivative of the COSMOSPACE equation with respect to temperature are in \(\left(\frac{\partial \Gamma}{\partial T} \right)_{v,\mathbf{z}}\), or \(\left(\frac{\partial^2 \Gamma}{\partial T^2} \right)_{v,\mathbf{z}}\), respectively and the exact same linear equation system as obtained by \textsc{Yan} \cite{yan2024robust} through an optimization approach has to be solved. It is worth emphasizing that all derivatives in \autoref{eq:47} and \autoref{eq:48} can be calculated analytically. In summary, an accurate estimation of mixing properties requires good description of both the two-phase split (i.e., the phase diagram) and the temperature dependence of fugacity coefficients and phase equilibrium. 

\hypertarget{results}{
\section{Results and Discussions}\label{Results and Discussions}}

\subsection{Pure Components}\label{Pure Components Results}

\subsubsection{Comparison with COSMO-SAC-Phi}

\begin{figure}[!b]
  \subcaptionbox*{\centering } [.5\linewidth]{%
    \includegraphics[width=\linewidth, clip]{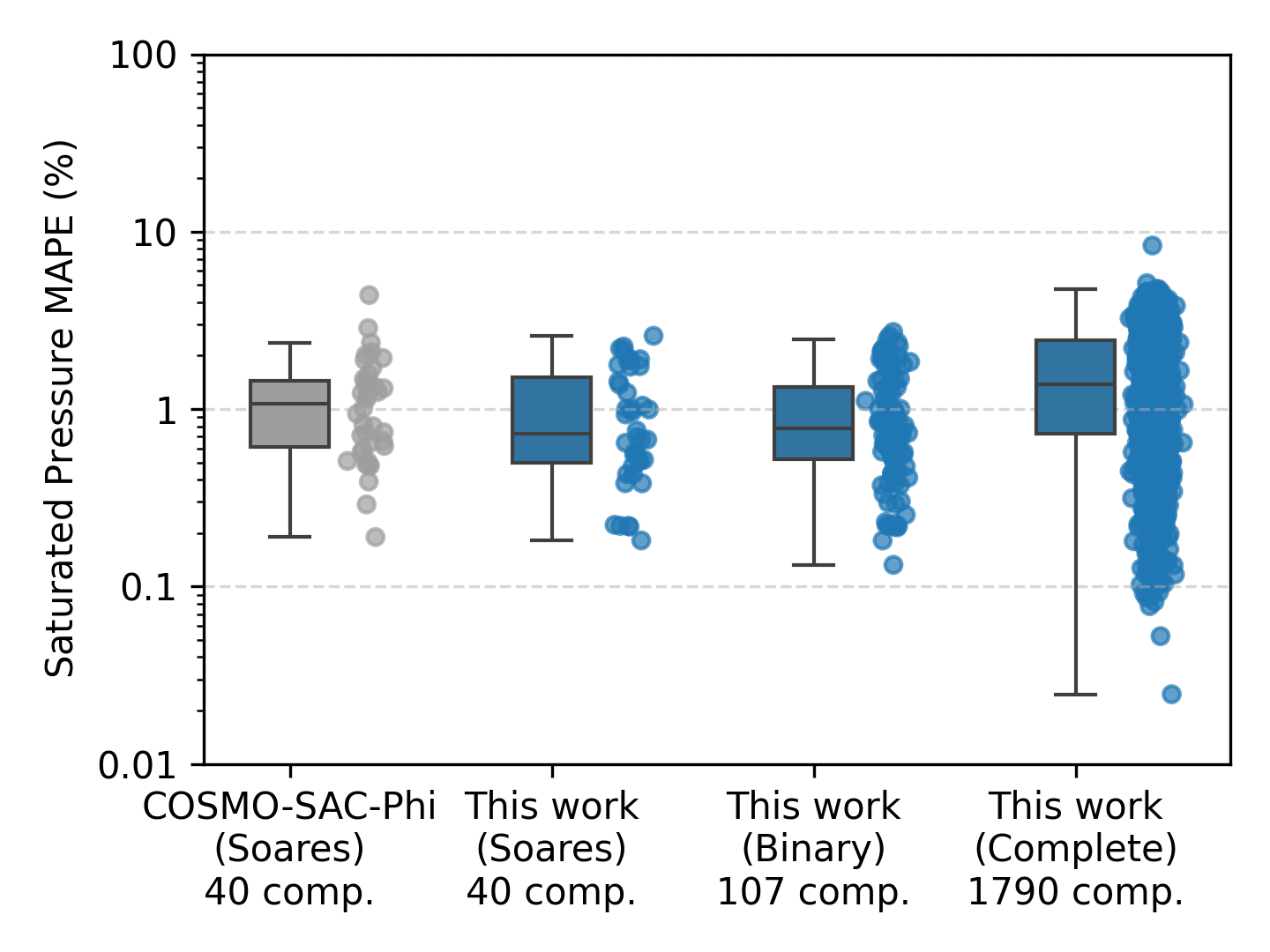}%
  }%
  \hfill
  \subcaptionbox*{\centering }[.5\linewidth]{%
    \includegraphics[width=\linewidth, clip]{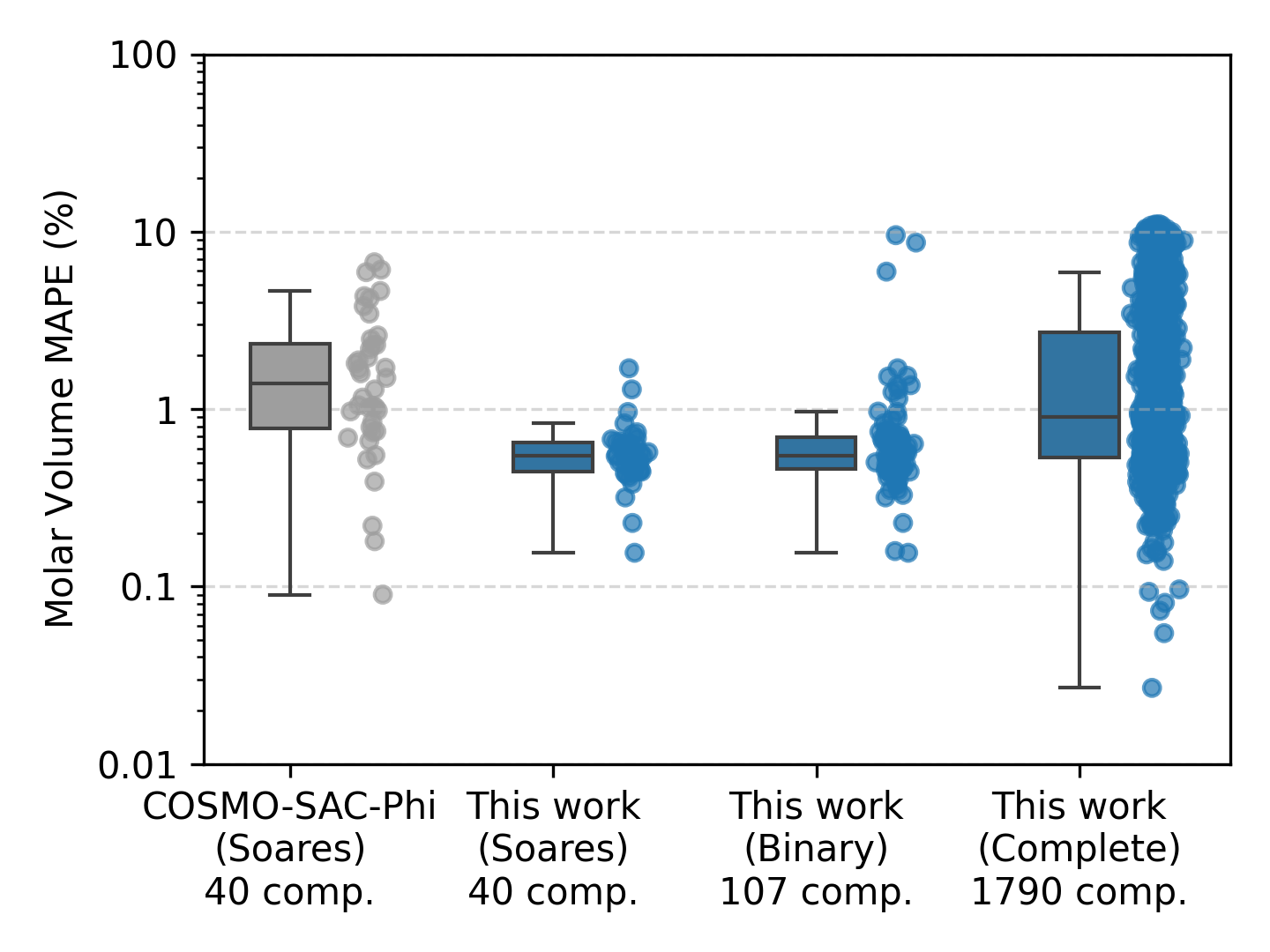}%
  }%
\caption{Comparison of the resulting MAPE of vapor pressure and molar volume from openCOSMO-RS-Phi across different sets of components and COSMO-SAC-Phi.}
\label{fig:pure_comp_MAPE_Soares}
\end{figure}

The performance is first assessed against the original COSMO-SAC-Phi publication.
As shown in \autoref{fig:pure_comp_MAPE_Soares}, comparison for the 40 components reported in the original publication indicates that COSMO-SAC-Phi and openCOSMO-RS-Phi yield similar results for vapor-pressure prediction, whereas the latter provides a substantial improvement in molar-volume calculations. Three molecular datasets are presented in this figure: 'Soares' refers to the set of molecules for which parameters are available for COSMO-SAC-Phi \cite{de_p_soares_pairwise_2019}; 'Binary' denotes the set of compounds used in the binary evaluation proposed by \textsc{Jaubert et al.} \cite{jaubert_benchmark_2020}, which is discussed further below; and 'Complete' corresponds to the full set of molecules parameterized in the present study, also shown in \autoref{fig:pure_comp_MAPE_EoS}. The full set of molecules appears twice: first, to enable comparison of the different datasets within openCOSMO-RS-Phi, and later, to facilitate comparison with other EoS.

\begin{figure}[!t]
  \centering
  \subcaptionbox*{\centering (a)}[.49\linewidth]{%
    \includegraphics[
      width=\linewidth,
      height=5cm,
      keepaspectratio
    ]{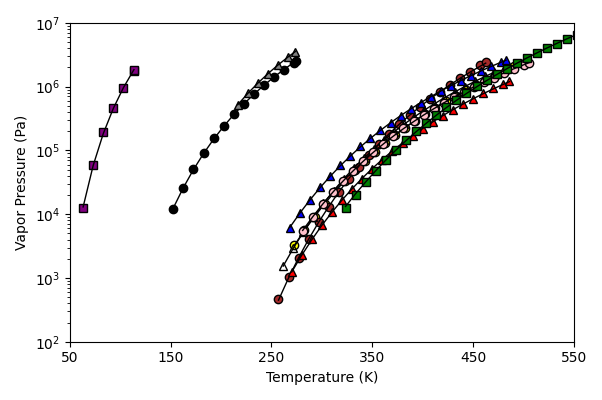}%
  }%
  \hfill
  \subcaptionbox*{\centering (b)}[.49\linewidth]{%
    \includegraphics[
      width=\linewidth,
      height=5cm,
      keepaspectratio
    ]{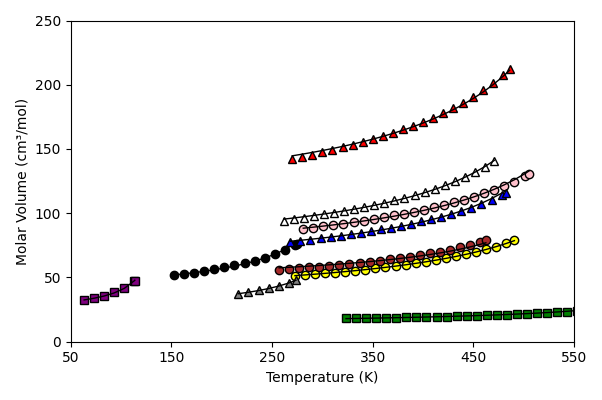}%
  }%

  \caption{Vapor pressure (a) and molar volume (b) for 10 exemplary components.}
  \label{fig:pure_diagrams}
\end{figure}

%\begin{figure}
%    \centering
%    \includegraphics[width=1\textwidth]{Figures/pure_cmp_diagrams.png}
%    \caption{Vapor Pressure versus temperature diagram and molar volume versus temperature for 10 exemplary components.}
%    \label{fig:pure_diagrams}
%\end{figure}
To further illustrate model performance, ten of the 107 compounds in the 'Binary' dataset from \autoref{fig:pure_comp_MAPE_Soares} were selected to present the calculated vapor pressure and liquid molar volume. This selection covers pairings across all nine BACs analyzed later, while representing the largest possible number of binary systems (10.5\%) with this limited number of components. A visual comparison between experimental data and calculated results is shown in \autoref{fig:pure_diagrams}. The close agreement indicates the robustness and validity of the new method.

\clearpage

\renewcommand{\arraystretch}{0.8}
\begin{table}[!htbp]
    \centering
    \resizebox{\textwidth}{!}{%
    \begin{tabular}{|l|cccc|cc|cc|cc|}
    \hline
        \multirow{3}{*}{Component} & \multicolumn{4}{c|}{openCOSMO-RS-Phi} & \multicolumn{2}{c|}{openCOSMO-RS-Phi} & \multicolumn{2}{c|}{improved CSP (2021)} & \multicolumn{2}{c|}{CSP (2019)} \\ \cline{2-11}
        ~ & $b_i$ & $\delta_m^0$ & $\delta_m^t$ & $b_{h,i}$ & $\text{MAPE}_P$ & $\text{MAPE}_v$ & $\text{MAPE}_P$ & $\text{MAPE}_v$ & $\text{MAPE}_P$ & $\text{MAPE}_v$ \\
        ~ & \AA $^3$ & $^\frac{\text{kcal}}{\text{mol}}$ & $10^2$ K & \AA $^3$ & \% & \% & \% & \% & \% & \% \\ \hline
        Methane & 50.81 & 0.594 & 3.186 & 10.68 & 2.12 & 0.68 & 0.28 & 0.42 & 1.94 & 1.29 \\ 
        Ethane & 73.49 & 0.803 & 3.549 & 12.68 & 1.78 & 0.67 & 0.47 & 0.72 & 1.48 & 2.30 \\ 
        Propane & 99.34 & 0.845 & 3.572 & 14.66 & 1.24 & 0.63 & 0.33 & 0.49 & 1.01 & 2.29 \\ 
        Butane & 126.31 & 0.860 & 3.657 & 15.66 & 0.67 & 0.54 & 0.18 & 0.43 & 0.51 & 1.16 \\ 
        Pentane & 153.60 & 0.891 & 3.529 & 16.80 & 0.38 & 0.44 & 0.22 & 0.47 & 0.29 & 0.52 \\ 
        2-Methylpentane & 181.83 & 0.887 & 3.444 & 17.75 & 0.22 & 0.45 & 2.24 & 1.66 & 2.10 & 4.33 \\ 
        Hexane & 180.60 & 0.915 & 3.365 & 17.71 & 0.65 & 0.41 & 0.42 & 0.26 & 0.49 & 0.97 \\ 
        Octane & 238.62 & 0.993 & 2.932 & 18.76 & 1.43 & 0.96 & 0.38 & 0.12 & 1.57 & 0.22 \\ 
        Decane & 301.48 & 1.099 & 2.545 & 19.45 & 2.25 & 1.70 & 1.13 & 0.49 & 1.21 & 5.91 \\ 
        Cyclohexane & 153.59 & 1.034 & 4.423 & 16.98 & 0.76 & 0.51 & 0.27 & 0.26 & 0.57 & 2.48 \\ 
        Ethene & 65.57 & 0.667 & 3.960 & 11.95 & 1.75 & 0.66 & 0.14 & 0.40 & 1.40 & 1.02 \\ 
        1-Butene & 117.13 & 0.865 & 3.534 & 15.36 & 1.37 & 0.65 & 0.41 & 0.50 & 0.51 & 0.55 \\ 
        1-Hexene & 174.28 & 0.916 & 3.357 & 16.77 & 0.52 & 0.53 & 1.83 & 0.79 & 1.89 & 1.68 \\ 
        Cyclohexene & 143.65 & 1.046 & 4.596 & 16.70 & 0.93 & 0.54 & ~ & ~ & 1.33 & 0.66 \\ 
        Benzene & 126.16 & 1.083 & 4.782 & 15.88 & 1.00 & 0.55 & 0.38 & 0.36 & 0.59 & 1.81 \\ 
        Pyridine & 117.23 & 1.203 & 4.998 & 16.09 & 0.51 & 0.57 & 0.71 & 2.43 & 0.71 & 1.50 \\ 
        Toluene & 154.48 & 1.069 & 4.498 & 17.01 & 0.22 & 0.45 & 1.19 & 0.54 & 1.23 & 0.39 \\ 
        Acetone & 98.69 & 1.048 & 3.982 & 15.87 & 0.22 & 0.44 & 0.70 & 0.73 & 0.80 & 0.85 \\ 
        2-Butanone & 124.25 & 1.039 & 3.820 & 16.97 & 0.52 & 0.44 & ~ & ~ & 0.62 & 0.98 \\ 
        Dimethyl ether & 82.90 & 0.914 & 3.787 & 13.29 & 0.99 & 0.62 & 0.25 & 0.18 & 0.64 & 2.60 \\ 
        Ethyl ether & 137.91 & 0.926 & 3.276 & 16.19 & 0.56 & 0.48 & 0.30 & 0.55 & 0.39 & 0.79 \\ 
        Tetrahydrofuran & 113.87 & 1.113 & 4.502 & 15.12 & 1.01 & 0.56 & 1.12 & 1.19 & 1.25 & 3.80 \\ 
        Methyl acetate & 108.56 & 0.996 & 3.595 & 14.82 & 0.43 & 0.43 & 0.39 & 0.13 & 0.48 & 0.09 \\ 
        Ethyl acetate & 137.42 & 1.017 & 3.146 & 15.98 & 0.38 & 0.45 & 0.41 & 0.86 & 0.48 & 1.71 \\ 
        Trichloromethane & 111.60 & 1.062 & 4.491 & 14.18 & 0.97 & 0.54 & 1.11 & 0.87 & 1.31 & 0.74 \\ 
        Carbon tetrachloride & 136.62 & 0.982 & 4.730 & 15.09 & 1.05 & 0.54 & 0.27 & 0.28 & 0.66 & 1.95 \\ 
        Methanol & 56.09 & 1.351 & 3.325 & 8.71 & 0.57 & 0.38 & 0.55 & 0.24 & 0.82 & 3.45 \\ 
        Ethanol & 85.74 & 1.541 & 2.258 & 9.14 & 1.73 & 1.29 & 0.81 & 0.17 & 0.74 & 4.63 \\ 
        1-Butanol & 136.21 & 2.204 & 1.253 & 12.73 & 0.67 & 0.50 & 1.70 & 0.33 & 1.34 & 4.20 \\ 
        1-Pentanol & 160.76 & 1.817 & 1.525 & 14.19 & 0.43 & 0.45 & 2.93 & 0.52 & 2.37 & 0.75 \\ 
        Methylamine & 57.96 & 1.021 & 3.967 & 10.44 & 1.88 & 0.62 & 1.30 & 0.33 & 1.34 & 6.10 \\ 
        N,N-Diethylethanamine & 199.01 & 0.904 & 3.464 & 17.55 & 0.70 & 0.57 & 1.86 & 0.07 & 1.14 & 2.18 \\
        Dimethylformamide & 111.18 & 1.304 & 4.690 & 17.91 & 0.22 & 0.32 & 4.45 & 0.82 & 4.39 & 1.02 \\ 
        Acetic acid & 80.19 & 0.768 & 6.236 & 11.84 & 0.55 & 0.23 & 2.57 & 6.40 & 2.02 & 6.72 \\ 
        Nitrogen & 45.72 & 0.403 & 1.994 & 9.76 & 2.59 & 0.83 & 0.41 & 0.76 & 1.71 & 1.04 \\ 
        Carbon dioxide & 46.92 & 0.886 & 3.331 & 9.21 & 0.18 & 0.15 & 0.11 & 0.16 & 0.19 & 0.18 \\ 
        Hydrogen sulfide & 48.42 & 0.837 & 6.062 & 9.74 & 1.91 & 0.65 & 0.21 & 0.40 & 1.42 & 1.59 \\ 
        Ammonia & 31.30 & 0.799 & 5.308 & 6.90 & 2.20 & 0.74 & 2.62 & 0.67 & 2.86 & 0.69 \\ 
        Dimethyl sulfoxide & 107.43 & 1.451 & 4.859 & 16.84 & 0.48 & 0.61 & 1.21 & 1.62 & 0.94 & 1.88 \\ 
        Water & 24.11 & 0.716 & 5.087 & 5.17 & 1.91 & 0.72 & 0.79 & 0.27 & 0.73 & 1.05 \\ \hline
        \multicolumn{5}{|r|}{average}  & 1.00 & 0.59 & 0.96 & 0.73 & 1.19 & 1.95 \\
        \multicolumn{5}{|r|}{standard deviation}  & 0.56 & 0.16 & 0.72 & 0.53 & 0.58 & 1.28 \\ \hline
    \end{tabular}
    }
    \caption{Comparison between MAPE from openCOSMO-RS-Phi, improved COSMO-SAC-Phi, and COSMO-SAC-Phi, for a few selection of components.}
    \label{tab:03}
\end{table}
\clearpage

The corresponding numerical results are summarized in \autoref{tab:03}, where COSMO-SAC-Phi is denoted as 'CSP (2019)'. The table also includes a comparison with the improved version of COSMO-SAC-Phi containing one additional parameter, denoted as 'improved CSP (2021)' \cite{zini_improved_2021}. For the selected compounds, the new model outperforms the improved version without requiring a fifth adjustable parameter. It gives lower average and maximum errors in molar volume, lower values for both MAPE metrics, and greater consistency, as reflected by the smaller standard deviation. A complete table containing all pure components parameterized in this study is provided in the Supporting Information.

\subsubsection{Derivative properties and critical points description}
The preceding section presented a detailed assessment of the ability of openCOSMO-RS-Phi to reproduce liquid densities and saturated vapor pressures, i.e., the two properties used for the regression of the pure-component parameters. To provide a more comprehensive evaluation of the equation of state's performance, the accuracy for the molar enthalpy of vaporization (\(\Delta h_i^\text{LV}\), liquid heat capacity at constant pressure evaluated at saturation (\(c_{\text{P,liq}}^{sat}\), critical temperature (\(T_c\)) and critical pressure (\(P_c\)) is examined in the following. The analysis is restricted to the 107 components included in the benchmark dataset of \textsc{Jaubert et al.} for mixtures.

\begin{figure}[!htbp]
  \subcaptionbox*{\centering (a)}[.33\linewidth]{%
    \includegraphics[width=\linewidth,height=7.1cm,keepaspectratio]{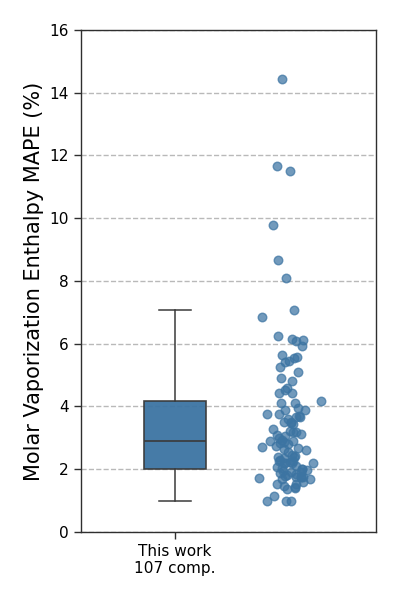}%
  }%
  \hfill
  \subcaptionbox*{\centering (b)}[.67\linewidth]{%
    \includegraphics[width=\linewidth,height=6.9cm,keepaspectratio]{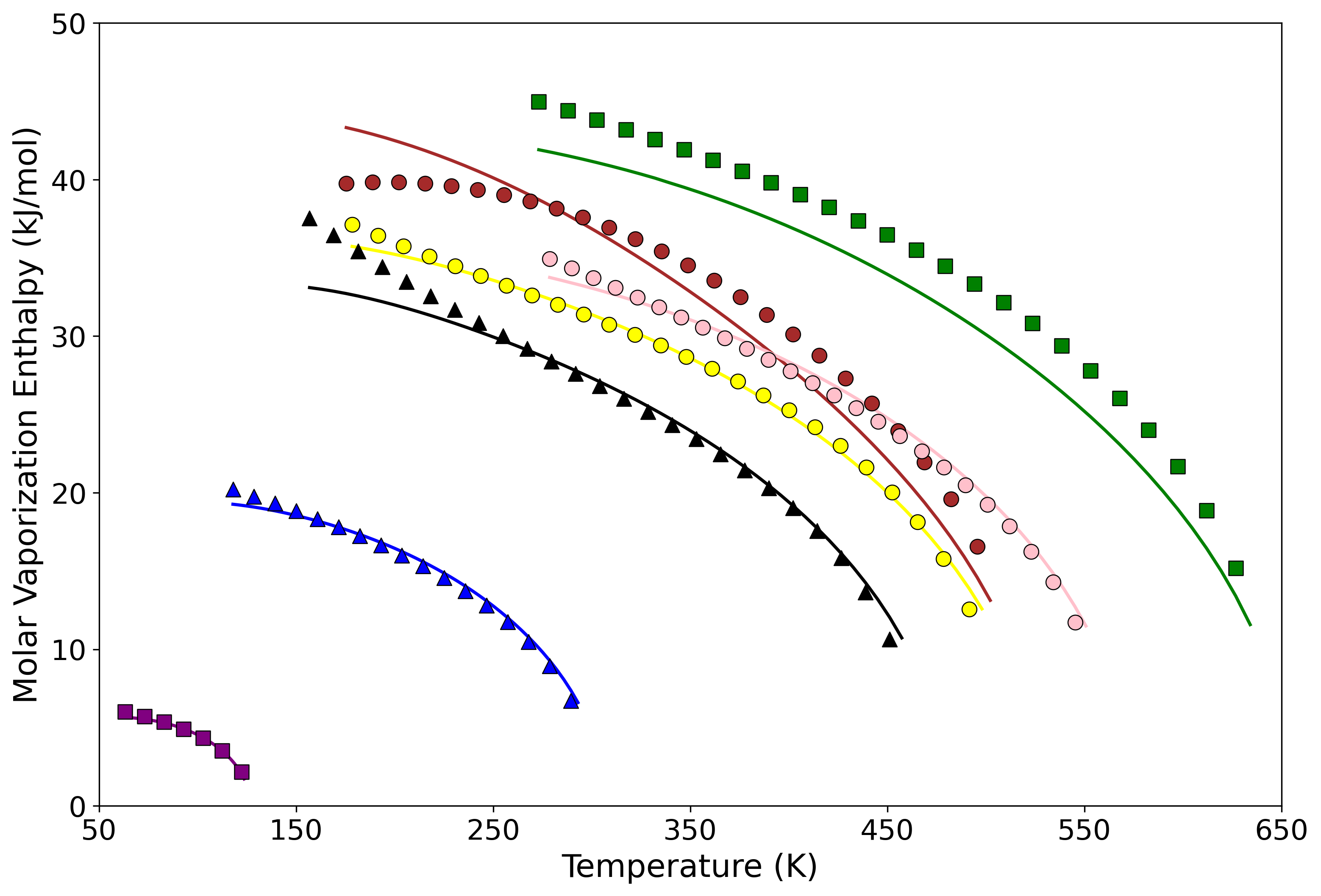}%
  }%
\caption{(a) Resulting MAPE for the molar enthalpy change of vaporization, evaluated across the 107 components of the benchmarking database. Acetic acid with a MAPE \(> 50\%\) is excluded from the scatter plot. (b) Temperature dependence of the vaporization enthalpy for seven selected components: {\color[HTML]{ffff00} $\bullet$} Acetone, {\color[HTML]{a52a2a} $\bullet$} Methanol, {$\blacktriangle$} Diethyl ether, {\color[HTML]{ffc0cb} $\bullet$} Benzene,  {\color[HTML]{0000fe} $\blacktriangle$} Trifluoromethane,  {\color[HTML]{008000} $\blacksquare$} Water, {\color[HTML]{800080} $\blacksquare$} Nitrogen. Points are DIPPR data, and lines are values given using openCOSMO-RS-Phi. For the sake of clarity, not all 50 DIPPR points used are shown.}
\label{fig:hvap_results}
\end{figure}

\paragraph{Pure-component vaporization enthalpies} 
The enthalpy change on vaporization is identical to the difference between the molar residual enthalpy in the vapor (\(h_\text{vap}^\text{r}\)) and liquid (\(h_\text{liq}^\text{r}\)) phase of the pure component \(i\) in vapor-liquid equilibrium: \(\Delta h_i^\text{LV}(T,P_i^{sat})=h_{\text{vap}, i}^\text{r}(T,P_i^{sat}) - h_{\text{liq}, i}^\text{r}(T,P_i^{sat})\). \(\Delta h_i^\text{LV}(T,P_i^{sat})\) denotes the molar enthalpy of vaporization of pure \(i\) at temperature \(T\). One relevance of enthalpies of vaporization of pure components lies in the fact that, together with an accurate description of VLE's, they determine the accuracy of an EoS in predicting mixing enthalpies as eludicated in \cite{qian2013enthalpy}.
Similar to the procedure described in \cite{ramirez2020parameterization}, a total of 50 enthalpies of vaporization were calculated as reference data for each of the 107 components using the DIPPR equation 106 (correlation parameters for all components are included in the DIPPR database 801) by decomposing the temperature interval \([T_\text{min}^\text{DIPPR}, \min \{T_\text{max}^\text{DIPPR}; 0.98 \cdot T_\text{c,exp} \}]\) equidistantly. This results in a total of 5350 data points. Subsequently, \(\Delta h_i^\text{LV}\) was calculated for the same temperatures using the EoS, and a MAPE was computed for each component. The resulting MAPE distribution is shown in \autoref{fig:hvap_results}. Across all 107 components, a mean of 3.97\(\%\) and a median of 2.90\(\%\) are obtained. Removing acetic acid from the assessment, for which an unusually high MAPE of about 50\(\%\) is achieved, reduces the average to 3.53\(\%\). Direct comparative values for the selected dataset are not available. For a larger number of components (1536), values of 1.92\(\%\) for tc-PR, 3.25\(\%\) for PC-SAFT, and 4.00\(\%\) for I-PC-SAFT are reported in \cite{pina-martinez_use_2021}. 

\begin{figure}[!htbp]
  \subcaptionbox*{\centering (a)}[.33\linewidth]{%
    \includegraphics[width=\linewidth,height=7.1cm,keepaspectratio]{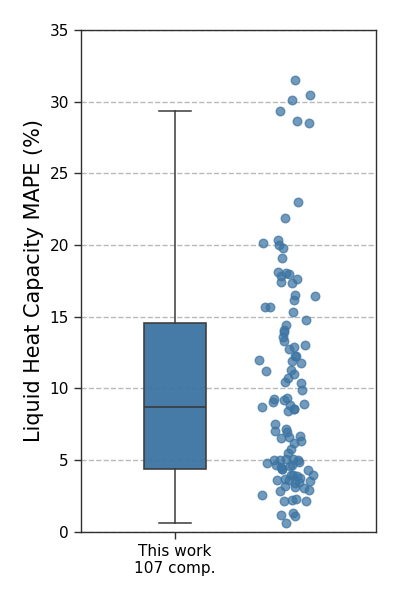}%
  }%
  \hfill
  \subcaptionbox*{\centering (b)}[.67\linewidth]{%
    \includegraphics[width=\linewidth,height=6.9cm,keepaspectratio]{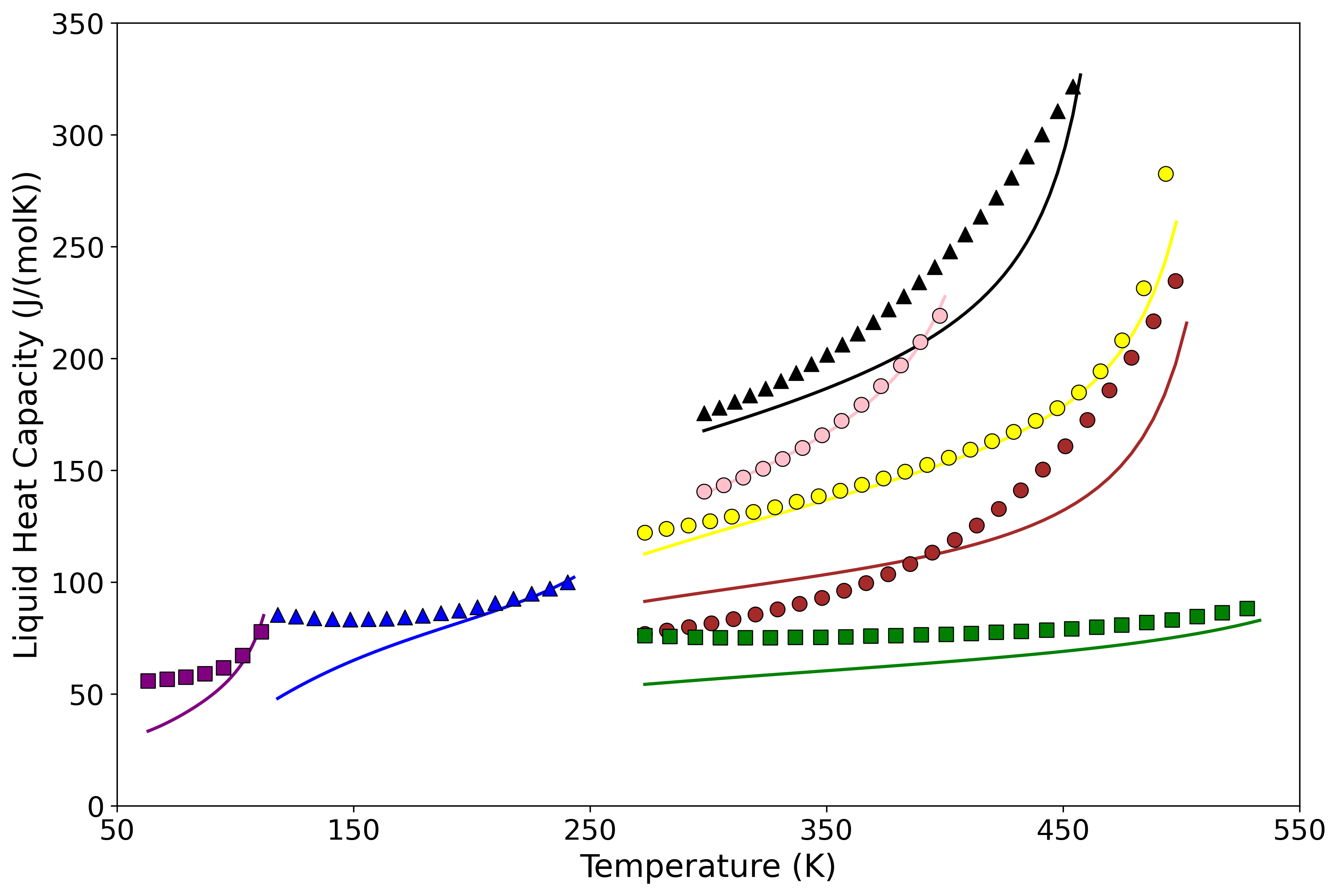}%
  }%
\caption{(a) Resulting MAPE for isobaric molar heat capacities of liquids at saturation, \(c_\text{P,liq}^\text{sat}\), evaluated across 107 components. The six compounds exhibiting the largest prediction errors, all with MAPE values of approximately \(30\%\), are (in descending order of MAPE): carbon tetrafluoride, xenon, chlorodifluoromethane, dichloromethane, ethane and cyclopentane. (b) Temperature dependence of the molar liquid heat capacity for seven selected components: {\color[HTML]{ffff00} $\bullet$} Acetone, {\color[HTML]{a52a2a} $\bullet$} Methanol, {$\blacktriangle$} Diethyl ether, {\color[HTML]{ffc0cb} $\bullet$} n-Butane,  {\color[HTML]{0000fe} $\blacktriangle$} Trifluoromethane,  {\color[HTML]{008000} $\blacksquare$} Water, {\color[HTML]{800080} $\blacksquare$} Nitrogen.  Points are DIPPR data, and lines are values given using openCOSMO-RS-Phi. For the sake of clarity, not all 50 DIPPR points used are shown.}
\label{fig:cpLsat_results}
\end{figure}

\paragraph{Liquid isobaric molar heat capacities at saturation} The liquid isobaric molar heat capacity at saturation, \(c_\text{P,liq}^{sat}\) is the sum of an ideal-gas and residual contribution: \(c_\text{P,liq}^\text{sat}(T,P^{sat}) = c_\text{P}^\text{IG}(T) + c_\text{P,liq}^\text{r}(T,P^{sat})\).
Since an EoS only predicts \(c_\text{P,liq}^\text{r}\), an accurate temperature-dependent model for the ideal gas molar heat capacity has to be coupled with the EoS. Here, a DIPPR correlation for ideal-gas heat capacity was employed. As for the evaluation of \(\Delta h_i^\text{LV}\), for each of the 107 components, \(c_\text{P,liq}^{sat}\) was calculated for 50 temperature points on the temperature domain \([\max\{T_{\text{min,}c_\text{P,IG}}^\text{DIPPR};T_{\text{min,}c_\text{P,liq}}^\text{DIPPR}\}, \min\{\min\{T_{\text{max,}c_\text{P,IG}}^\text{DIPPR};T_{\text{max,}c_\text{P,liq}}^\text{DIPPR}\}; 0.98 \cdot T_c\}\) and compared with the corresponding values obtained from the DIPPR correlations. The MAPE was again used as the performance metric. The resulting MAPE values are shown in \autoref{fig:cpLsat_results}a. The MAPE ranges from \(0.60\%\) for butane to \(31.48\%\) for carbon tetrafluoride, which is comparable to the spread reported for classical and ML-based tc-PR variants \cite{paes2026prediction}, although the present evaluation is based on a smaller dataset. The mean and median MAPE values are \(10.3\%\) and \(8.7\%\), respectively. \autoref{fig:cpLsat_results}b compares the predicted and reference saturated liquid molar heat capacities for seven selected real fluids. Finally, it should be pointed out that it is well-known that the obtained MAPE value depends on the chosen fitting strategy, particularly on whether energetic properties are included in the parameter regression, as \cite{ramirez2020parameterization} shows using PC-SAFT as an example.

\paragraph{Critical Properties}

\begin{figure}[!b]
    \centering
    \includegraphics[width=1\textwidth]{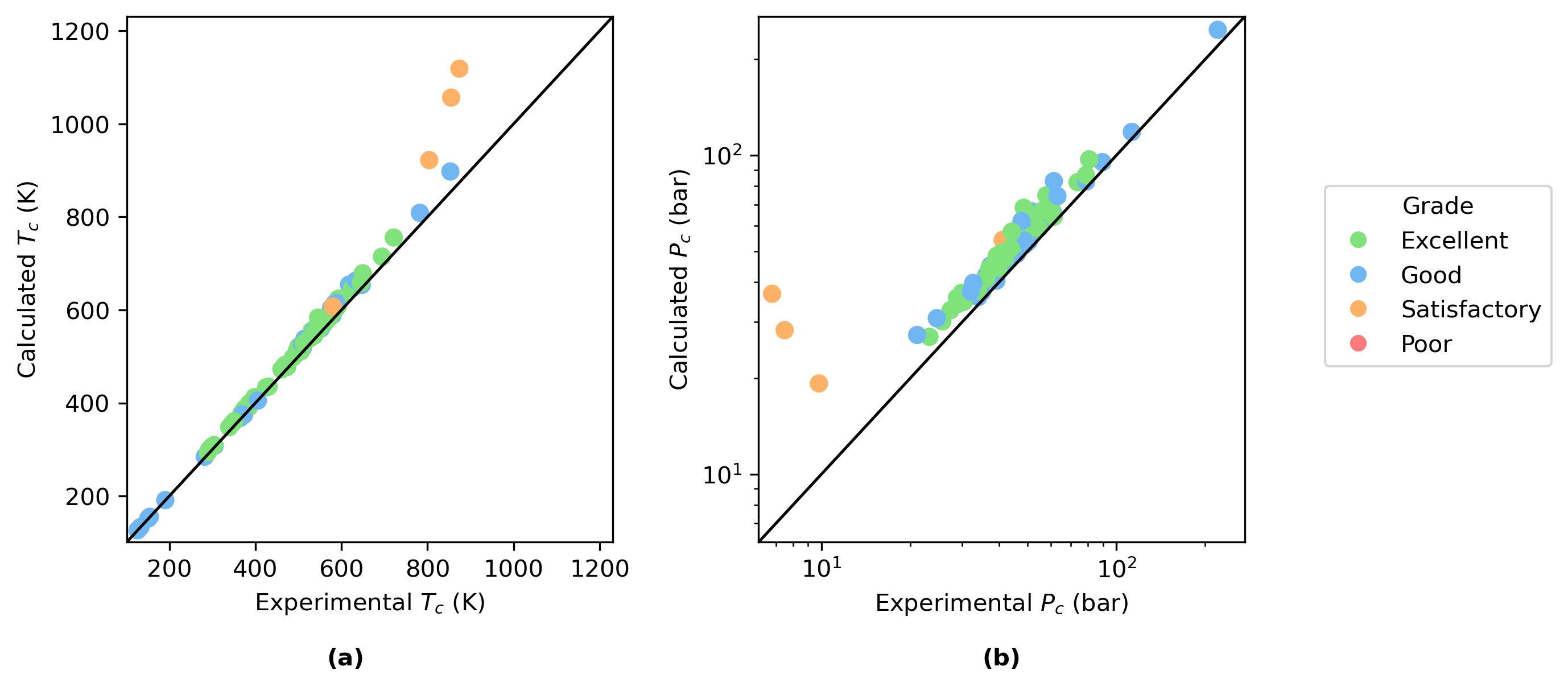}
    \caption{A parity plot of (a) the predicted critical temperature (K) from openCOSMO-RS-Phi compared to experimental critical temperature (K) obtained from DIPPR on a log-log scale and (b) the predicted critical pressure (bar) from openCOSMO-RS-Phi compared to experimental pressure temperature (bar) obtained from DIPPR. The three clearly visible outliers refer to hexatriacontane, n-dotriacontane, and n-tetracosane.}
    \label{Fig:Parity plots PC critical data}
\end{figure}

%The predictions of EoS are generally sensitive to the accuracy of the critical properties, particularly the critical temperature. Recently, \textsc{Pinegar et al.} demonstrated this relationship for vapor-pressure predictions using EoS \cite{pinegar2026accuracy}. <- this is for cubics.
To assess the performance of openCOSMO-RS-Phi in critical-point calculations, \autoref{Fig:Parity plots PC critical data} shows parity plots comparing predicted critical temperatures (\(T_c\)) and critical pressures (\(P_c\)) with experimental DIPPR data for all pure components included in the database of \textsc{Jaubert et al.} \\

The resulting MAPE values are 3.32~\% for \(T_c\) and 22.98~\% for \(P_c\), indicating substantially larger deviations for critical pressure. The largest errors in both properties are found for the long-chain alkanes \(n\)-hexatriacontane, \(n\)-dotriacontane, and \(n\)-tetracosane, which appear as clear outliers in \autoref{Fig:Parity plots PC critical data}. Overall, \(T_c\) is overestimated for most compounds, with ammonia as the only exception, whereas \(P_c\) is systematically overestimated for all components.

A similar trend has been reported for other EoS \cite{wilhelmsen2017thermodynamic}. For example, PC-SAFT and CPA have been reparameterized to improve agreement with \(T_c\) and \(P_c\) \cite{cismondi2005rescaling, palma2017re}, and it has also been shown that fitting PR parameters to vapor-pressure and density data, rather than deriving them from critical properties, leads to poorer reproduction of critical points \cite{voutsas2006vapor}.
\renewcommand{\arraystretch}{1.0}
\subsubsection{Comparison with other EoS}

\begin{figure}[!t]
    \centering
    \includegraphics[
        width=\textwidth,
        trim=0cm 2.5cm 0cm 3.5cm,
        clip
    ]{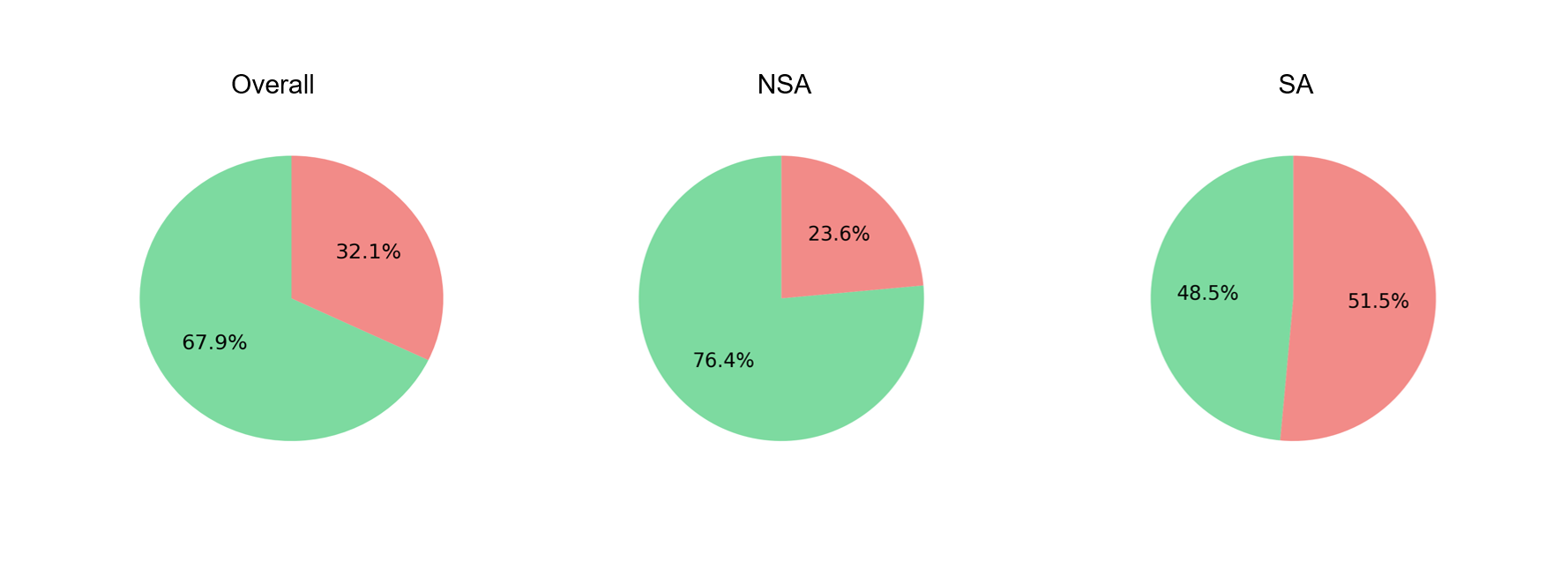}
    \caption{Ratio of {\color[HTML]{7ddaa0} $\bullet$} well- and {\color[HTML]{f28b88} $\bullet$} badly-modeled molecules for the entire dataset, non-self-associating (NSA), and self-associating (SA) fluids.}
    \label{fig:WB_pie_chart}
\end{figure}

The performance of openCOSMO-RS-Phi is further evaluated using the same analysis previously applied to other EoS \cite{pina-martinez_use_2021, ramirez-velez_assessing_2022}. As shown in \autoref{fig:WB_pie_chart}, the assessment considers both the complete dataset and its subdivision into non-self-associating (NSA) and self-associating (SA) components. Consistent with the studies of \textsc{Piña-Martinez et al.} and \textsc{Ramírez-Vélez et al.}, openCOSMO-RS-Phi shows lower internal performance for SA fluids than for NSA fluids.

However, the ratio of well-modeled to badly modeled compounds should not be interpreted directly as a measure of relative model performance. This classification is inherently model dependent and reflects only the internal consistency of a given EoS. Its value for comparisons between different models is therefore limited and should be interpreted with caution. A more detailed distribution of the MAPE for each property is provided in Appendix A3 of the Supporting Information.

\begin{figure}[!b]
  \subcaptionbox*{\centering } [.5\linewidth]{%
    \includegraphics[width=\linewidth, clip]{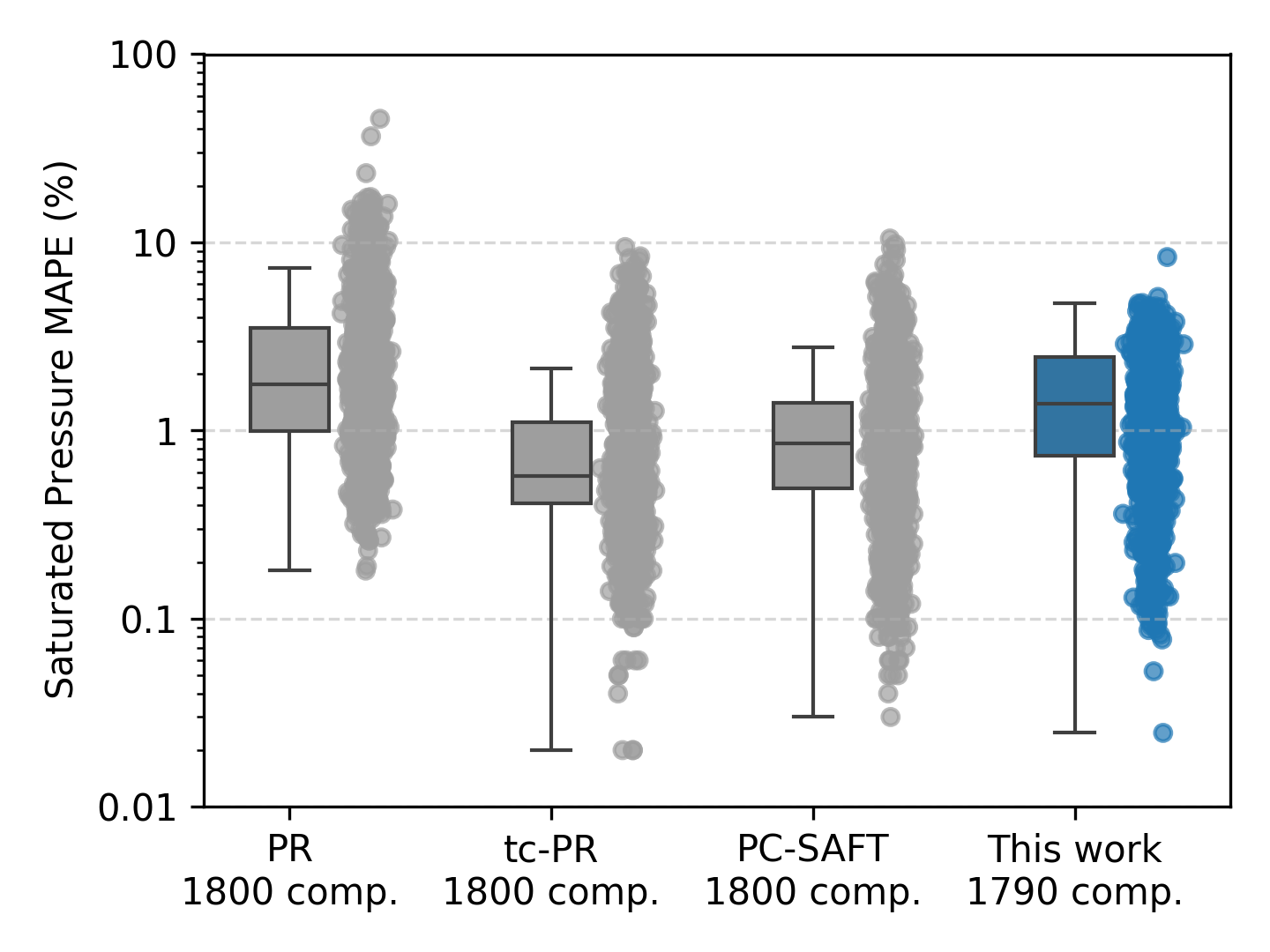}%
  }%
  \hfill
  \subcaptionbox*{\centering }[.5\linewidth]{%
    \includegraphics[width=\linewidth, clip]{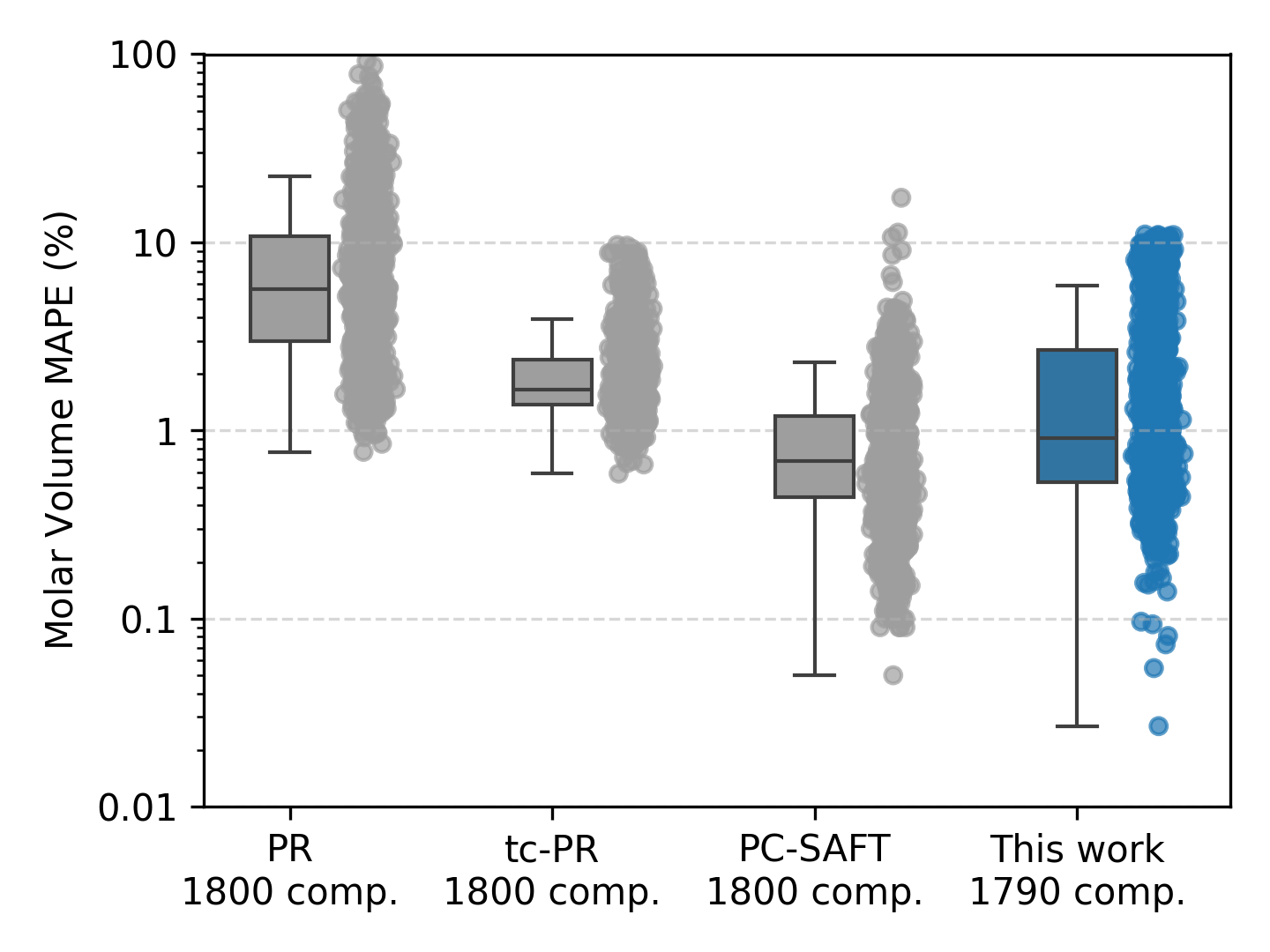}%
  }%
\caption{Comparison of the resulting MAPE of vapor pressure and molar volume from all components studied across different EoS.}
\label{fig:pure_comp_MAPE_EoS}
\end{figure}

The MAPE values for vapor pressure and molar volume across all components are presented in \autoref{fig:pure_comp_MAPE_EoS}, allowing a direct comparison of the overall performance of openCOSMO-RS-Phi with that of other EoS. For vapor pressure, all models show comparable behavior and generally satisfactory accuracy. For molar volume, in contrast, clearer differences between the models are observed.

\begin{figure}[!t]
  \subcaptionbox*{\centering (a)}[.5\linewidth]{%
    \includegraphics[width=\linewidth]{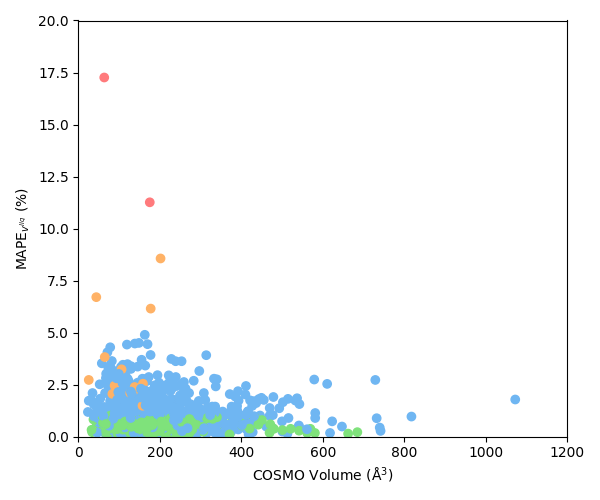}%
  }%
  \hfill
  \subcaptionbox*{\centering (b)}[.5\linewidth]{%
    \includegraphics[width=\linewidth]{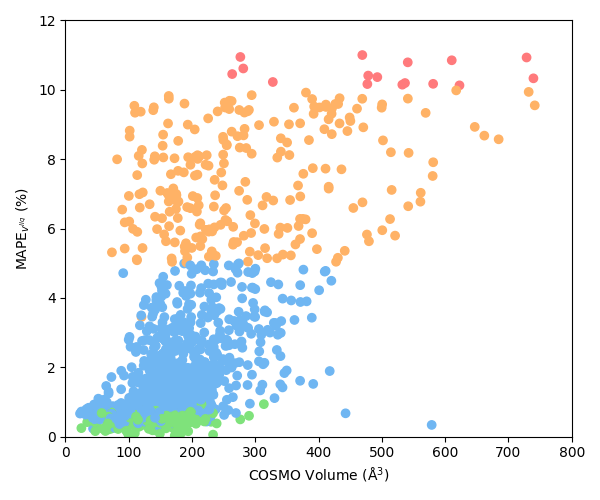}%
  }%
\caption{Correlation between COSMO volume and molar-volume MAPE for (a) PC-SAFT and (b) openCOSMO-RS-Phi, with points grouped according to an accuracy-based classification. {\color[HTML]{7fe27b} $\bullet$} excellent, {\color[HTML]{6fb6f2} $\bullet$} good, {\color[HTML]{ffb266} $\bullet$} satisfactory, {\color[HTML]{ff7a7c} $\bullet$} poor.}
\label{fig:MAPE_v_comparison}
\end{figure}

For molar volume, both PC-SAFT and openCOSMO-RS-Phi perform reasonably well; however, PC-SAFT shows higher accuracy in terms of both mean error and error distribution. This difference may be related to the treatment of repulsive interactions in openCOSMO-RS-Phi, which relies on the classical van der Waals formulation. Such an approach may limit predictive accuracy, particularly for larger molecules. A direct comparison between the two models is provided in \autoref{fig:MAPE_v_comparison}.

To further illustrate the capabilities of the EoS across different molecular groups, the proposed accuracy-based classification is applied to the chemical families defined in the DIPPR database. The performance classification is based jointly on the MAPE values obtained for vapor pressure and liquid molar volume. A compound is assigned to a given class only if the errors for both properties satisfy the corresponding threshold, such that the classification is determined by the less accurately represented property. Compounds are classified as \textit{excellent} when both MAPEs are less than or equal to \(1\,\%\), \textit{good} when both are less than or equal to \(5\,\%\), \textit{satisfactory} when both are less than or equal to \(10\,\%\), and \textit{poor} when either MAPE exceeds \(10\,\%\). This classification has previously been used for PC-SAFT \cite{ramirez-velez_assessing_2022} and tc-PR \cite{pina-martinez_use_2021}, and is extended here to openCOSMO-RS-Phi, as summarized in \autoref{tab:02}. In this table, the abbreviations comp., exc., and sat. denote compounds, excellent, and satisfactory, respectively.

As shown in \autoref{tab:02}, the trends are consistent with those in \autoref{fig:pure_comp_MAPE_EoS}. The tc-PR model predominantly yields errors in the range of 1 to 5\% and is therefore mainly classified as good, whereas PC-SAFT achieves errors largely below 5\%, spanning both the good and excellent categories. openCOSMO-RS-Phi shows a substantial proportion of compounds classified as excellent and good, together with a smaller fraction categorized as satisfactory. This behavior can again be linked to the reduced accuracy of openCOSMO-RS-Phi for larger molecules, as illustrated in \autoref{fig:MAPE_v_comparison}.

\begin{table}[!t]
    \centering
    \resizebox{\textwidth}{!}{%
    \begin{tabular}{l|c|cccc|cccc|cccc|}
    \hline
        \multirow{2}{*}{Chemical Family} & \multirow{2}{*}{\parbox[c]{1.2cm}{Number \\of comp.}} & \multicolumn{4}{c|}{tc-PR} & \multicolumn{4}{c|}{PC-SAFT} & \multicolumn{4}{c|}{openCOSMO-RS-Phi} \\ \cline{3-14}
        ~ & ~ & Exc. & Good & Sat. & Poor & Exc. & Good & Sat. & Poor & Exc. & Good & Sat. & Poor \\ \hline
        Alcohols & 130 & 2 & 119 & 9 & 0 & 13 & 110 & 7 & 0 & 39 & 59 & 29 & 3 \\ 
        Alkanes & 163 & 3 & 158 & 2 & 0 & 131 & 32 & 0 & 0 & 89 & 64 & 10 & 0 \\ 
        Alkenes & 122 & 0 & 121 & 1 & 0 & 94 & 28 & 0 & 0 & 72 & 45 & 4 & 1 \\ 
        Alkynes & 18 & 1 & 15 & 2 & 0 & 12 & 6 & 0 & 0 & 7 & 11 & 0 & 0 \\ 
        Amines/amides & 103 & 2 & 95 & 6 & 0 & 43 & 59 & 1 & 0 & 28 & 59 & 16 & 0 \\ 
        Aromatics & 150 & 1 & 147 & 2 & 0 & 75 & 75 & 0 & 0 & 35 & 105 & 10 & 0 \\ 
        Esters/ethers & 238 & 0 & 228 & 10 & 0 & 112 & 124 & 2 & 0 & 60 & 109 & 60 & 9 \\ 
        Halogen compounds & 232 & 9 & 214 & 9 & 0 & 112 & 118 & 2 & 0 & 130 & 95 & 7 & 0 \\ 
        Inorganic compounds & 72 & 0 & 58 & 14 & 0 & 24 & 41 & 5 & 2 & 19 & 44 & 1 & 0 \\ 
        Ketones/aldehydes & 80 & 2 & 77 & 1 & 0 & 35 & 45 & 0 & 0 & 31 & 47 & 2 & 0 \\ 
        Nitriles & 31 & 1 & 23 & 7 & 0 & 1 & 27 & 2 & 1 & 9 & 18 & 4 & 0 \\ 
        Nitrogen compounds & 98 & 5 & 86 & 7 & 0 & 22 & 72 & 4 & 0 & 28 & 54 & 16 & 0 \\ 
        Organic acids & 82 & 4 & 76 & 2 & 0 & 11 & 71 & 0 & 0 & 4 & 32 & 44 & 2 \\ 
        Organic salts & 13 & 0 & 8 & 5 & 0 & 5 & 7 & 1 & 0 & 3 & 8 & 2 & 0 \\ 
        Other compounds & 32 & 0 & 30 & 2 & 0 & 16 & 15 & 1 & 0 & 9 & 23 & 0 & 0 \\ 
        Peroxides & 11 & 0 & 10 & 1 & 0 & 3 & 8 & 0 & 0 & 2 & 4 & 5 & 0 \\ 
        Polyfunctional compounds & 70 & 1 & 63 & 6 & 0 & 24 & 43 & 3 & 0 & 7 & 42 & 21 & 0 \\ 
        Silicon compounds & 61 & 2 & 54 & 5 & 0 & 22 & 39 & 0 & 0 & 19 & 26 & 13 & 1 \\ 
        Sulfur compounds & 94 & 2 & 90 & 2 & 0 & 66 & 28 & 0 & 0 & 40 & 53 & 1 & 0 \\ \hline
        Total & 1800 & 35 & 1672 & 93 & 0 & 821 & 948 & 28 & 3 & 631 & 898 & 245 & 16 \\ \hline
    \end{tabular}
    }
    \caption{Accuracy classification based on vapor pressure and molar volume predictions for compounds grouped by chemical family using tc-PR, PC-SAFT, and openCOSMO-RS-Phi. The classification system is explained in the text.}
    \label{tab:02}
\end{table}

\subsubsection{Adjusted Parameter Evaluation}

\begin{figure}[!b]
  \subcaptionbox*{\centering (a)}[.5\linewidth]{%
    \includegraphics[width=\linewidth]{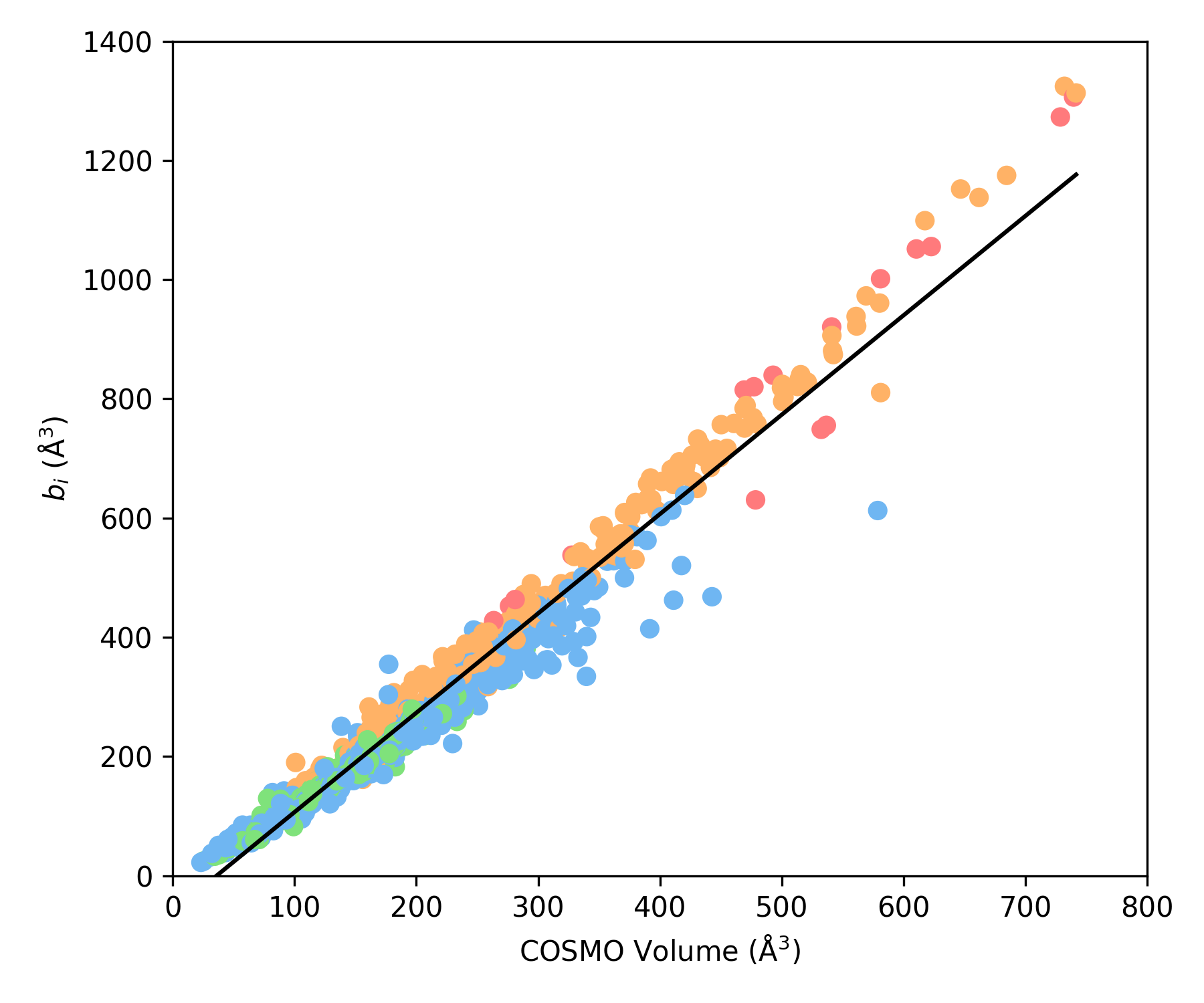}%
  }%
  \hfill
  \subcaptionbox*{\centering (b)}[.5\linewidth]{%
    \includegraphics[width=\linewidth]{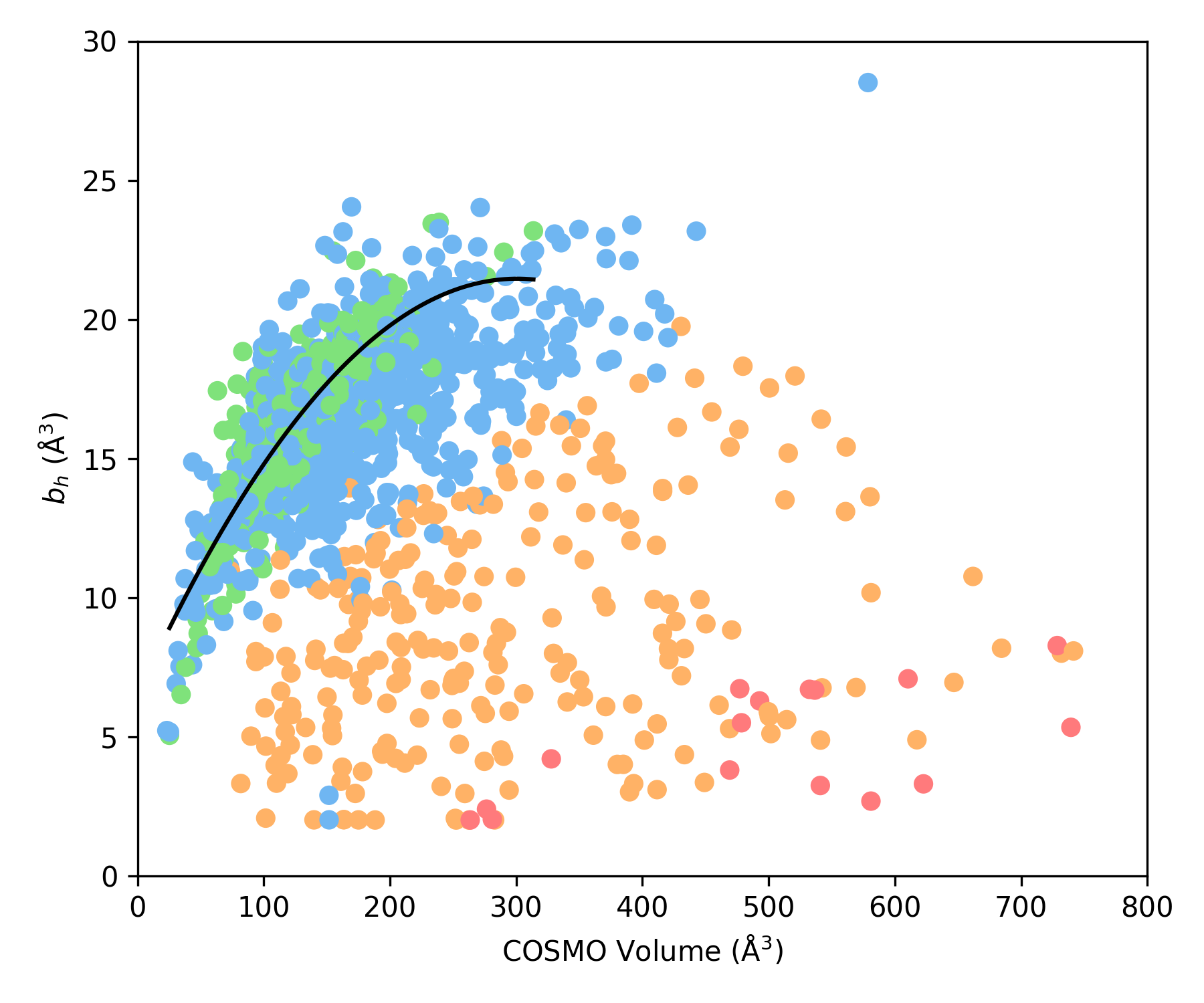}%
  }%
\caption{Molar cavity (a) and  hole molar cavity (b) as a function of COSMO volumes. {\color[HTML]{7fe27b} $\bullet$} excellent,  {\color[HTML]{6fb6f2} $\bullet$} good, {\color[HTML]{ffb266} $\bullet$} satisfactory,  {\color[HTML]{ff7a7c} $\bullet$} poor.}
\label{fig:bi_&_bh}
\end{figure}

An examination of the correlations between the adjusted parameters and molecular size reveals trends similar to those reported for COSMO-SAC-Phi. In particular, a linear relationship is observed between the adjusted parameter \(b_i\) and the COSMO volume, as shown in \autoref{fig:bi_&_bh} (a). \autoref{fig:bi_&_bh} (b) presents the corresponding relationship between the hole molar cavity, representing free volume, and the COSMO volume. Distinct regions can be identified. Analyzing only the components classified as excellent correlations can be done fot thes shown parameters. For the molecular volume this correlation seems to hold true for most solvents. However, for the parameter \(b_h\) this correlation is much weaker.

The model shows limitations in accurately describing large molecules and the effects associated with them. In particular, the combination of increasing molecular size and high mean absolute percentage error (MAPE) in molar-volume predictions leads to lower values of \(b_h\). This behavior can be attributed to the selected hard-sphere term, which relies on a simplified representation of repulsive interactions. Such an approach becomes less suitable for larger and more structurally complex molecules, for which size, shape, and packing effects are not adequately captured.

A more detailed analysis of the variables contributing to this behavior is provided in Figure SI14 in the Supporting Information, complementing the trends already discussed in \autoref{fig:MAPE_v_comparison} and \autoref{fig:bi_&_bh} (b).

No clear global correlation is observed for \(\delta_m^0\) and \(\delta_m^t\). These parameters are plotted against COSMO volume in Figure SI15 in the Supporting Information, where the influence of local minima, previously described in the three-step minimization approach, becomes evident. This behavior is consistent with the role of \(\delta_m^0\) and \(\delta_m^t\) as effective fitting parameters, since they absorb contributions from molecular effects that openCOSMO-RS cannot describe explicitly. In this sense, all phenomena not properly captured by the model become embedded in these parameters during the fitting procedure.

Nevertheless, clear trends emerge within specific chemical subgroups for \(\delta_m^0\) and \(\delta_m^t\), as shown in Figure SI16 and Figure SI17 in the Supporting Information for \(n\)-alcohols and \(n\)-alkanes, respectively, as a function of COSMO volume. For both groups, moderately accurate functional relationships can be identified, providing a useful description of the behavior of the systems under investigation.
\subsection{Binary Mixtures}\label{Binary Mixtures Results}

As outlined in the Methodology section, the performance of openCOSMO-RS-Phi for binary-mixture thermodynamic data is assessed using the grading procedure proposed by \textsc{Jaubert et al.} \cite{jaubert_benchmark_2020}, including the modifications introduced by \textsc{Pi\~na-Martinez et al.} \cite{pina2021optimal}. The corresponding results are presented in this section, beginning with an overview and a comparison with other models. Since the performance of the EoS depends on both the property and the BAC of the system considered, a more detailed analysis is carried out for each BAC and property. Representative example plots are included to illustrate model behavior. For the sake of comparability with previous benchmark studies, only systems that were also illustrated in those studies are plotted.

\subsubsection{Overall Benchmark}

Within the applied benchmark procedure, the model performance is quantified by the MAPE and the SR, which accounts for OM points. \autoref{tab:04} reports the MAPEs for ten properties - liquid-phase composition (\(x\)), vapor-phase (or second liquid-phase) composition (\(y\)), three-phase pressure (\(P_\text{LLV}\)), three-phase composition (\(z_\text{LLV}\)), critical pressure (\(P_\text{c}\)), critical composition (\(x_\text{c}\)), azeotropic pressure (\(P_\text{azeo}\)), azeotropic composition (\(x_\text{azeo}\)), enthalpy of mixing (\(h^\text{M}\)), and heat capacity of mixing (\(c_\text{P}^\text{M}\)) - for each BAC, together with the corresponding SR values. \\
These MAPE and SR values are subsequently converted into marks (score over 20) for each BAC and property. Then, Mark20 is calculated by category of association (N-A, S-A, C-A, S-A + C-A). The overall score is obtained as the average of the four category scores. For openCOSMO-RS-Phi, a final score of 7.3/20 is obtained, corresponding to fair performance. The resulting marks are summarized in \autoref{tab:05}. \\
By penalizing low SR values, the modified grading scheme provides a more realistic measure of predictive capability than the original procedure. This is especially relevant for openCOSMO-RS-Phi, since several SR values are low (see \autoref{tab:04}). \\
For example, pronounced difficulties are observed for VLLE and azeotropic data of \(\text{BAC}_8\). When SR is taken into account, the marks decrease from 17.1 (\(x_\text{LLV}\)) and 19.7 (\(P_\text{LLV}\)) to 1.4 and 1.6, respectively, and from 3.3 (\(x_\text{azeo}\)) and 13.7 (\(P_\text{azeo}\)) to 1.1 and 4.6.

\begin{table}[!t]
    \centering
    \resizebox{\textwidth}{!}{%
    \begin{tabular}{l|ccccccccccccccc|}
    \hline
        \multirow{2}{*}{BAC} & \multicolumn{15}{c|}{MAPE and corresponding SR on:}  \\ \cline{2-16}
        ~  & \(x\) & \(\text{SR}_x\) & \(y\) & \(\text{SR}_y \) & \(P_\text{LLV}\) & \(z_\text{LLV}\) &  \(\text{SR}_{\text{LLV}}\) & \(P_\text{c}\) & \(x_\text{c}\) &  \(\text{SR}_{c}\) & \(P_\text{azeo}\) & \(x_\text{azeo}\) &  \(\text{SR}_{\text{azeo}}\)  & \(h^\text{M}\) & \(c_\text{P}^\text{M}\) \\ \hline
        1 (NA-NA) & 23.9 & 0.93 & 13.8 & 0.93 & 3.3 & 17.5 & 0.5 & 22.8 & 38.3 & 0.81 & 4.1 & 4.6 & 1.0 & 42.9 & 88.9 \\
        2 (HA-NA) & 25.8 & 0.81 & 18.1 & 0.81 & & & & 28.0 & 49.0 & 0.65 & 8.9 & 27.6 & 0.68 & 46.8 & 199.6 \\
        3 (HD-NA) & 40.0 & 0.68 & 23.1 & 0.68 & & & & 14.5 & 49.4 & 0.57 & 21.0 & 35.8 & 0.41 & 52.2 & 117.9 \\
        4 & 19.4 & 0.90 & 16.2 & 0.90 & & & & 21.9 & 53.0 & 0.87 & 5.3 & 26.0 & 0.29 & 31.0 & 196.9 \\
        5 (SA-NA) & 44.3 & 0.64 & 14.7 & 0.64 & 3.7 & 23.6 & 1.0 & 19.6 & 85.4 & 0.47 & 4.4 & 12.2 & 1.0 & 45.3 & 61.9 \\
        6 (HD-HA) & 36.3 & 0.90 & 30.2 & 0.90 & & & & 11.8 & 60.1 & 0.83 & 10.9 & 18.8 & 1.0 & 40.9 & 86.6 \\
        7 (SA-HD) & 39.3 & 0.70 & 21.7 & 0.70 & & & & & & & 3.9 & 22.7 & 0.6 & 65.6 & 49.3 \\
        8 (SA-HA) & 47.0 & 0.76 & 22.4 & 0.76 & 0.5 & 5.9 & 0.08 & 20.5 & 43.1 & 0.76 & 12.5 & 33.4 & 0.33 & 61.7 & 88.9 \\
        9 (SA-SA) & 33.0 & 0.91 & 21.8 & 0.91 & 7.1 & 6.0 & 1.0 & 29.8 & 55.6 & 0.77 & 6.9 & 20.3 & 1.0 & 57.3 & 75.4 \\
        \hline
    \end{tabular}
     }
    \caption{Overview of the MAPE between Experimental Data and Model Calculations along with the Corresponding SRs (openCOSMO-RS-Phi without BIPs).}
    \label{tab:04}
\end{table}

\begin{table}[!ht]
    \centering
    \resizebox{\textwidth}{!}{%
    \begin{tabular}{l|cccccccccc|c|c|c|}
    \hline
        \multirow{2}{*}{BAC} & \multicolumn{10}{c|}{mark on:} & \multirow{2}{*}{\parbox[c]{0.8cm}{BAC \\mark}} & \multirow{2}{*}{\parbox[c]{1.2cm}{category \\mark}} & \multirow{2}{*}{\parbox[c]{1cm}{final \\mark}}  \\ \cline{2-11}
        ~  & \(x\) & \(y\) & \(P_\text{LLV}\) & \(z_\text{LLV}\) & \(P_\text{c}\) & \(x_\text{c}\) & \(P_\text{azeo}\) & \(x_\text{azeo}\) & \(h^\text{M}\) & \(c_\text{P}^\text{M}\) & ~ & ~ & ~ \\ \hline
        1 (NA-NA) & 7.5 & 12.2 & 9.2 & 5.6 & 2.3 & 0.7 & 18.0 & 17.7 & 9.3 & 11.1 & 9.4 & \multirow{4}{*}{5.8} &  \multirow{9}{*}{\textbf{7.3}/20}  \\ 
        2 (HA-NA) & 5.7 & 8.9 & & & 0.0 & 0.0 & 10.6 & 4.2 & 8.3 & 0.0 & 4.7 & & \\
        3 (HD-NA) & 0.0 & 5.8 & & & 5.2 & 0.0 & 3.9 & 0.9 & 7.0 & 8.2 & 3.9 & & \\
        4 (HD/HA-HD/HA) & 9.3 & 10.7 & & & 3.1 & 0.0 & 5.0 & 2.0 & 12.2 & 0.3 & 5.3 & & \\ \cline{13-13}
        5 (SA-NA) & 0.0 & 8.1 & 18.2 & 8.2 & 2.5 & 0.0 & 17.8 & 13.9 & 8.7 & 13.8 & 9.1 & 9.1 & \\ \cline{13-13}
        6 (HD-HA) & 1.6 & 4.4 & & & 9.2 & 0.0 & 14.5 & 10.6 & 9.8 & 11.3 & 7.7 & 7.7 & \\ \cline{13-13}
        7 (SA-HD) & 0.3 & 6.4 & & & & & 10.8 & 5.2 & 3.6 & 15.1 & 6.9 & &  \\
        8 (SA-HA) & 0.0 & 6.7 & 1.6 & 1.4 & 3.5 & 0.0 & 4.6 & 1.1 & 4.6 & 11.1 & 3.5 & \multirow{2}{*}{6.4}  & \\
        9 (SA-SA) & 3.2 & 8.3 & 16.5 & 17.0 & 0.0 & 0.0 & 16.6 & 9.8 & 5.7 & 12.5 & 8.9 & & \\
        \hline
        \textbf{overall} & 2.1 & 7.3 & 12.1 & 7.7 & 4.0 & 0.0 & 13.1 & 9.0 & 8.1 & 10.7 & & & \\
        \hline
    \end{tabular}
    }
    \caption{Rating of the Nine BACs To Finally Grade the Model (openCOSMO-RS-Phi with no BIPs).}
    \label{tab:05}
\end{table}

\begin{figure}[!b]
    \centering
    \includegraphics[width=1\textwidth]{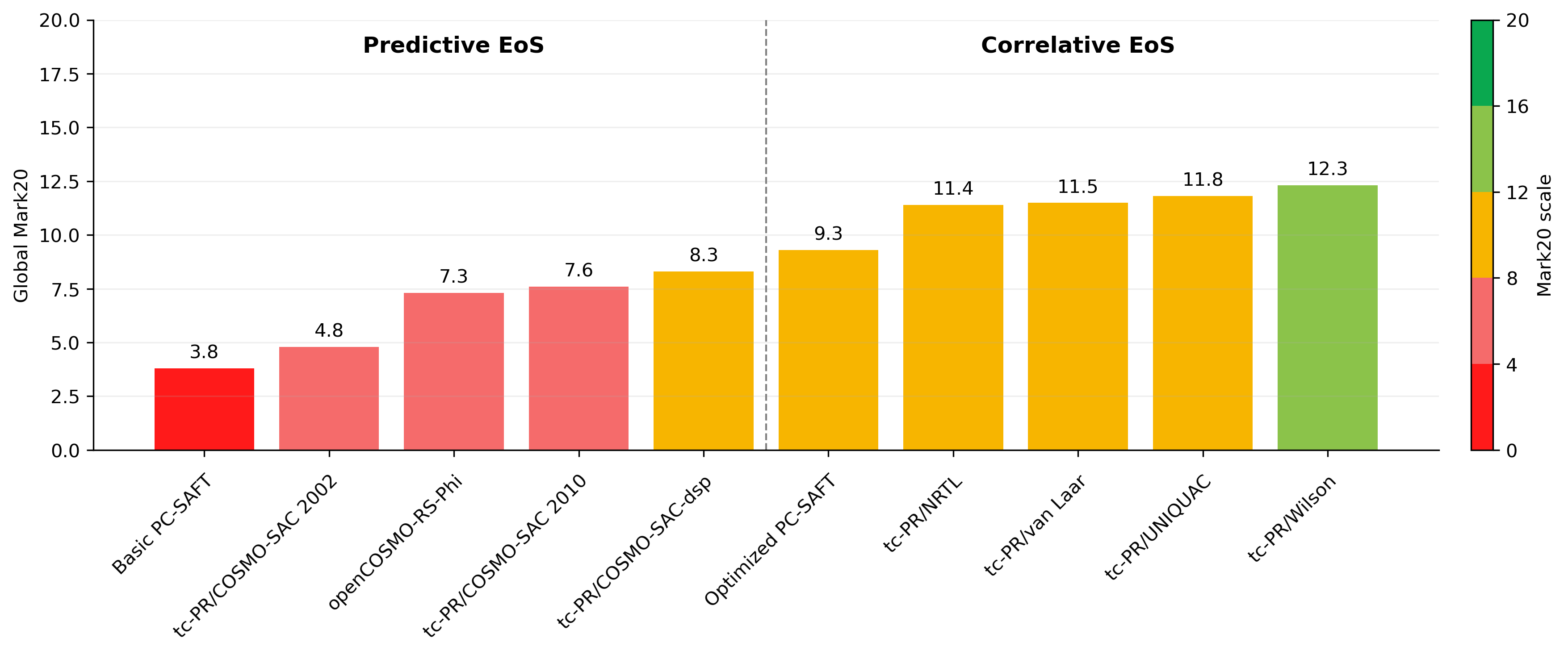}
    \caption{Comparison of the global marks of different EoSs following the methodology proposed in \cite{jaubert_benchmark_2020} and \cite{pina2021optimal}.}
    \label{fig:16}
\end{figure}

For a more comprehensive assessment of openCOSMO-RS-Phi, \autoref{fig:16} compares its global mark with those of other thermodynamic models benchmarked against the database of \textsc{Jaubert et al.} using the same methodology and evaluation criteria. These models include predictive EoS without binary interaction parameters (BIPs), such as basic PC-SAFT \cite{nikolaidis2021assessment} and tc-PR combined with various COSMO-SAC versions \cite{paes2026using}, as well as correlative EoS with BIPs regressed to experimental data, such as optimized PC-SAFT \cite{nikolaidis2023effect} and tc-PR coupled with activity-coefficient models including Wilson, NRTL, UNIQUAC, and van Laar \cite{pina2021optimal}. 
It should be noted that, for tc-PR combined with COSMO-SAC-2002, COSMO-SAC-2010, and COSMO-SAC-dsp, the calculations were performed for only 160 out of the 200 systems. Consequently, comparisons with these models are not exact. In addition to the global scores shown in \autoref{fig:16}, detailed comparisons of all models in terms of individual properties are provided in the Supporting Information. 

Among the predictive EoS, openCOSMO-RS-Phi performs significantly better than basic (pure-component) PC-SAFT and tc-PR combined with COSMO-SAC-2002. However, it performs worse than tc-PR combined with COSMO-SAC-2010 and tc-PR-COSMO-SAC-dsp, primarily because of a less accurate description of two-phase compositions (\(x\), \(y\)) and critical compositions (\(x_\text{c}\)). While openCOSMO-RS-Phi employs the COSMO-RS framework to compute the attractive pressure through a pseudo-mixture approach, tc-PR-COSMO-SAC-dsp incorporates the COSMO-based activity-coefficient model only into the equation system used to determine the mixture attraction parameter \(a\). Other differences that may affect performance include the global parameters, the determination of pure-component parameters (regression versus estimation from critical data), the absence of volume translation in openCOSMO-RS-Phi, differences in the mixing rules for the co-volume parameter \(b_i\), and variations in the temperature dependence of the interaction-energy contributions. \\
As shown in \autoref{fig:16}, predictive models generally perform worse than correlative EoS, which can be attributed to the absence of binary interaction parameters. However, this limitation is offset by their broader applicability, including the ability to predict previously unseen mixtures. For systems with sufficient experimental data, the introduction of binary parameters, for example in the dispersion term as proposed by \textsc{Soares et al.} \cite{de_p_soares_pairwise_2019}, is expected to significantly improve performance. For example, incorporating \(k_{ij}\) values and induced association parameters increased the PC-SAFT score from 3.8 to 9.3 \cite{nikolaidis2023effect}. Further potential for improvement arises from the aspects discussed above. The aim of this work was to benchmark a fully predictive base version of openCOSMO-RS-Phi and thereby provide a reference for future developments of the EoS.

\paragraph{Benchmark Across BACs}

At the BAC level, the highest score is obtained for the \(\text{BAC}_1\) group (9.4/20), which comprises mixtures of two non-associating compounds. However, because of the lower scores for \(\text{BAC}_2\)-\(\text{BAC}_4\), the aggregated non-associating (N-A) category (\(\text{BAC}_1\)-\(\text{BAC}_4\)) represents the weakest-performing category overall. For non-associating systems with at least one non-polar component, the relative importance of the dispersion contribution is higher than that of the misfit and hydrogen-bonding terms. The comparatively poor performance may therefore be attributed to the fact that the global parameters of openCOSMO-RS-Phi were regressed without explicitly accounting for dispersion contributions, which are included only in the EoS formulation. In addition, \textsc{Zini et al.} reported that the dispersion term used in this implementation lacks a clear theoretical foundation and therefore proposed an alternative term \cite{zini_improved_2021}. \\
The second-highest BAC-level score, and the highest category score after averaging, is obtained for the \(\text{BAC}_5\) group, with \(\text{mark}_\text{SA} = 9.1\). This favorable result is mainly due to the accurate representation of three-phase line data, azeotropic behavior, and energetic properties. Notably, this score is the second-highest reported for \(\text{BAC}_5\) among all predictive and correlative EoS evaluated to date. The \(\text{BAC}_5\) group consists of mixtures of non-associating and self-associating components, such as alkane + alcohol systems. Interestingly, basic PC-SAFT also shows its best performance in this category, which has been attributed to its explicit treatment of hydrogen-bonding interactions. A similar explanation likely applies to openCOSMO-RS-Phi: in the presence of self-associating components, the relative importance of dispersion interactions decreases, favoring models with an accurate description of association effects. \\
The second-best category score is obtained for the cross-associating (C-A) group, comprising \(\text{BAC}_6\) systems. Here, openCOSMO-RS-Phi performs significantly better than PC-SAFT and tc-PR/COSMO-SAC-2002, but worse than tc-PR/COSMO-SAC-2010 and tc-PR/COSMO-SAC-dsp. These systems are typically characterized by negative deviations from ideality. \textsc{Paes et al.} showed that inclusion of a dispersion term reduces systematic errors for \(\text{BAC}_6\) systems \cite{paes2026using}. A more accurate representation of hydrogen bonding further improves model performance. In the present implementation of openCOSMO-RS-Phi, hydrogen bonding is described using a simplified approach with a single global \(c_\text{hb}\) parameter and no temperature dependence. A more refined treatment of hydrogen bonding may therefore improve the results further. \\
The results for the fourth category, comprising mixtures with both self- and cross-association (\(\text{BAC}_7\)-\(\text{BAC}_9\)), are mixed. While openCOSMO-RS-Phi shows the best performance among predictive EoS for \(\text{BAC}_9\) systems, it performs comparatively poorly for \(\text{BAC}_7\) and \(\text{BAC}_8\). The lowest score across all BACs is obtained for \(\text{BAC}_8\), with a value of 3.5/20. For most EoS benchmarked to date, however, the lowest performance has typically been observed for \(\text{BAC}_6\). \\
Finally, it is noteworthy, and expected, that similar qualitative trends, such as the comparatively good description of complex systems involving full or partial self-association, were also reported by \textsc{Nikolaidis et al.} in the benchmark of PC-SAFT \cite{nikolaidis2021assessment}. This can be attributed to the fact that both approaches explicitly account for hydrogen-bonding interactions, in contrast to cubic EoS such as Peng-Robinson.

\subsubsection{Phase Equilibria}

\paragraph{Two-phase data}

\begin{figure}[p]
\centering

\subcaptionbox*{\centering (a)}[0.33\textwidth]{%
  \includegraphics[width=\linewidth,height=0.21735\textwidth]{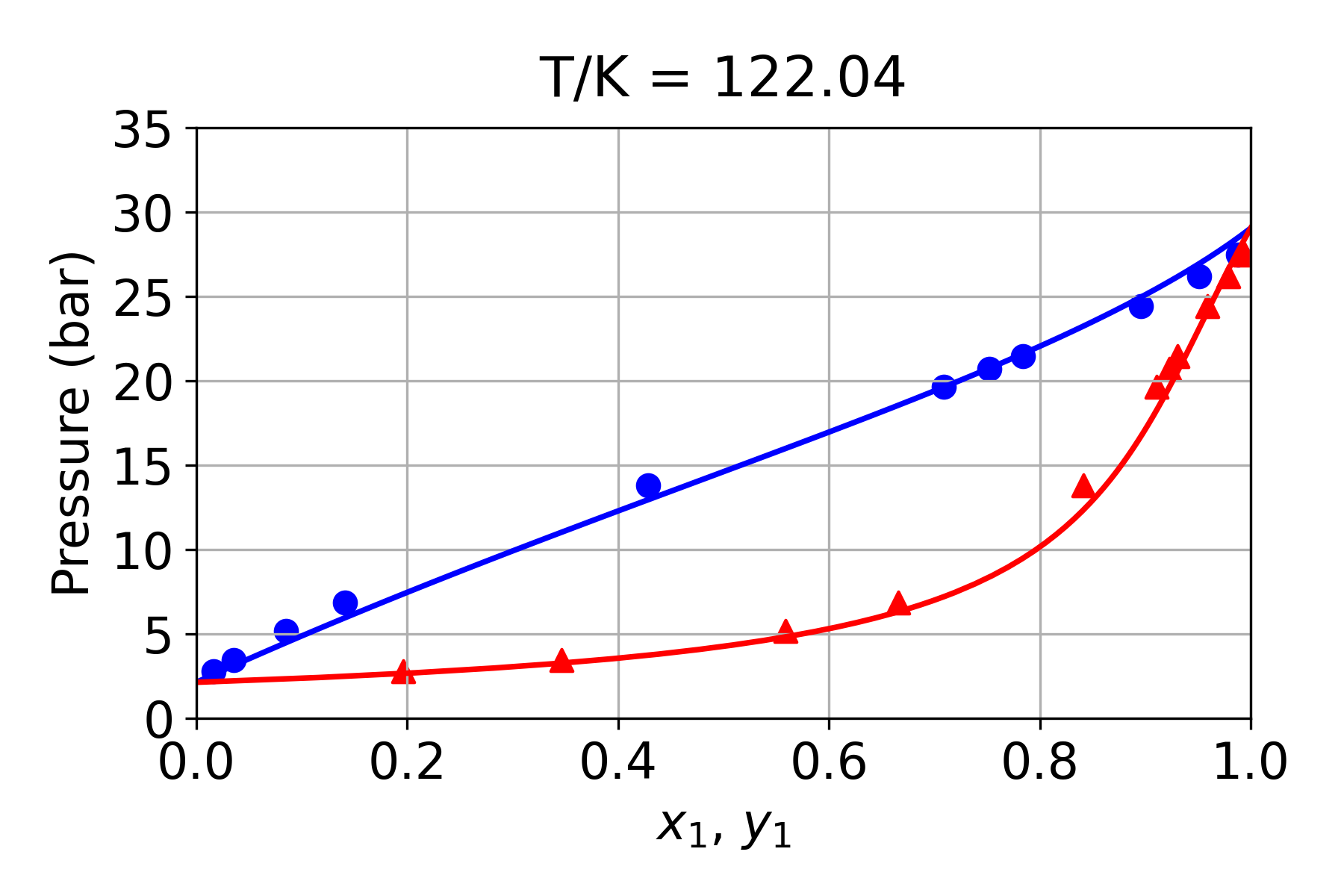}%
}\hfill
\subcaptionbox*{\centering (b)}[0.33\textwidth]{%
  \includegraphics[width=\linewidth,height=0.21735\textwidth]{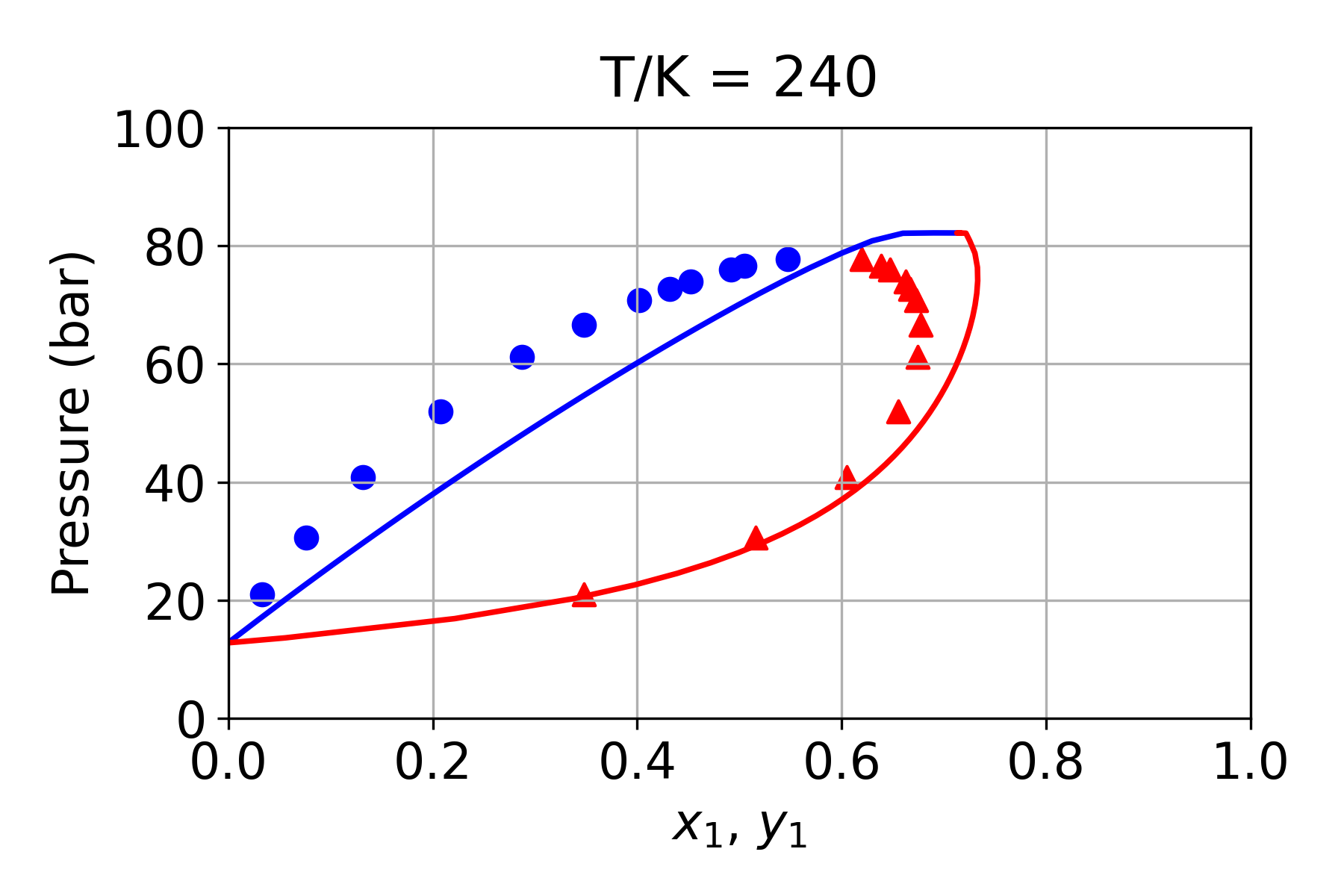}%
}\hfill
\subcaptionbox*{\centering (c)}[0.33\textwidth]{%
  \includegraphics[width=\linewidth,height=0.21735\textwidth]{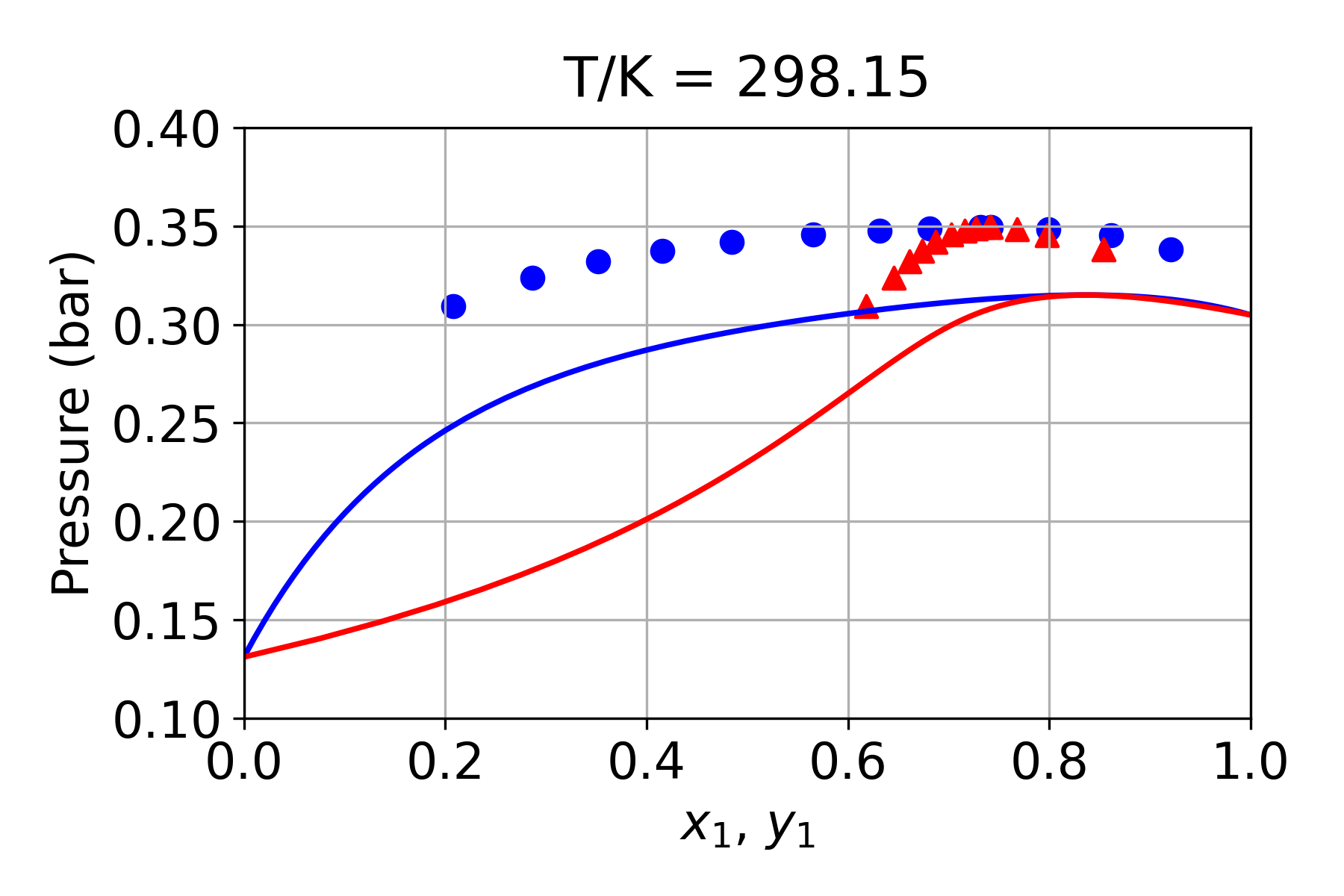}%
}

\subcaptionbox*{\centering (d)}[0.33\textwidth]{%
  \includegraphics[width=\linewidth,height=0.21735\textwidth]{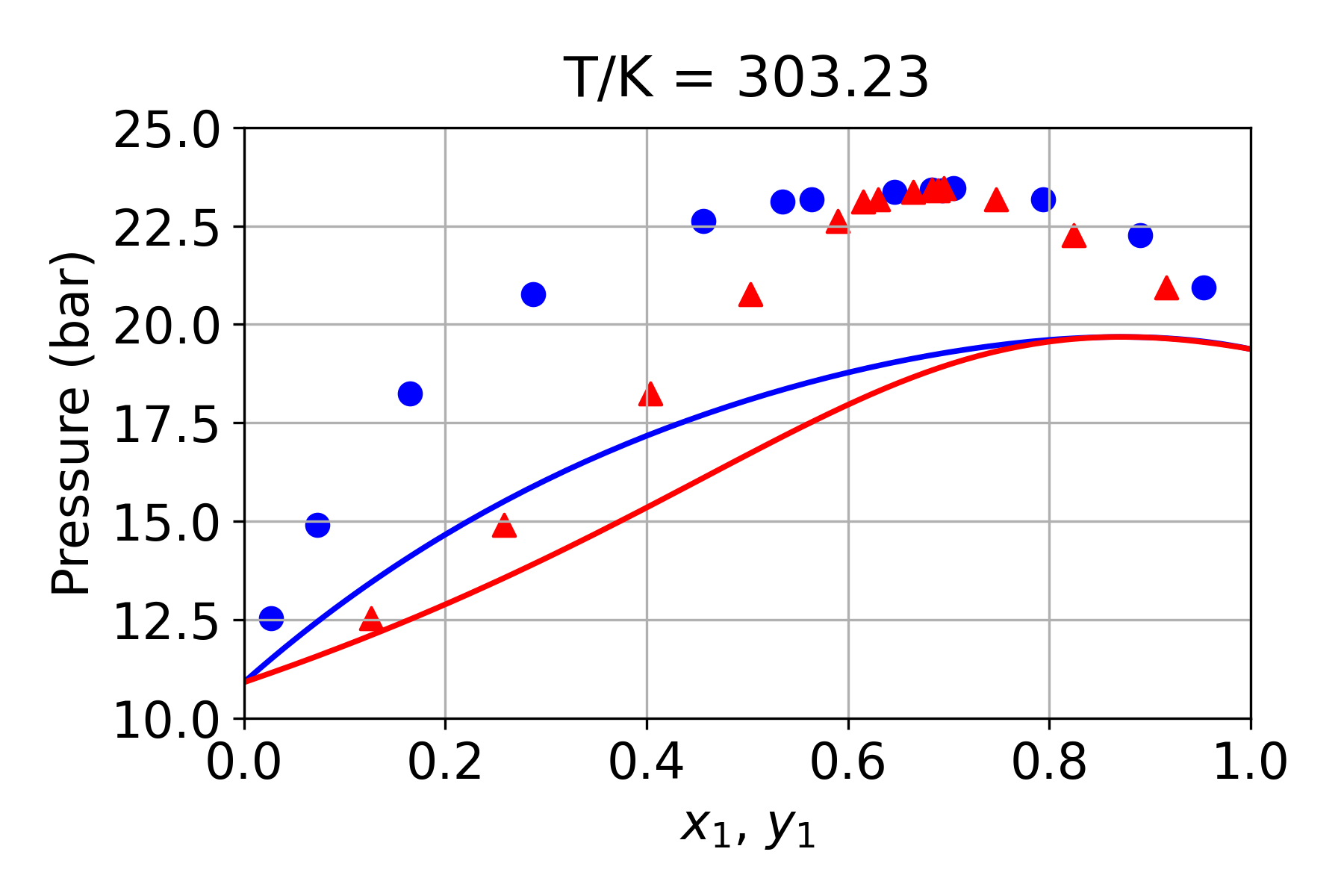}%
}\hfill
\subcaptionbox*{\centering (e)}[0.33\textwidth]{%
  \includegraphics[width=\linewidth,height=0.21735\textwidth]{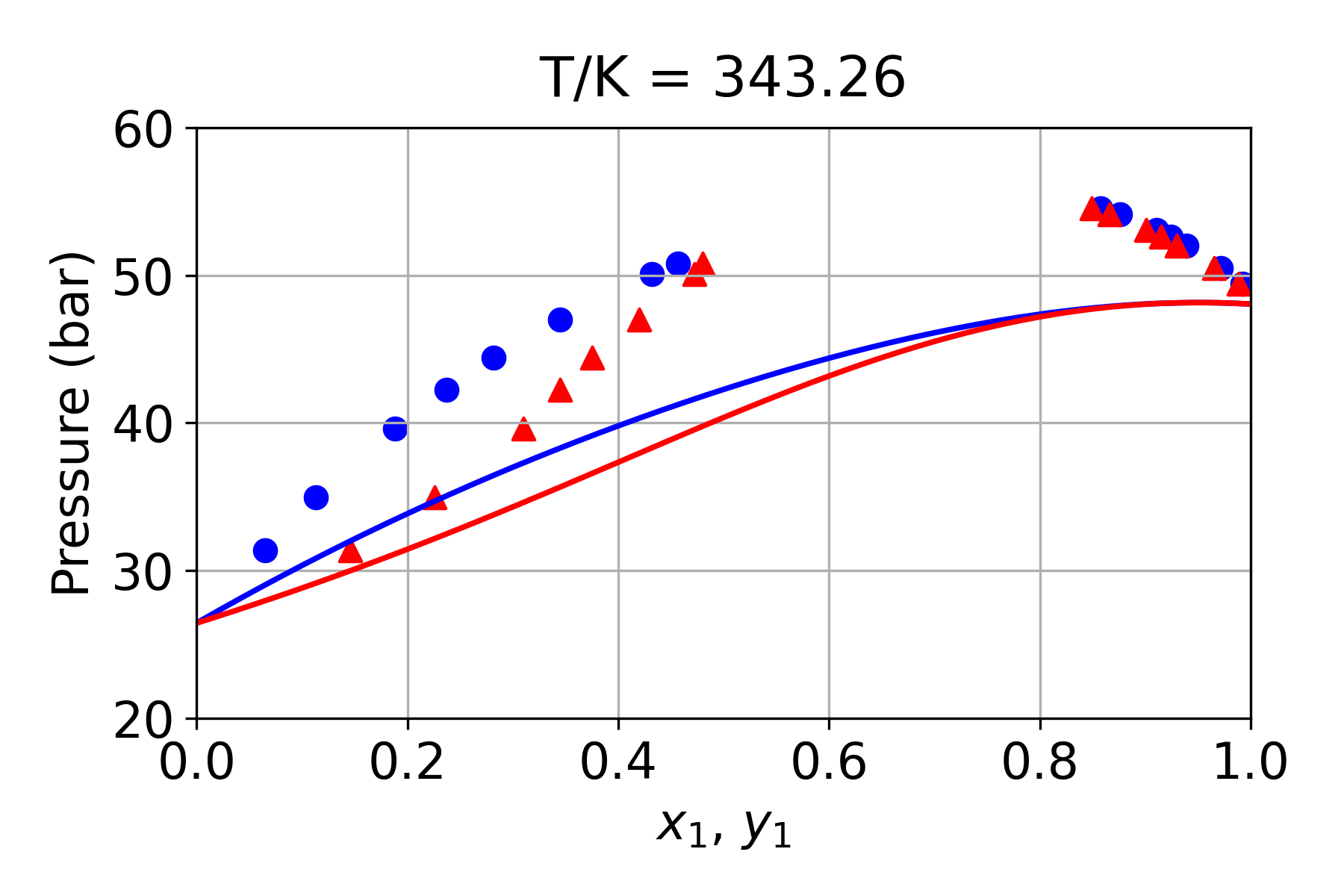}%
}\hfill
\subcaptionbox*{\centering (f)}[0.33\textwidth]{%
  \includegraphics[width=\linewidth,height=0.21735\textwidth]{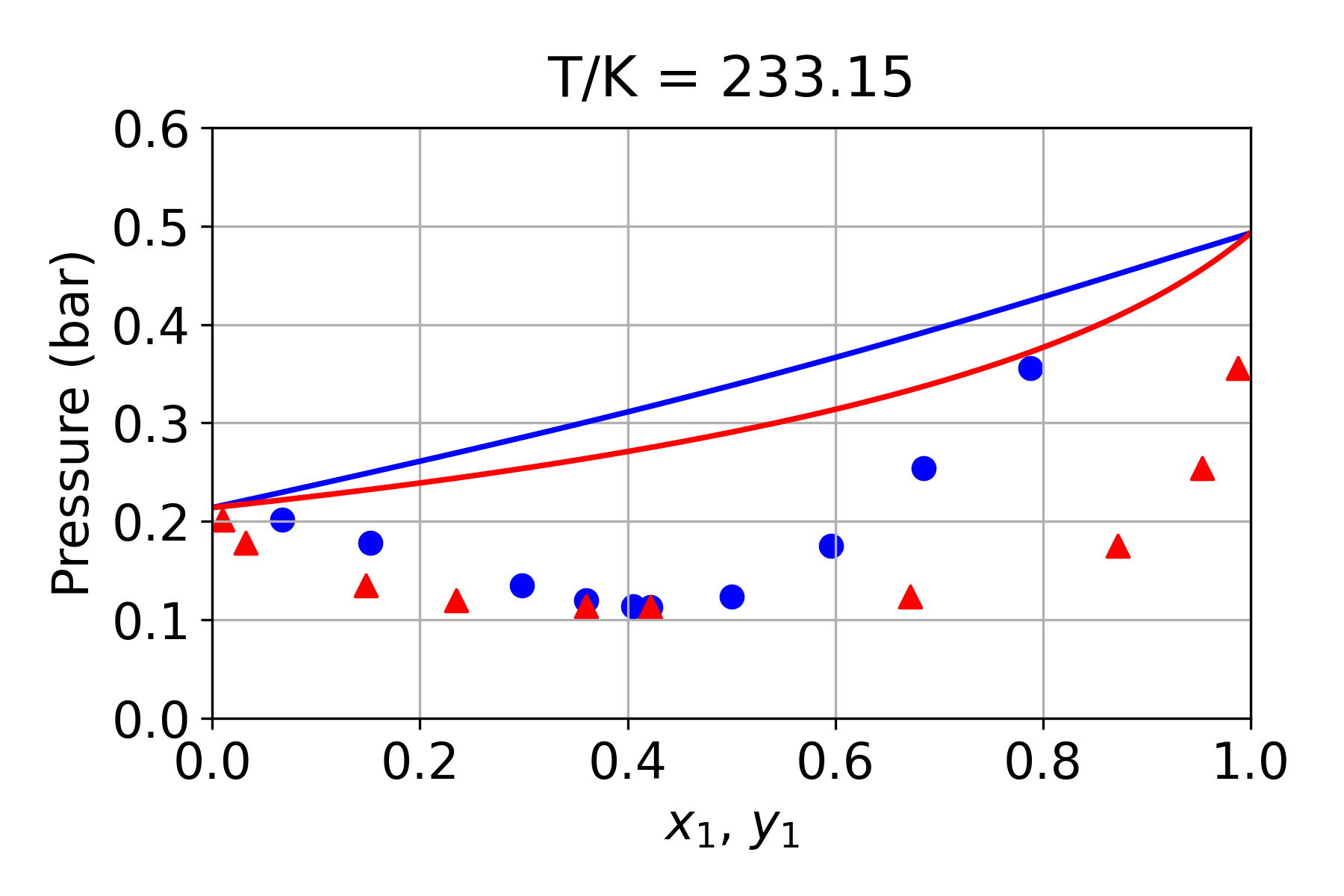}%
}

\subcaptionbox*{\centering (g)}[0.33\textwidth]{%
  \includegraphics[width=\linewidth,height=0.21735\textwidth]{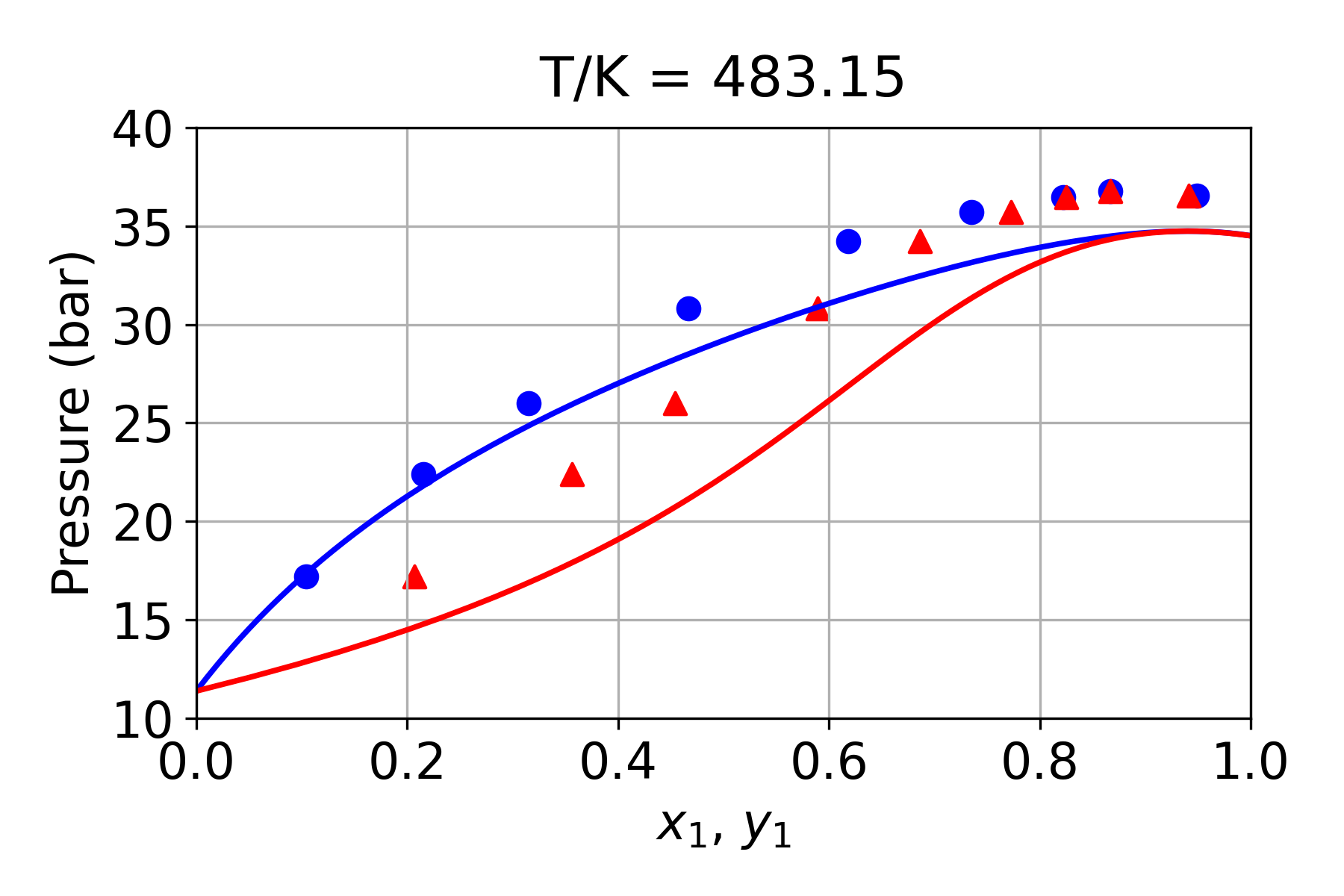}%
}\hfill
\subcaptionbox*{\centering (h)}[0.33\textwidth]{%
  \includegraphics[width=\linewidth,height=0.21735\textwidth]{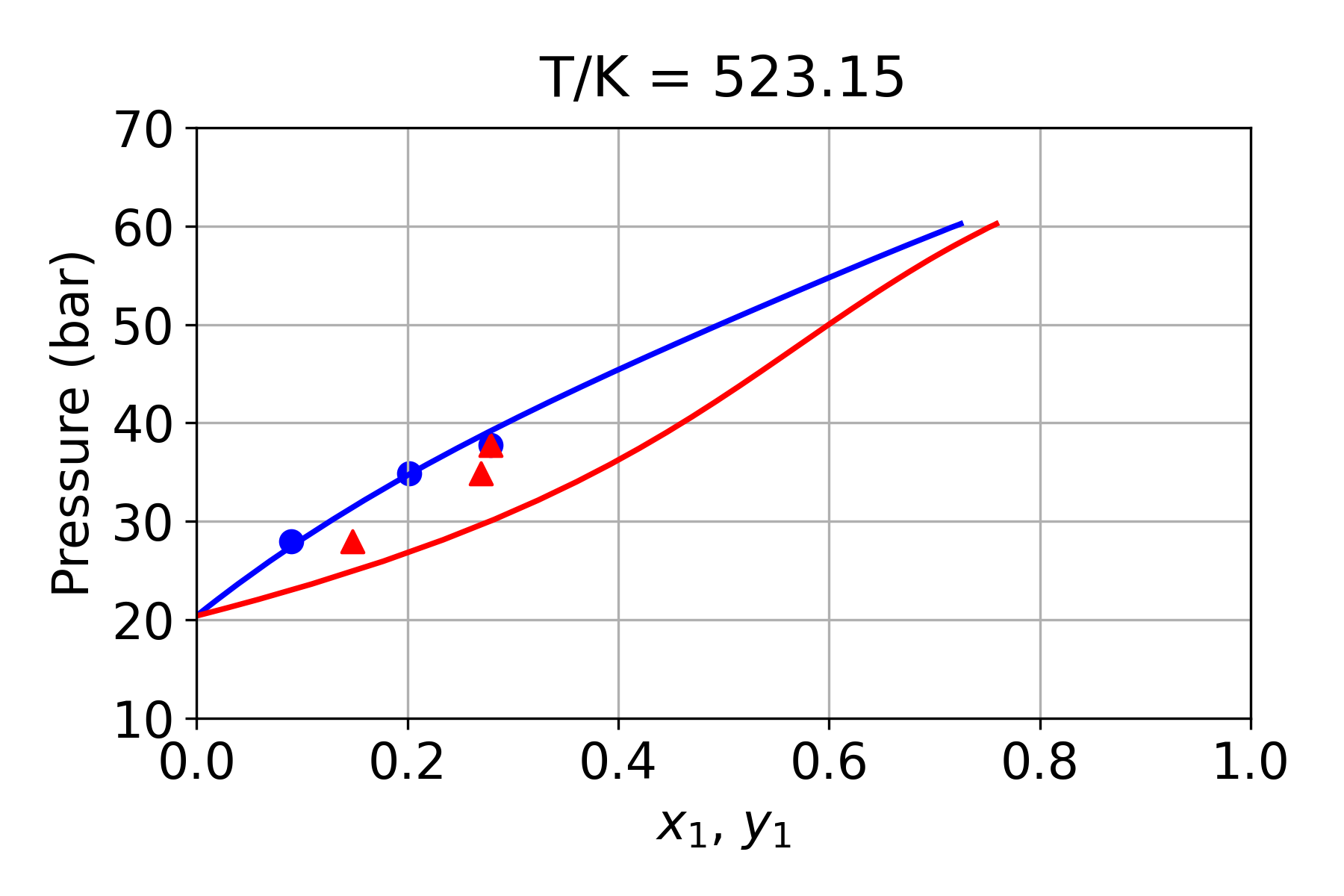}%
}\hfill
\subcaptionbox*{\centering (i)}[0.33\textwidth]{%
  \includegraphics[width=\linewidth,height=0.21735\textwidth]{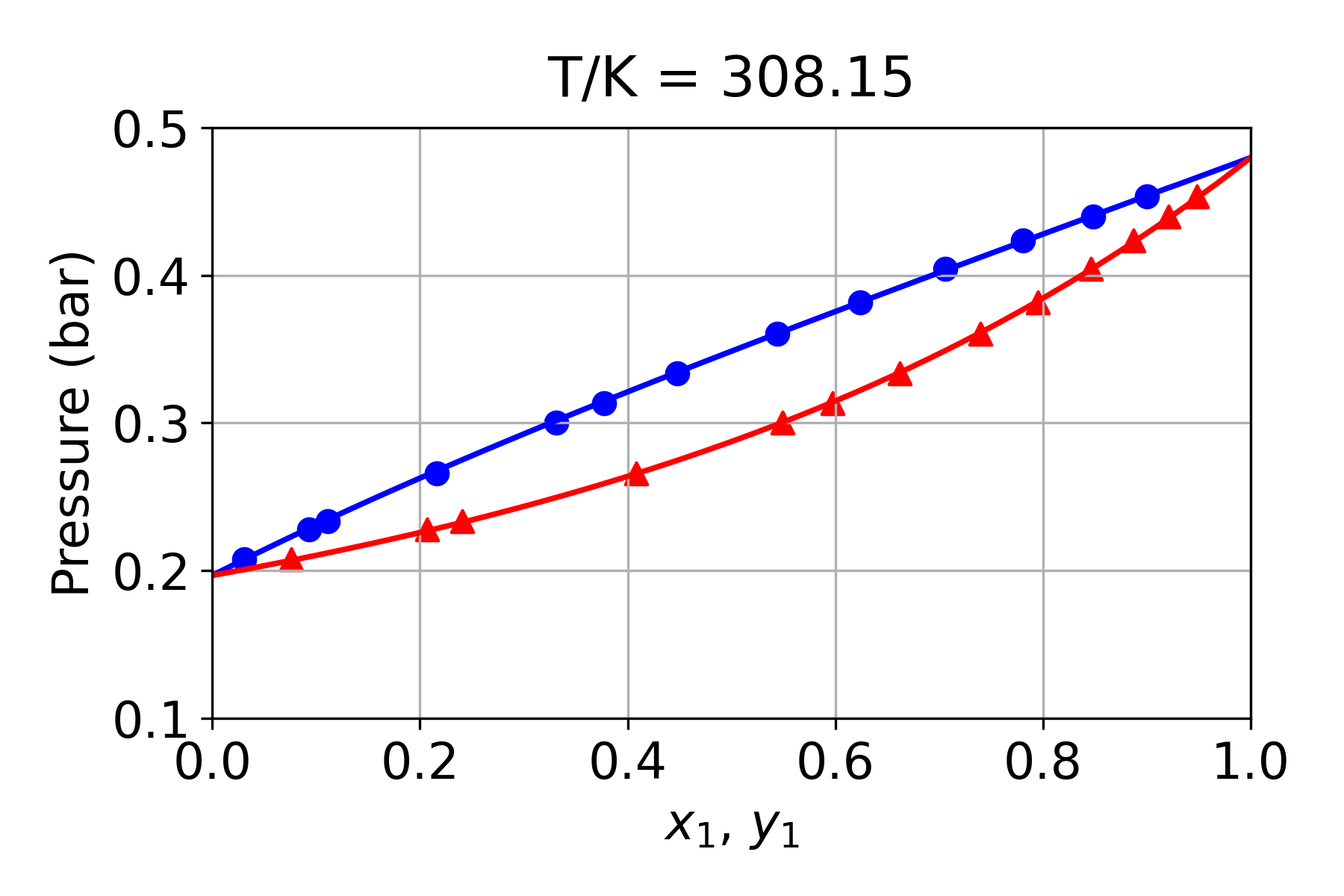}%
}

\subcaptionbox*{\centering (j)}[0.33\textwidth]{%
  \includegraphics[width=\linewidth,height=0.21735\textwidth]{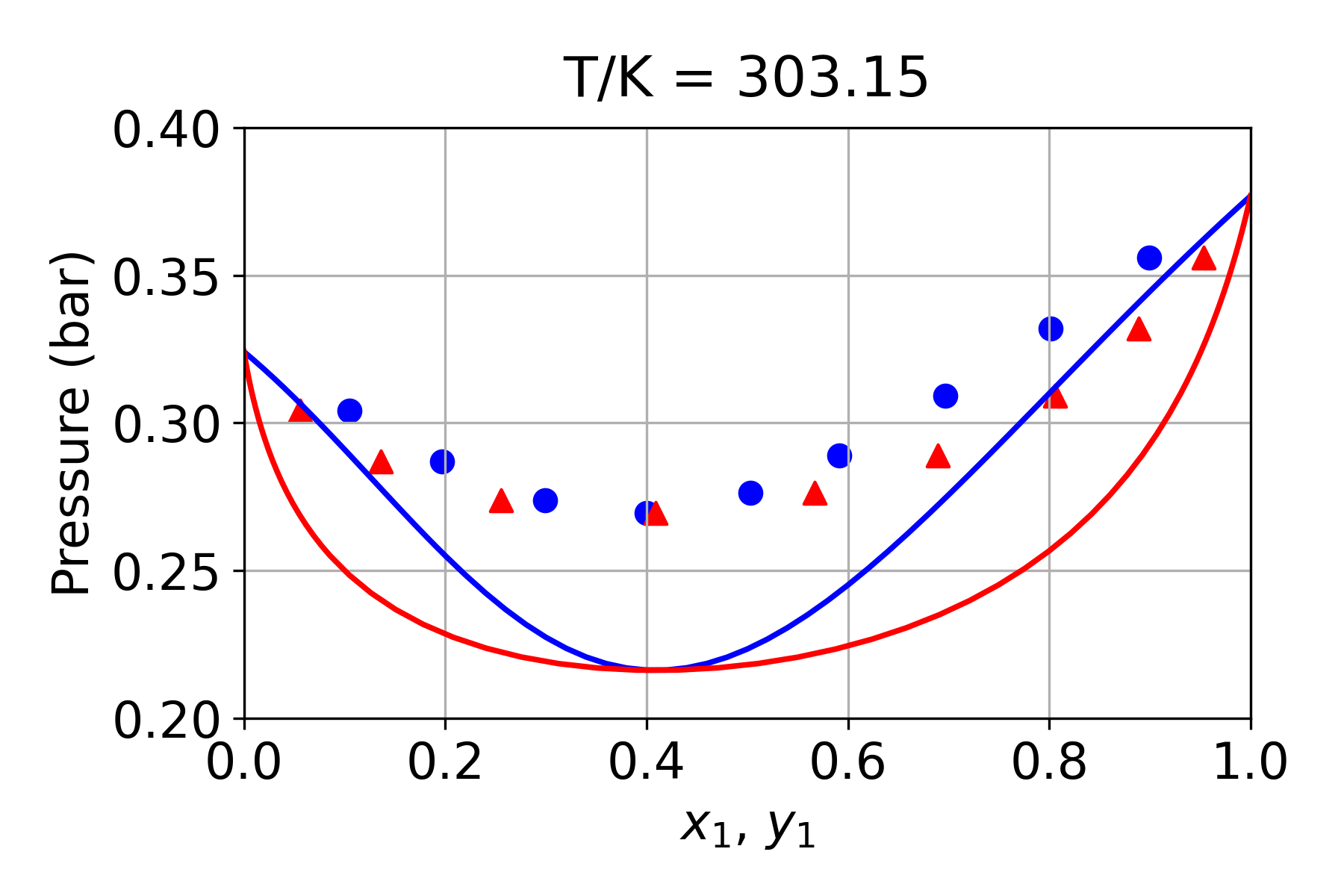}%
}\hfill
\subcaptionbox*{\centering (k)}[0.33\textwidth]{%
  \includegraphics[width=\linewidth,height=0.21735\textwidth]{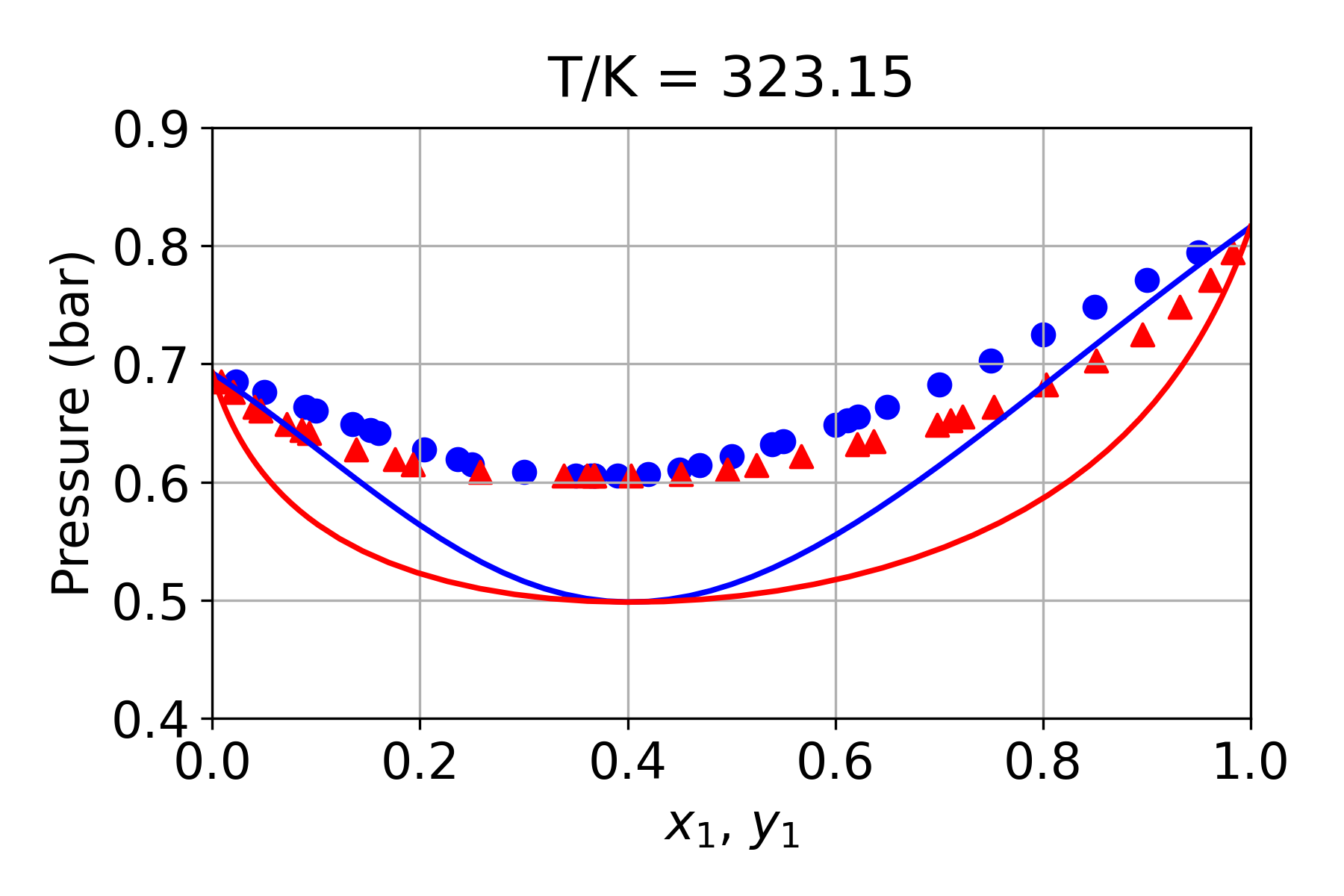}%
}\hfill
\subcaptionbox*{\centering (l)}[0.33\textwidth]{%
  \includegraphics[width=\linewidth,height=0.21735\textwidth]{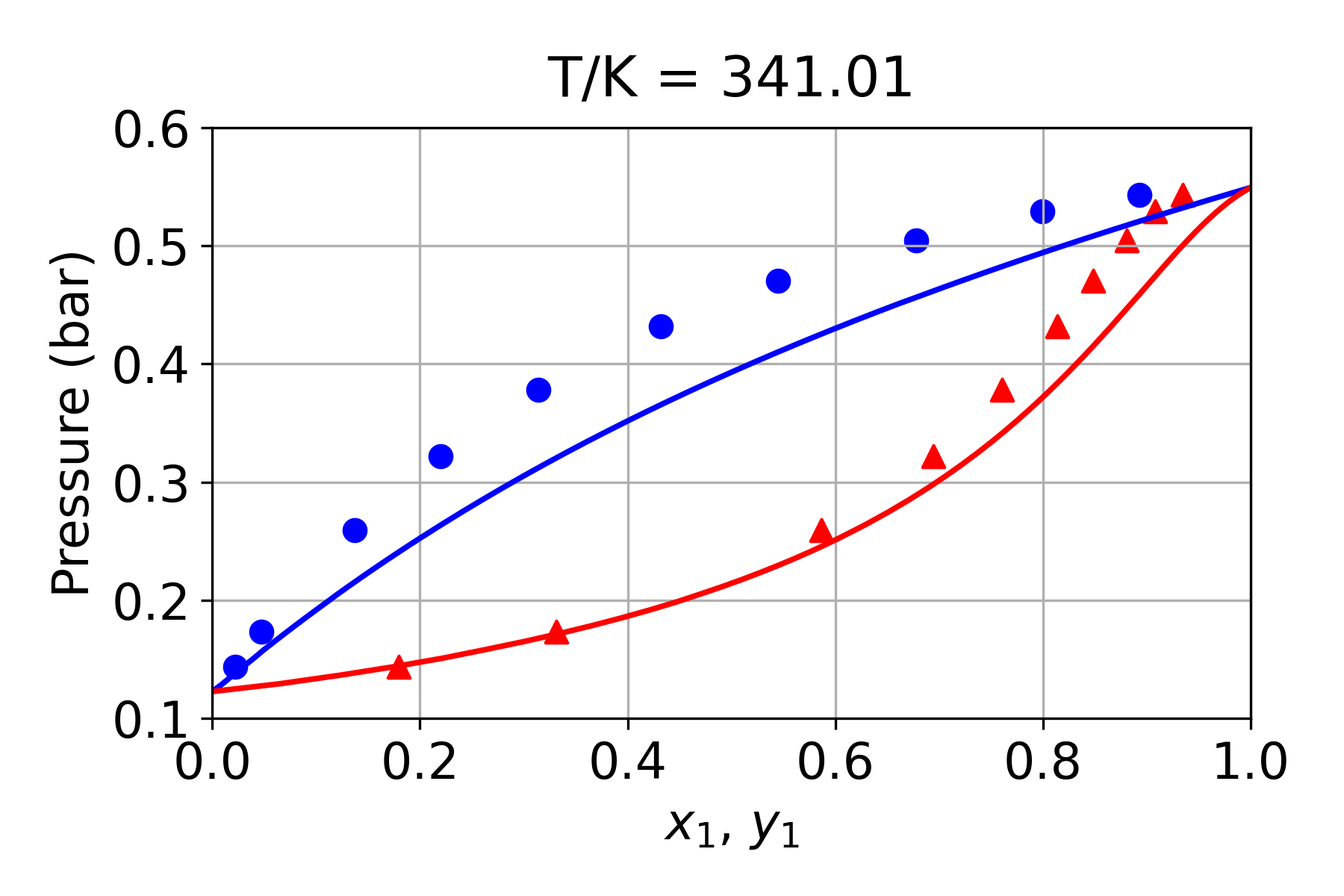}%
}

\subcaptionbox*{\centering (m)}[0.33\textwidth]{%
  \includegraphics[width=\linewidth,height=0.21735\textwidth]{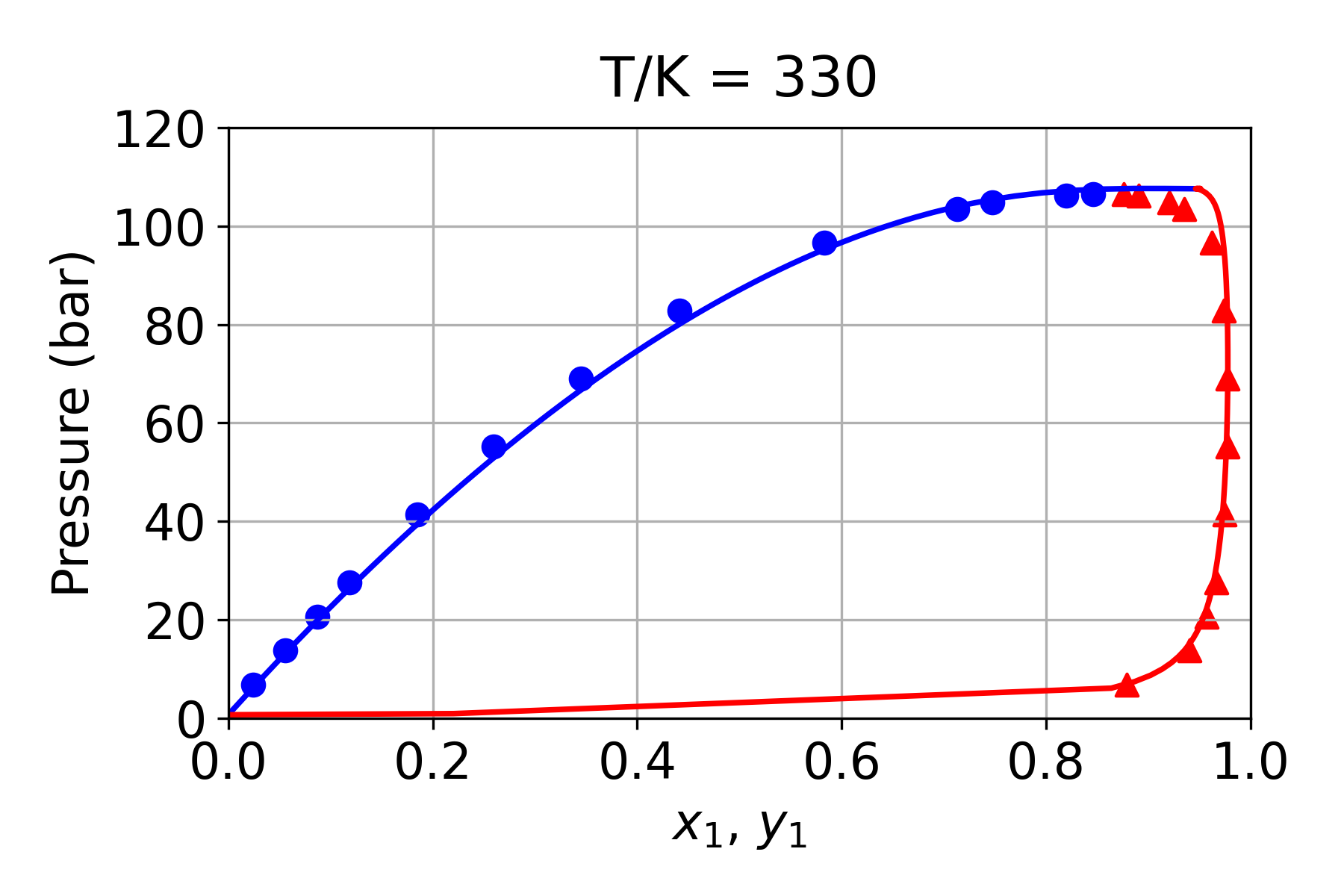}%
}\hfill
\subcaptionbox*{\centering (n)}[0.33\textwidth]{%
  \includegraphics[width=\linewidth,height=0.21735\textwidth]{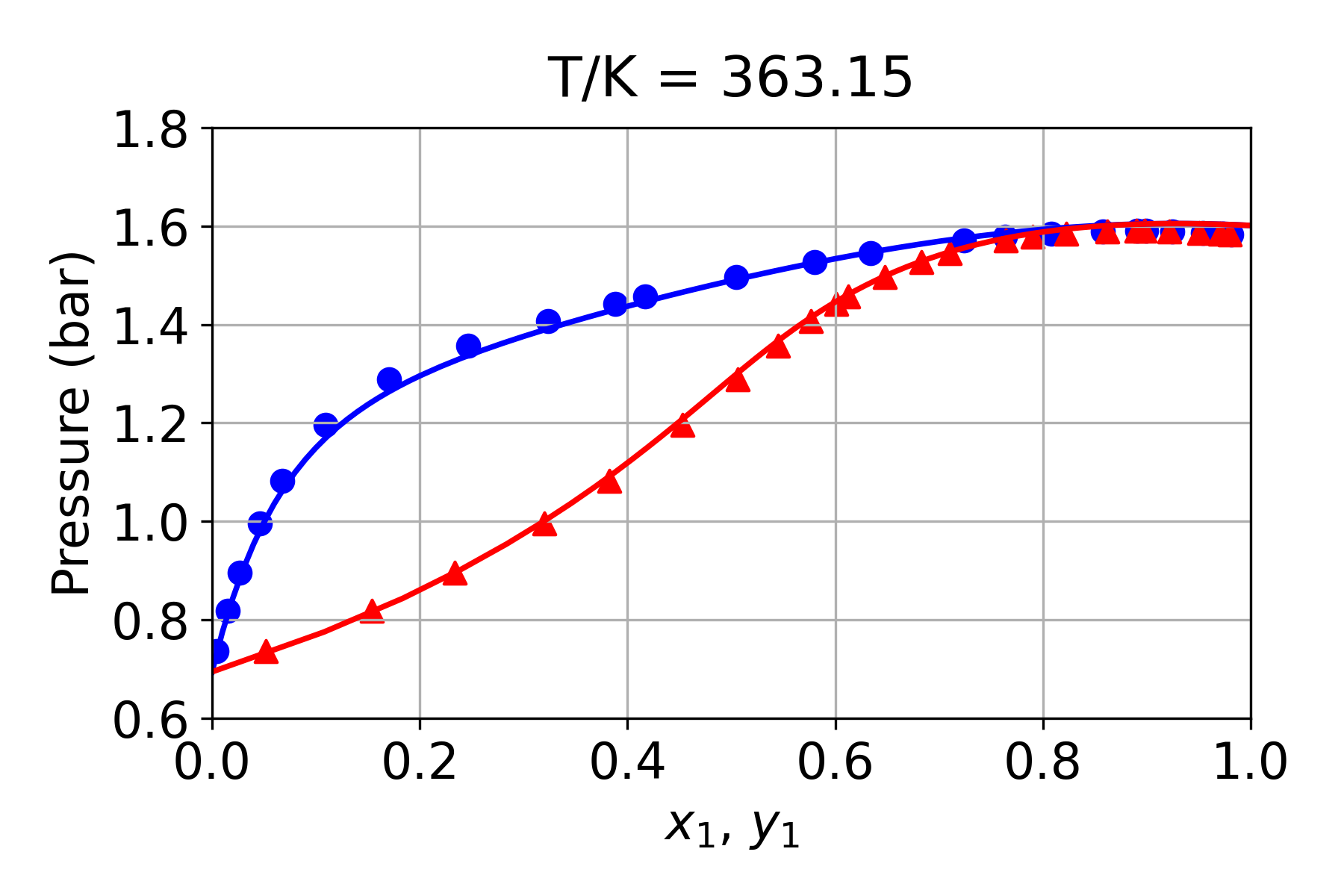}%
}\hfill
\subcaptionbox*{\centering (o)}[0.33\textwidth]{%
  \includegraphics[width=\linewidth,height=0.21735\textwidth]{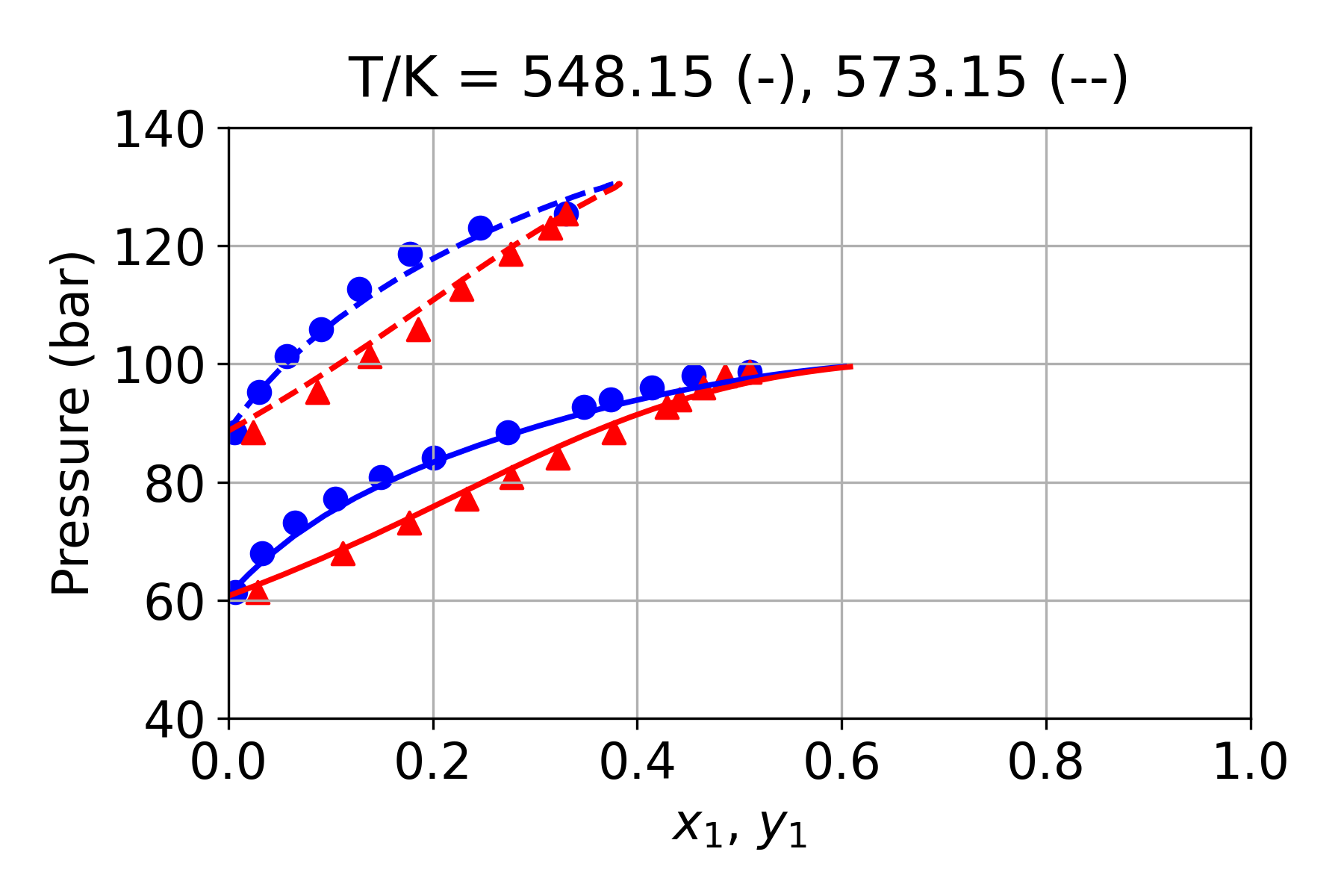}%
}

\caption{P–x–y diagrams for binary mixtures exhibiting vapor-liquid equilibrium at the specified temperature, calculated with openCOSMO-RS-Phi: {\color{blue} $\bullet$}: \(x_1^\text{exp}\), \; {\color{red} $\blacktriangle$}: \(y_1^\text{exp}\), \; {\color{blue} \rule[0.5ex]{1.5em}{0.6pt}}: \(x_1^\text{calc}\), \; {\color{red} \rule[0.5ex]{1.5em}{0.6pt}}: \(y_1^\text{calc}\); (a) nitrogen (1) - methane (\(\text{BAC}_1\)); (b) methane (1) - carbon dioxide (\(\text{BAC}_2\)); (c) acetone (1) - cyclohexane (\(\text{BAC}_2\)); (d) and (e) difluoromethane (1) - propane (\(\text{BAC}_3\)); (f) dimethyl ether (1) - sulfur dioxide (\(\text{BAC}_4\)); (g) and (h) ethanol (1) - \(n\)-heptane (\(\text{BAC}_5\)); (i) diethylamine (1) - benzene (\(\text{BAC}_5\)); (j) and (k) acetone (1) - chloroform (\(\text{BAC}_6\)); (l) trichloroethylene (1) - 2-methoxyethanol (\(\text{BAC}_7\)); (m) carbon dioxide (1) - methanol (\(\text{BAC}_8\)); and (n) and (o) ethanol (1) - water (\(\text{BAC}_9\)).}
\label{Fig:VLE_plots_all}
\end{figure}

For the prediction of liquid-phase compositions (\(\mathbf{x}\)) at given \(P\) and \(T\), the lowest MAPE is obtained for \(\text{BAC}_4\) (19.4~\%), whereas the highest MAPE is observed for \(\text{BAC}_8\) (47.0~\%), indicating the poorest performance in this class. For vapor-phase compositions (\(\mathbf{y}\)), the lowest MAPE is achieved for \(\text{BAC}_1\) (13.8~\%), while the highest value (30.2~\%) is found for \(\text{BAC}_6\). \\
The SR for two-phase equilibria ranges from 64~\% to 93~\%, with an overall average of 0.83. These values are comparable to those reported for other predictive EoS (e.g., PC-SAFT: 74-80~\%, PR: 78~\%, SAFT-VR Mie: 85~\% \cite{nikolaidis2021assessment} \cite{li2024comparative}), whereas correlative models typically reach around 95~\% \cite{pina2021optimal}. Reduced SR values, particularly for \(\text{BAC}_5\), are a common limitation of predictive approaches and reflect difficulties in reproducing the correct qualitative phase behavior. \\
For openCOSMO-RS-Phi, these deficiencies are mainly linked to an underestimation of azeotropic pressures, especially for positive azeotropes. Despite this, the phase diagrams are generally reproduced qualitatively well. The P–x–y diagrams, including the VLE region, predicted by openCOSMO-RS-Phi for 15 different binary systems are presented in \autoref{Fig:VLE_plots_all}.

\paragraph{Azeotropic data}

openCOSMO-RS-Phi achieves a score of 13.1 for azeotropic pressure (\(P_\text{azeo}\)), indicating overall good performance. The corresponding MAPEs range from 3.9~\% (\(\text{BAC}_7\)) to 21.0~\% (\(\text{BAC}_3\)). The deviations arise almost exclusively from a systematic underestimation of azeotropic pressures; overestimations are rare and occur only for a few \(\text{BAC}_9\) systems. SR values span a wide range from 0.29 to 1.0. \\
Excellent performance is observed for mixtures belonging to \(\text{BAC}_1\) (18.0/20), \(\text{BAC}_5\) (17.8/20), and \(\text{BAC}_9\) (16.6/20), all with SR = 1. For \(\text{BAC}_9\), openCOSMO-RS-Phi yields the best result among all EoS evaluated to date. In contrast, the lower scores for \(\text{BAC}_3\), \(\text{BAC}_4\), and \(\text{BAC}_8\) are primarily caused by reduced SR values. \\
However, the statistical significance of these results is limited by the relatively small azeotropic dataset (225 data points across nine BACs). In particular, the mark for \(\text{BAC}_4\) is based on only seven data points, five of which correspond to the dimethyl ether + sulfur dioxide system at different temperatures. For this system, openCOSMO-RS-Phi predicts nearly ideal behavior instead of azeotropy, resulting in a low SR (29~\%) and a substantial reduction of the overall score despite good performance for the remaining systems. \\
The scores for azeotropic compositions (\(x_\text{azeo}\)) are consistently lower than those for \(P_\text{azeo}\), in line with observations for other EoS. The best performance is again obtained for \(\text{BAC}_1\), \(\text{BAC}_5\), \(\text{BAC}_6\), and \(\text{BAC}_9\), with openCOSMO-RS-Phi achieving the highest reported score for \(\text{BAC}_9\) (9.8). Larger deviations persist for \(\text{BAC}_3\), \(\text{BAC}_4\), and \(\text{BAC}_8\). Eight systems with azeotropic behavior are presented in \autoref{Fig:VLE_plots_all} (c), (d), (e), (f), (g), (j), (k), and (n).

\paragraph{VLLE}

\begin{figure}[!b]
  \subcaptionbox*{\centering (a)}[.3\linewidth]{%
    \includegraphics[width=\linewidth]{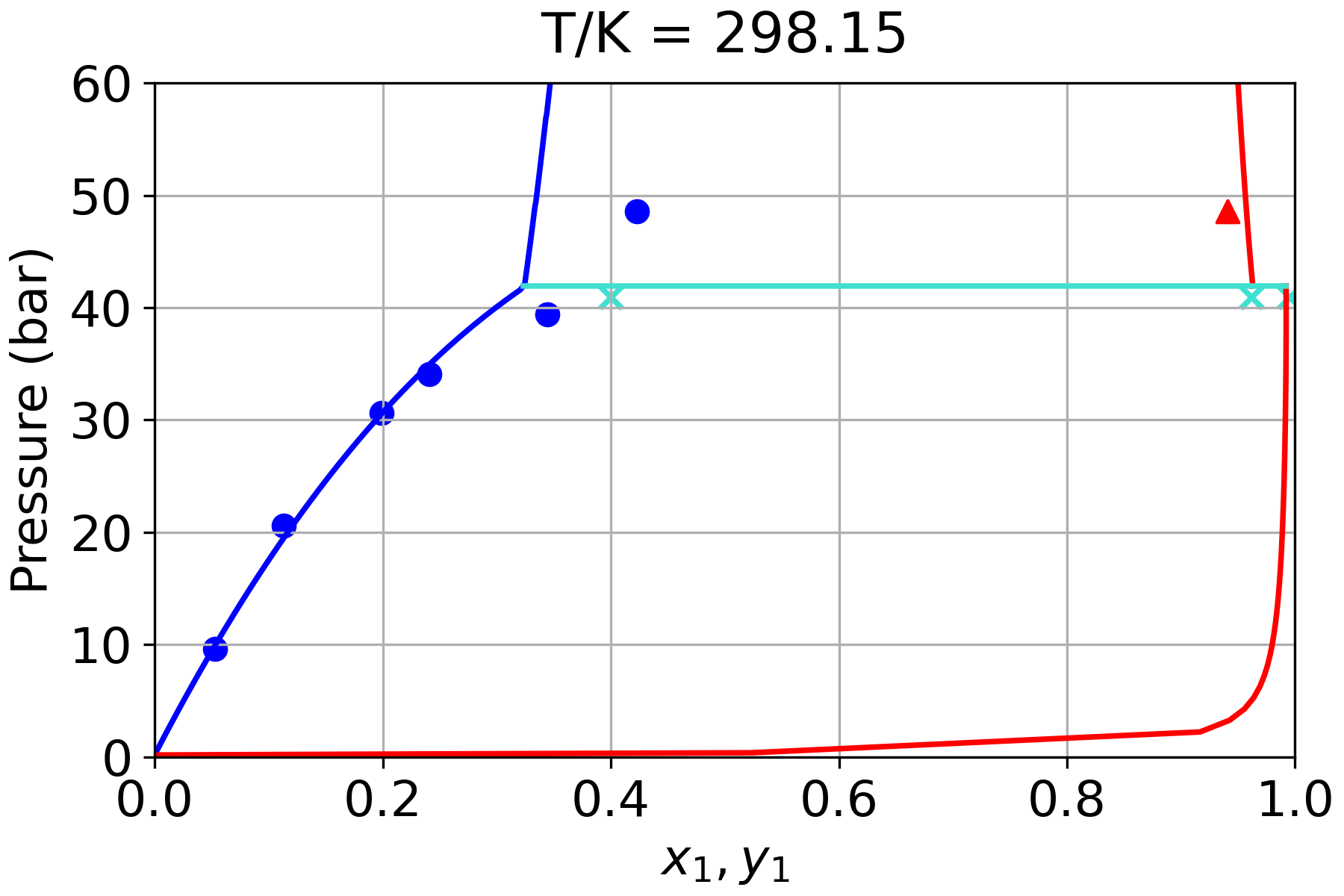}%
  }%
  \hfill
  \subcaptionbox*{\centering (b)}[.3\linewidth]{%
    \includegraphics[width=\linewidth]{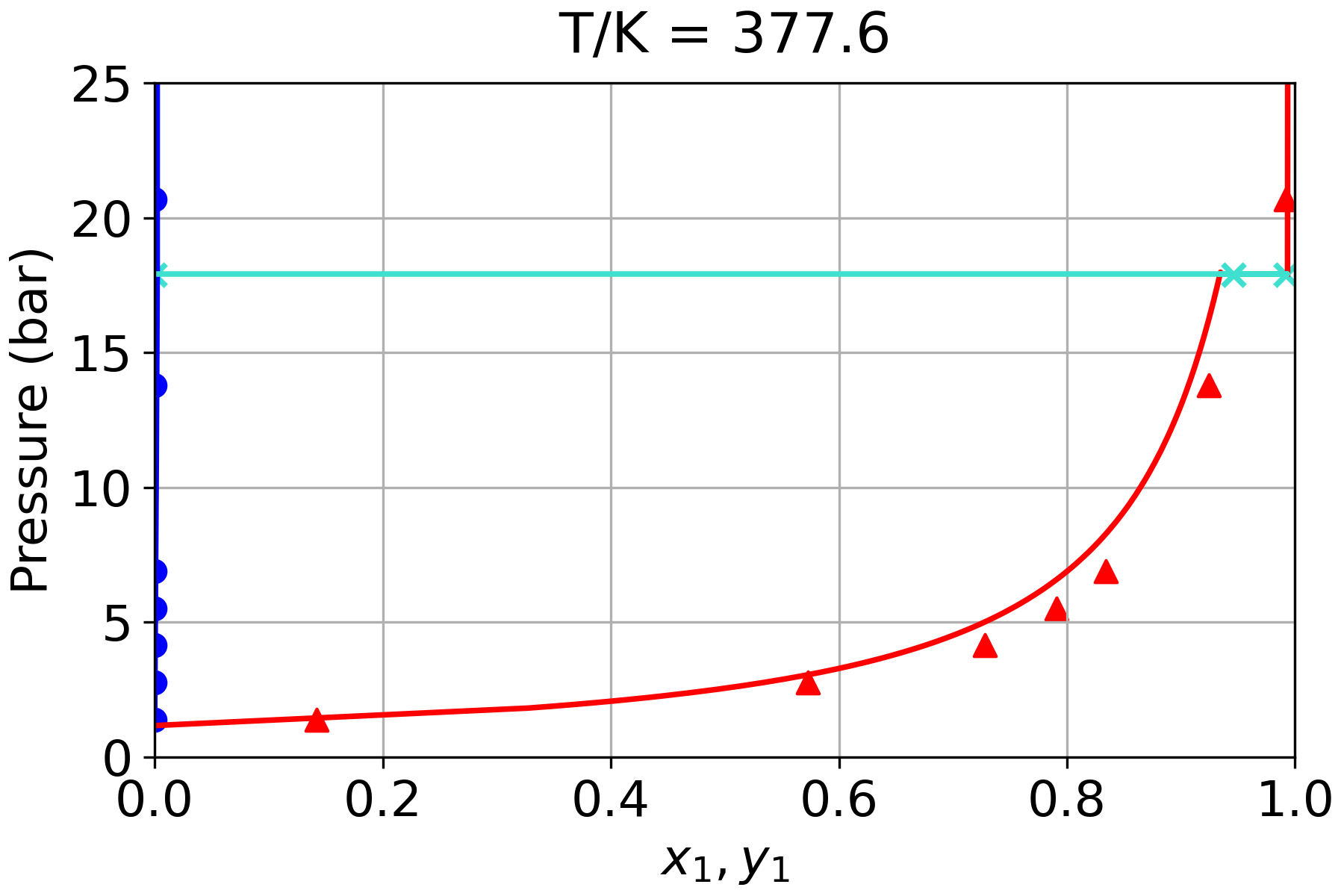}%
  }%
  \hfill
  \subcaptionbox*{\centering (c)}[.3\linewidth]{%
    \includegraphics[width=\linewidth]{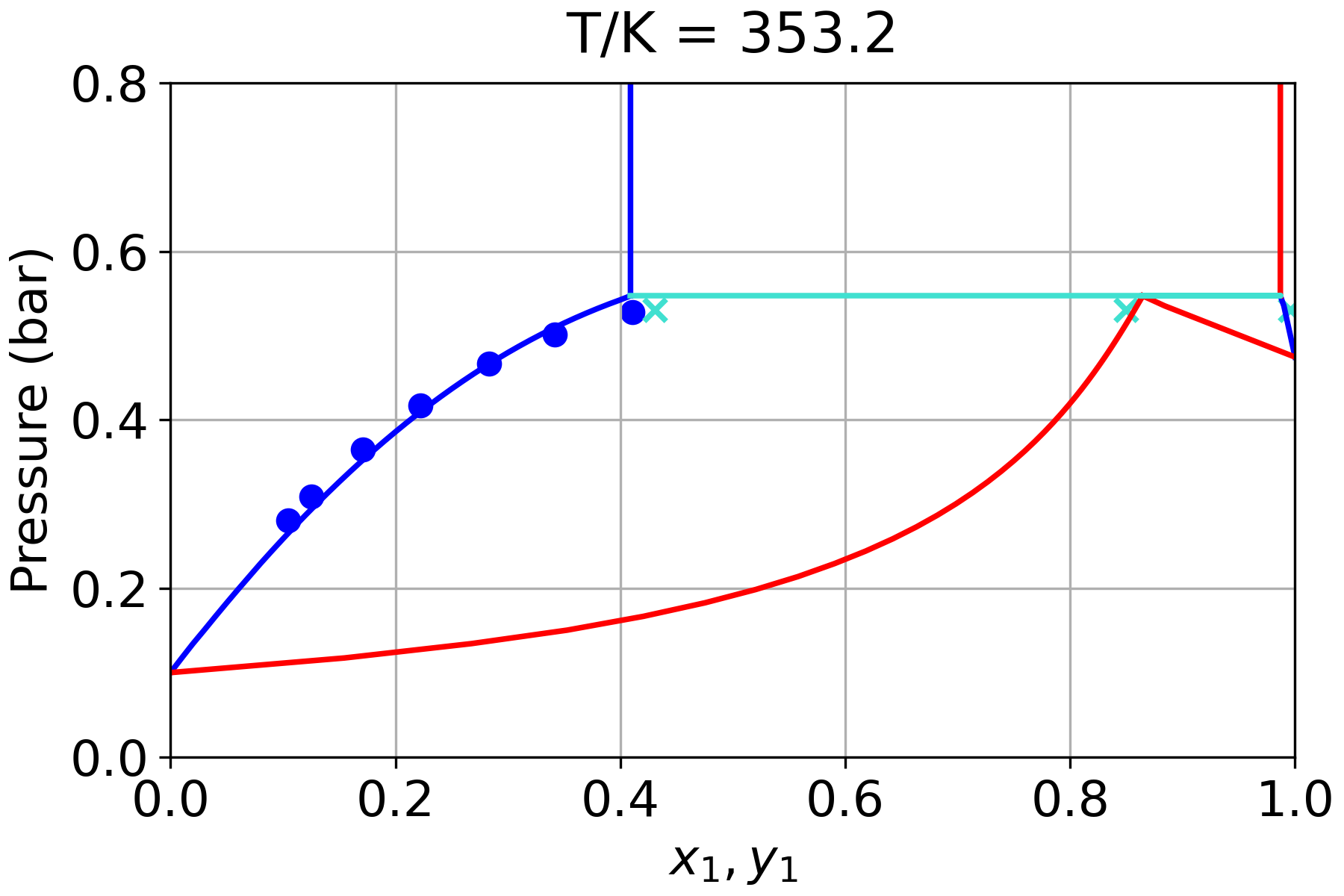}%
  }
\caption{P–x–y diagrams for three binary mixtures that exhibit a VLLE for the specified temperature calculated with openCOSMO-RS-Phi: {\color{blue} $\bullet$}: \(x_1^\text{exp}\) / \(x_1^{\alpha, \text{exp}}\) , \; {\color{red} $\blacktriangle$}: \(y_1^\text{exp}\) / \(x_1^{\beta, \text{exp}}\), \; {\color{turquoise} $\times$}: \(z_\text{LLV}^\text{exp}\), \; {\color{blue} \rule[0.5ex]{1.5em}{0.6pt}}: \(x_1^\text{calc}\) / / \(x_1^{\alpha, \text{calc}}\), \; {\color{red} \rule[0.5ex]{1.5em}{0.6pt}}: \(y_1^\text{calc}\) / / \(x_1^{\beta, \text{calc}}\), \; {\color{turquoise} \rule[0.5ex]{1.5em}{0.6pt}}: calculated three-phase line; (a) ethane (1) - methanol (\(\text{BAC}_5\)), (b) \(n\)-butane (1) - water (\(\text{BAC}_5\)), (c) water (1) - 1-pentanol (\(\text{BAC}_9\)).}
\label{Fig:VLLE plots}
\end{figure}

Three-phase line data are available for \(\text{BAC}_1\), \(\text{BAC}_5\), \(\text{BAC}_8\), and \(\text{BAC}_9\) systems. For both three-phase pressure (\(P_\text{LLV}\)) and composition (\(z_\text{LLV}\)), openCOSMO-RS-Phi performs significantly better than most predictive EoS and reaches a level comparable to that of correlative models. \\
A notable exception is the very low SR value for \(\text{BAC}_8\) (8~\%). This is mainly due to the composition of the dataset: 11 out of 12 experimental three-phase lines correspond to the dimethyl ether + water system, for which openCOSMO-RS-Phi fails to predict heteroazeotropic behavior. Although the remaining system, CO\(_2\) + water at 298.15~K, is described very accurately, the overall result is dominated by dimethyl ether + water. \\
In \autoref{Fig:VLLE plots}, P–x–y diagrams including both the VLE and LLE regions predicted by openCOSMO-RS-Phi for three binary systems are presented.
%This observation also highlights a limitation of previous benchmarks. In some studies, cases with SR = 0 (e.g., \(\text{BAC}_1\) for tc-PR/COSMO-SAC-2002 and -2010, or \(\text{BAC}_9\) for basic PC-SAFT) were effectively excluded from the averaging procedure, even though a score of 0 could formally be assigned. While this omission has only a minor impact on the global score, it introduces an inconsistency in the evaluation. In the present work, such cases are explicitly accounted for. Consequently, the low score for \(\text{BAC}_8\), associated with an SR of 8~\%, is retained in the overall assessment. \\

\paragraph{Critical data}

\begin{figure}[h!]
    \subcaptionbox*{\centering (a)}[0.45\textwidth]{%
  \includegraphics[width=\linewidth,height=0.45\textwidth]{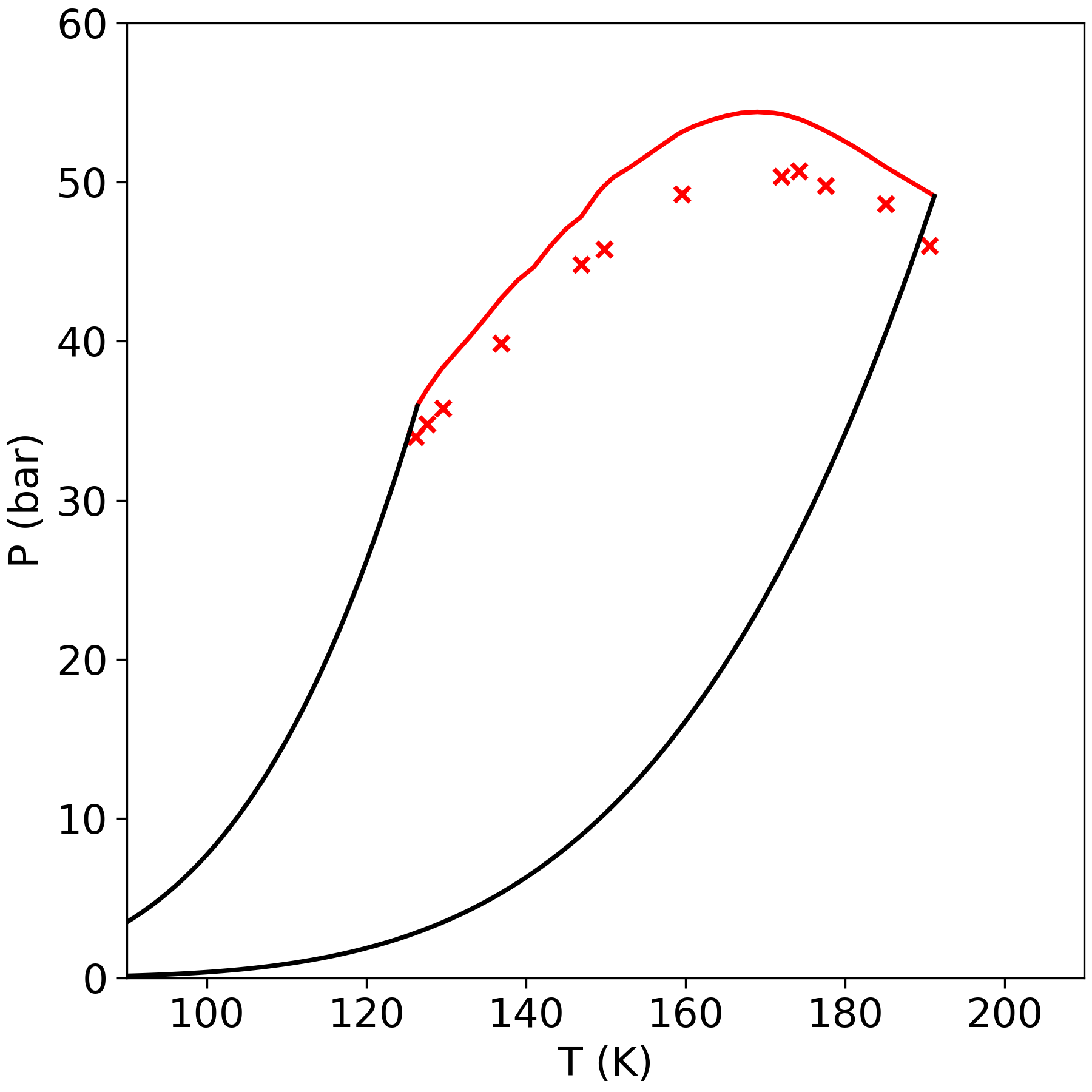}%
}\hfill
\subcaptionbox*{\centering (b)}[0.45\textwidth]{%
  \includegraphics[width=\linewidth,height=0.45\textwidth]{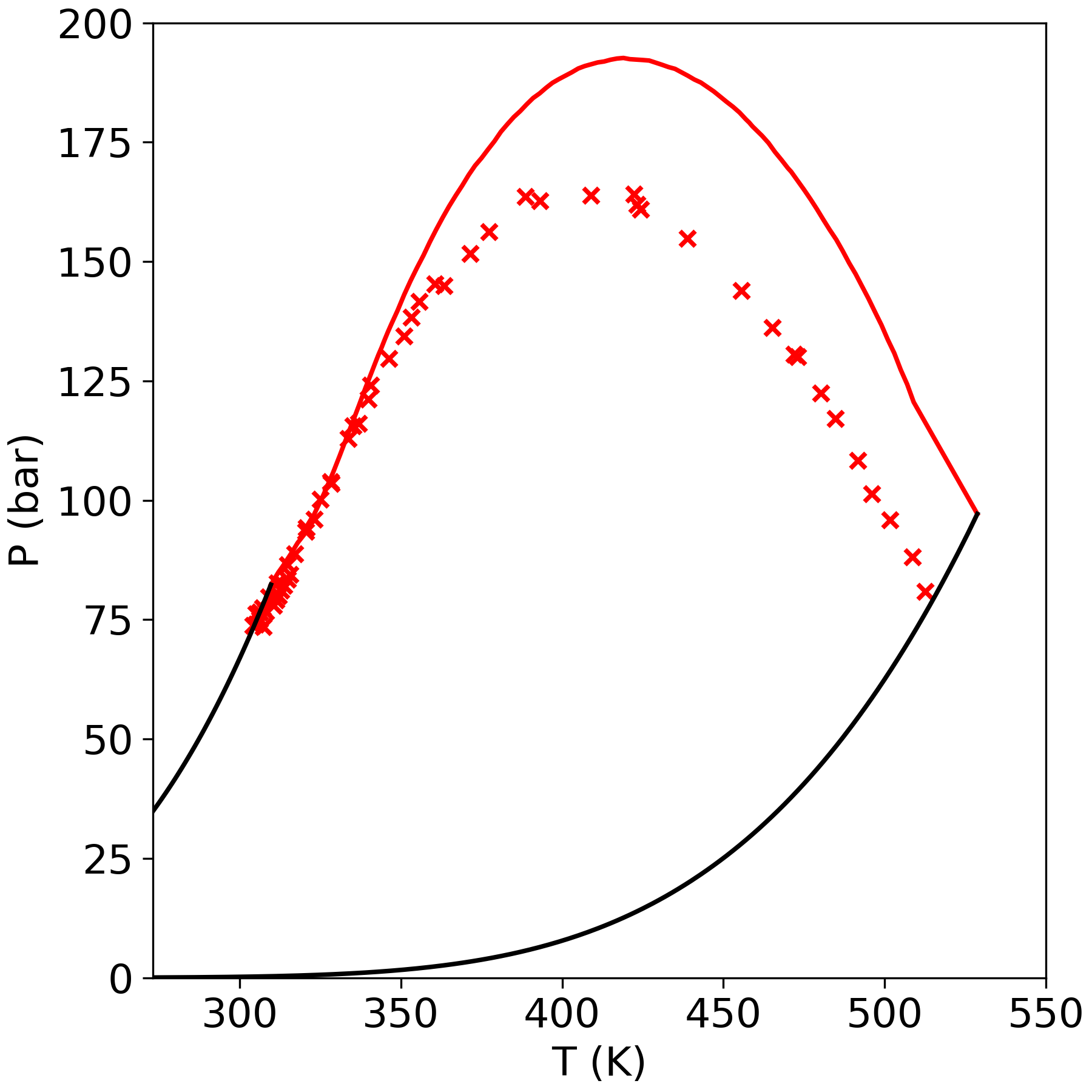}%
}
    \caption{Global phase equilibrium diagram (GPED) for two binary systems calculated with openCOSMO-RS-Phi: (a) \(\text{N}_2\) (1) - methane (\(\text{BAC}_1\)), (b) carbon dioxide (1) - methanol (\(\text{BAC}_8\)). It includes the vapor-pressure curves of the pure components (black lines), experimental critical points (\textcolor{red}{\(\times\)}) and the calculated critical line (red line). The critical line is evaluated on a temperature grid with \(\Delta T = 2 \ \text{K}\) and mixture critical points are determined using a tracing procedure as outlined in \autoref{Thermodynamic Calculations}.}
    \label{fig:CLplot}
\end{figure}

A clear weakness of openCOSMO-RS-Phi is the description of mixture critical points. For the critical composition (\(x_\text{c}\)), the overall score (Mark20) is 0.0, resulting from MAPEs above 40~\% for all BACs except \(\text{BAC}_1\). \\
A similar trend is observed for the critical pressure (\(P_\text{c}\)), although openCOSMO-RS-Phi performs somewhat better than the other predictive EoS, achieving a score of 4.0. The large deviations are mainly caused by an inaccurate representation of the bubble and dew curves near the critical region, typically leading to an overestimation of the critical pressure. This behavior is illustrated in \autoref{fig:CLplot}, which shows the critical line for the non-associating binary system \(\text{N}_2\) + \(\text{CH}_4\). \\
In addition to the MAPE, the SR values provide further insight. For \(\text{BAC}_5\) systems, only 47~\% of the critical points are reproduced, whereas \(\text{BAC}_4\) shows the highest SR (87~\%). Although openCOSMO-RS-Phi attains slightly higher average SR values than PC-SAFT or COSMO-based tc-PR EoS, these values still indicate significant deficiencies. As seen in \autoref{Fig:Parity plots PC critical data}, the model systematically overestimates critical temperatures of pure components, such that no critical point is predicted for mixtures at temperatures slightly above \(\min(T_{c,1}, T_{c,2})\).

\subsubsection{Derivative Properties}

\begin{figure}
  \subcaptionbox*{\centering (a)}[.3\linewidth]{%
    \includegraphics[width=\linewidth]{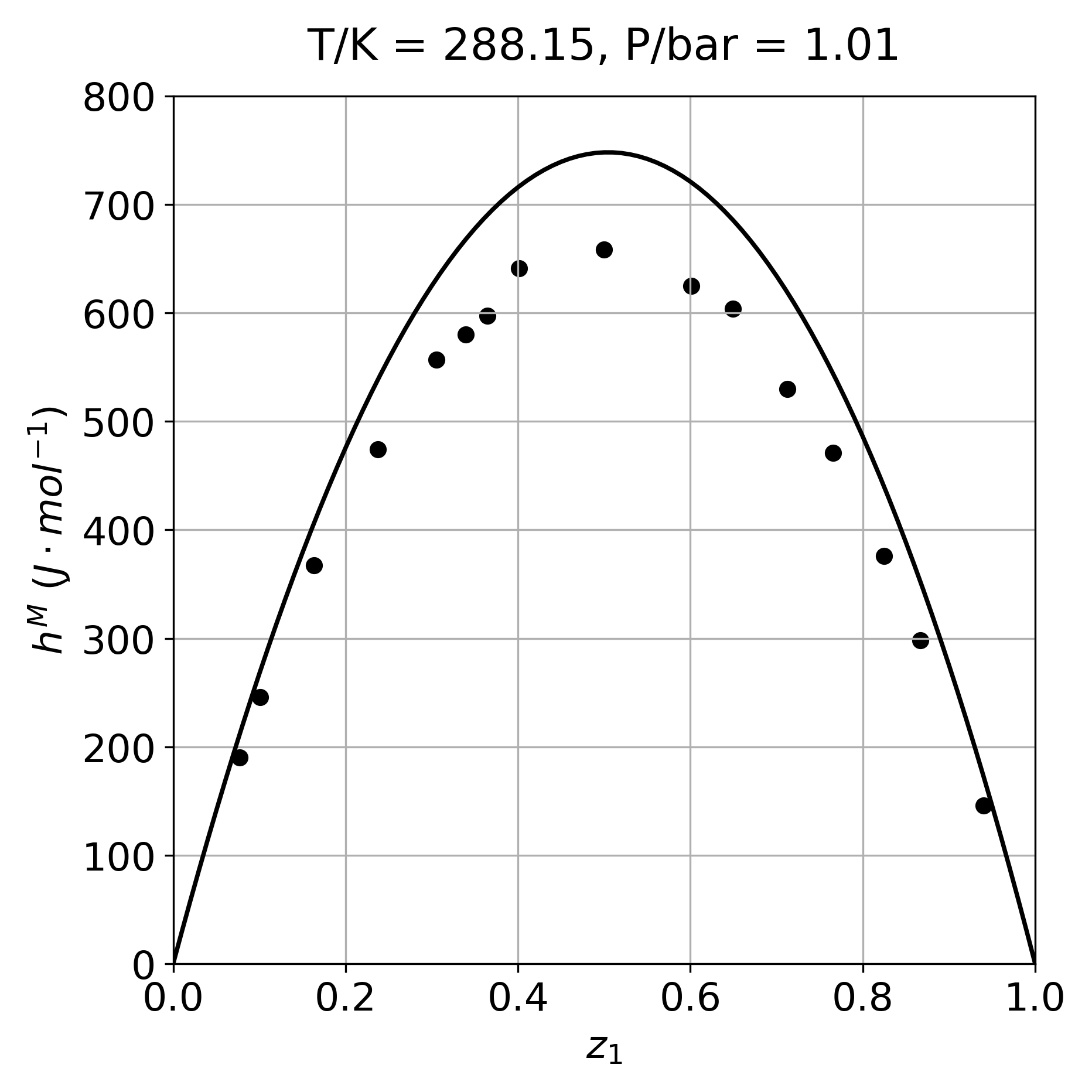}%
  }%
  \hfill
  \subcaptionbox*{\centering (b)}[.3\linewidth]{%
    \includegraphics[width=\linewidth]{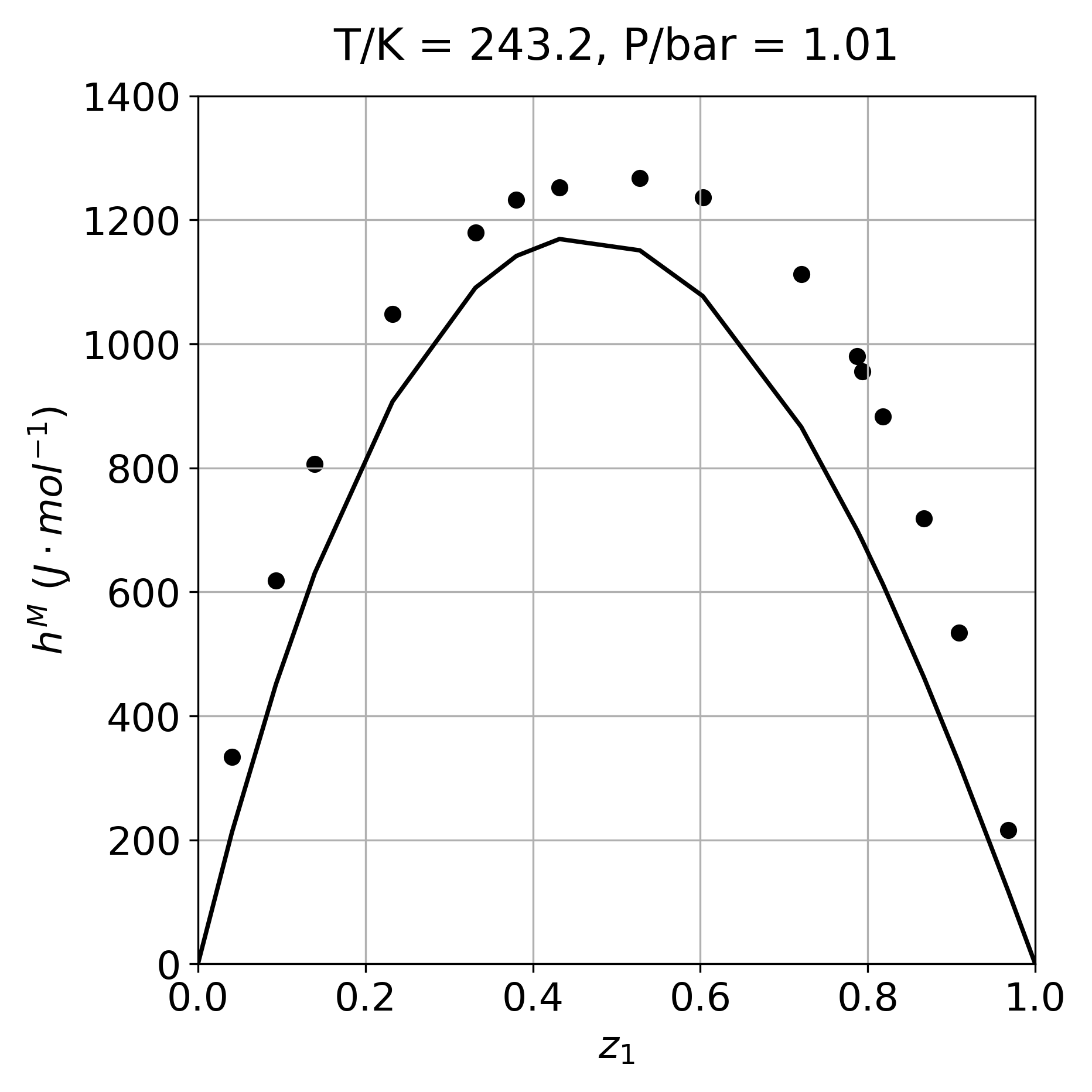}%
  }%
  \hfill
  \subcaptionbox*{\centering (c)}[.3\linewidth]{%
    \includegraphics[width=\linewidth]{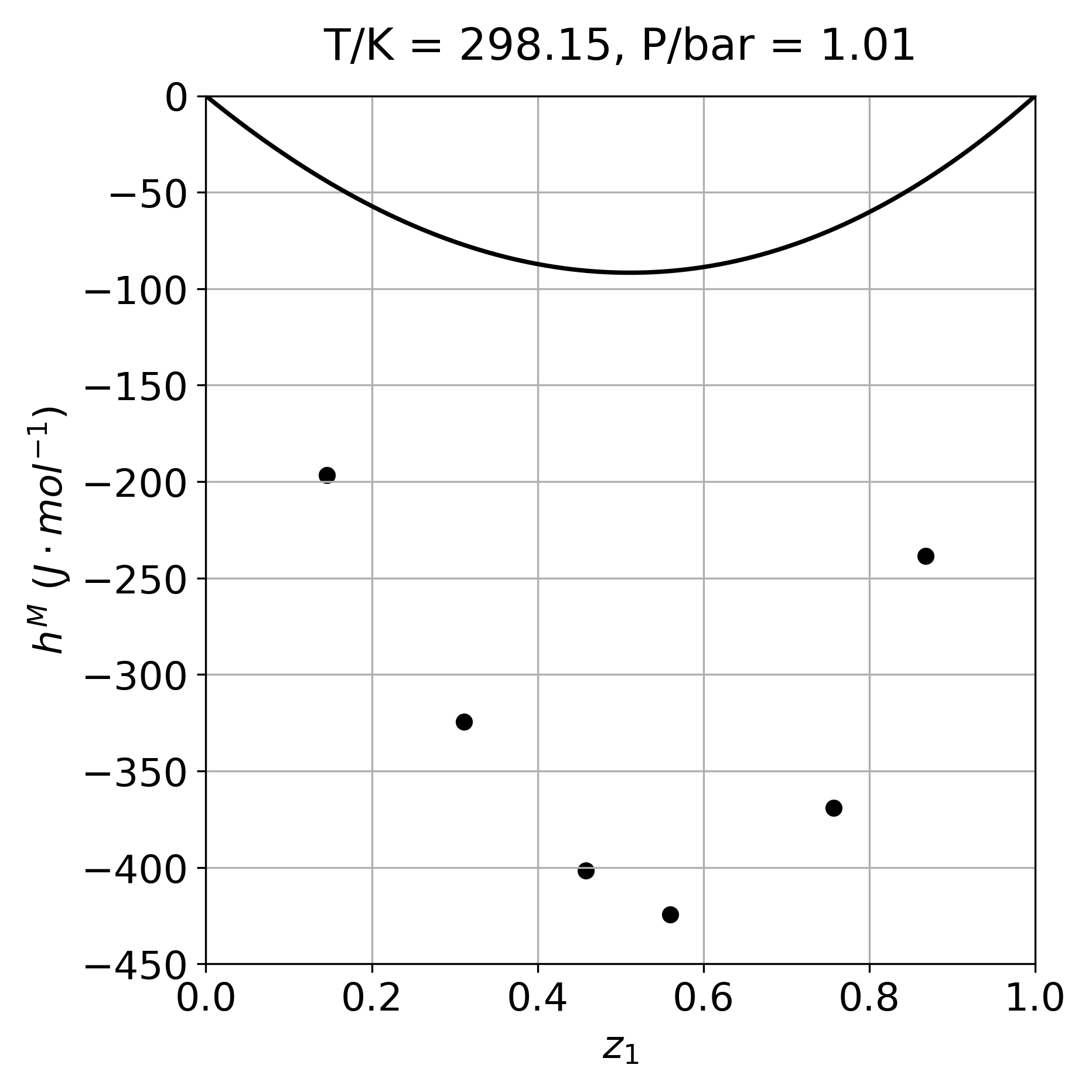}%
  } 
  \subcaptionbox*{\centering (d)}[.3\linewidth]{%
    \includegraphics[width=\linewidth]{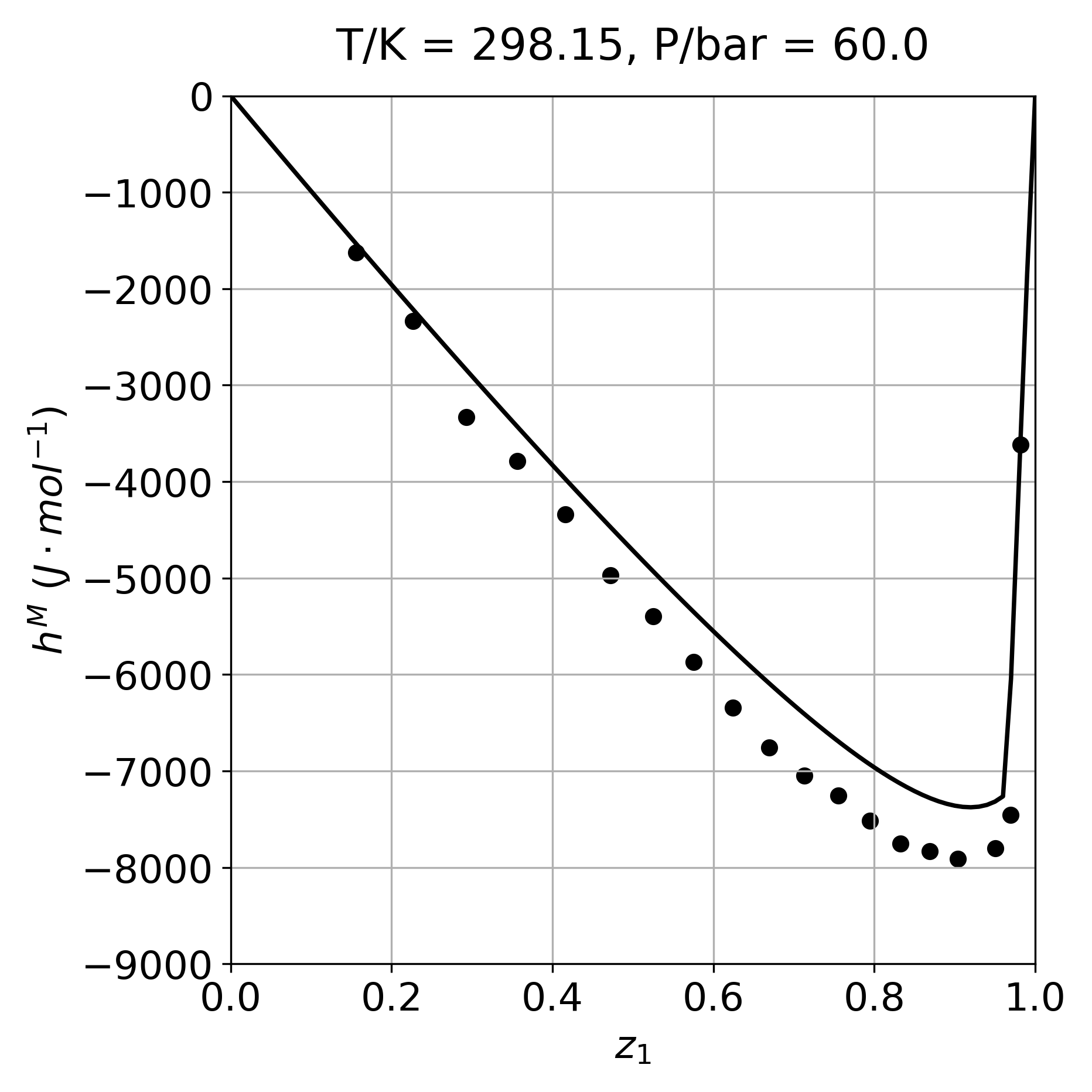}%
  }%
  \hfill
  \subcaptionbox*{\centering (e)}[.3\linewidth]{%
    \includegraphics[width=\linewidth]{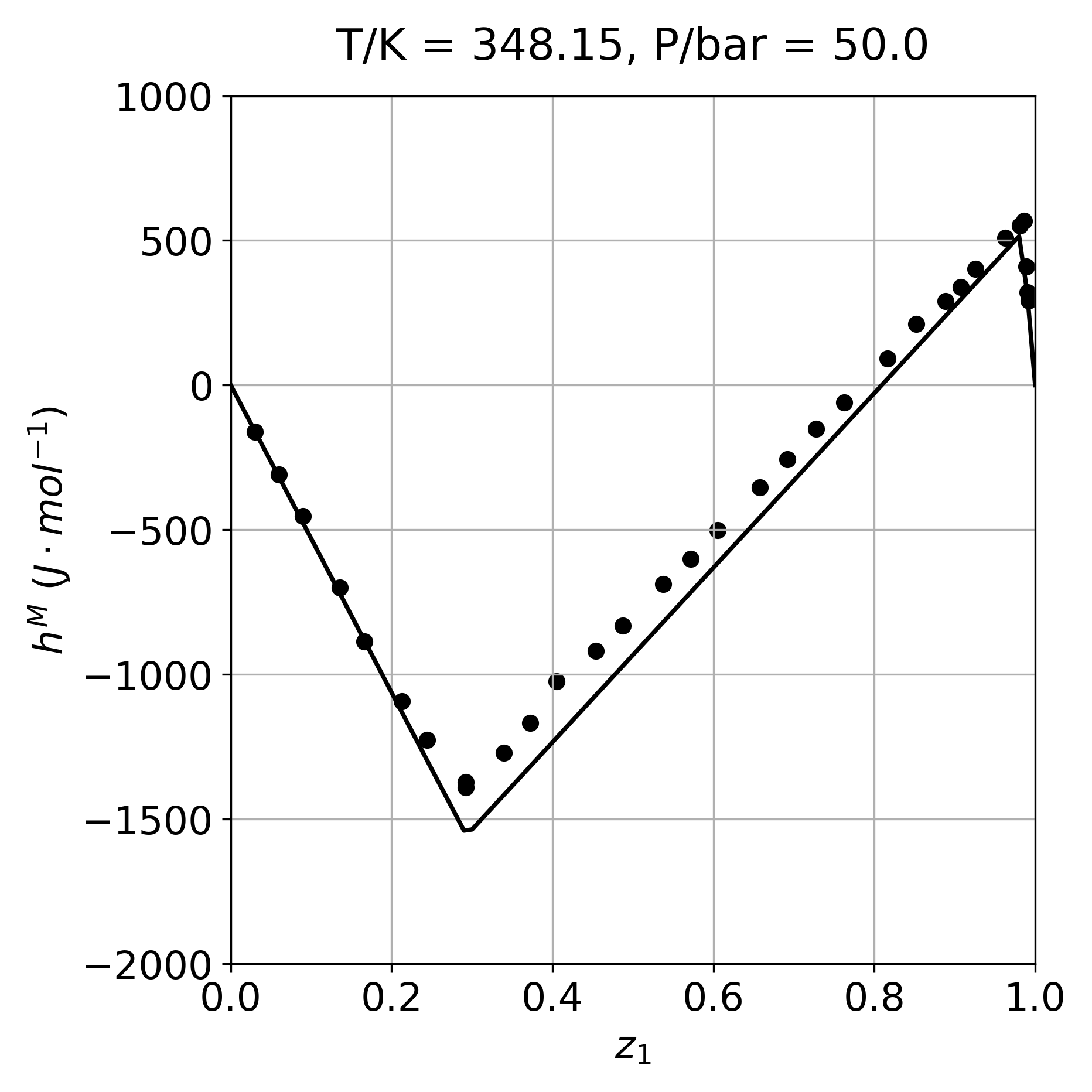}%
  }%
  \hfill
  \subcaptionbox*{\centering (f)}[.3\linewidth]{%
    \includegraphics[width=\linewidth]{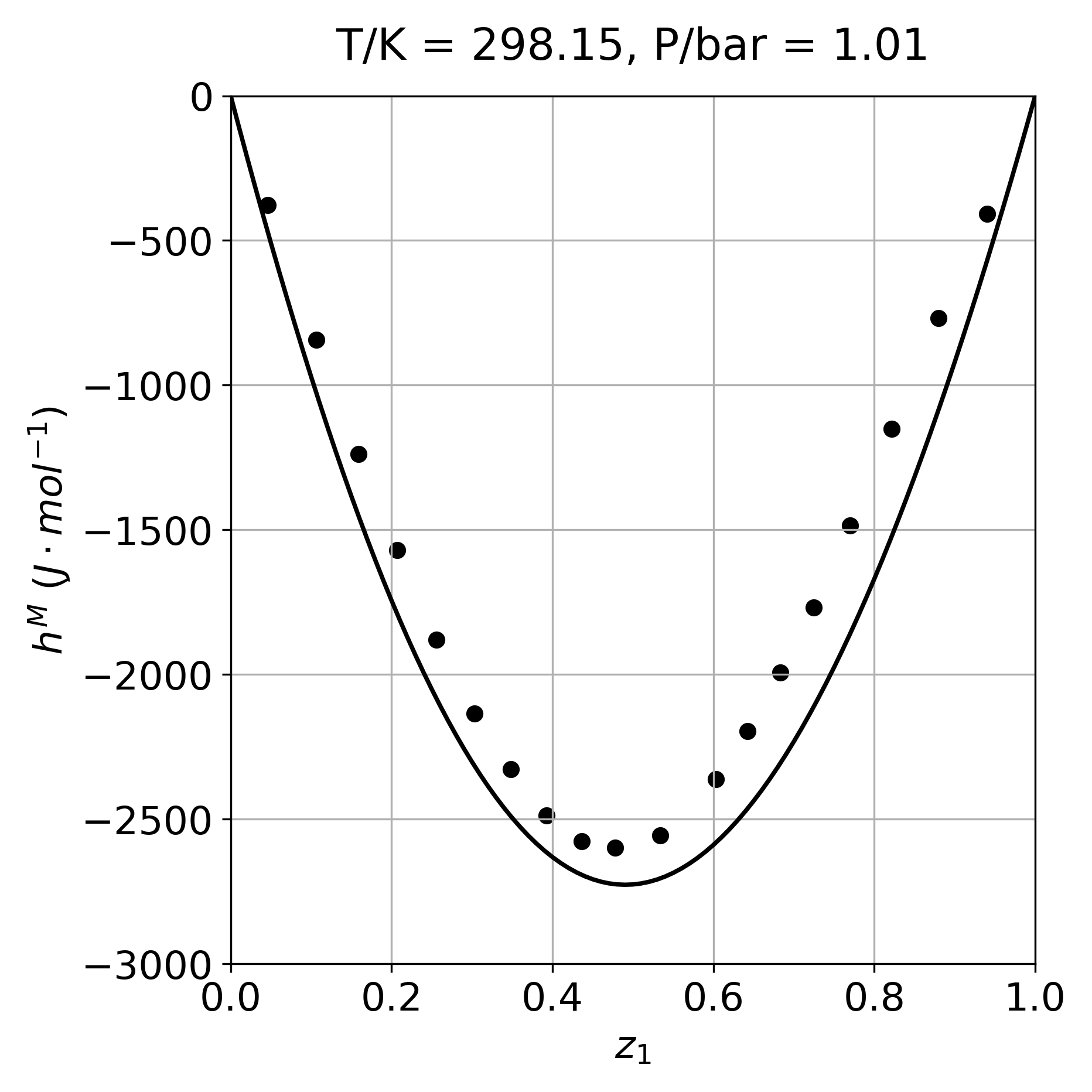}%
  }
  \subcaptionbox*{\centering (g)}[.3\linewidth]{%
    \includegraphics[width=\linewidth]{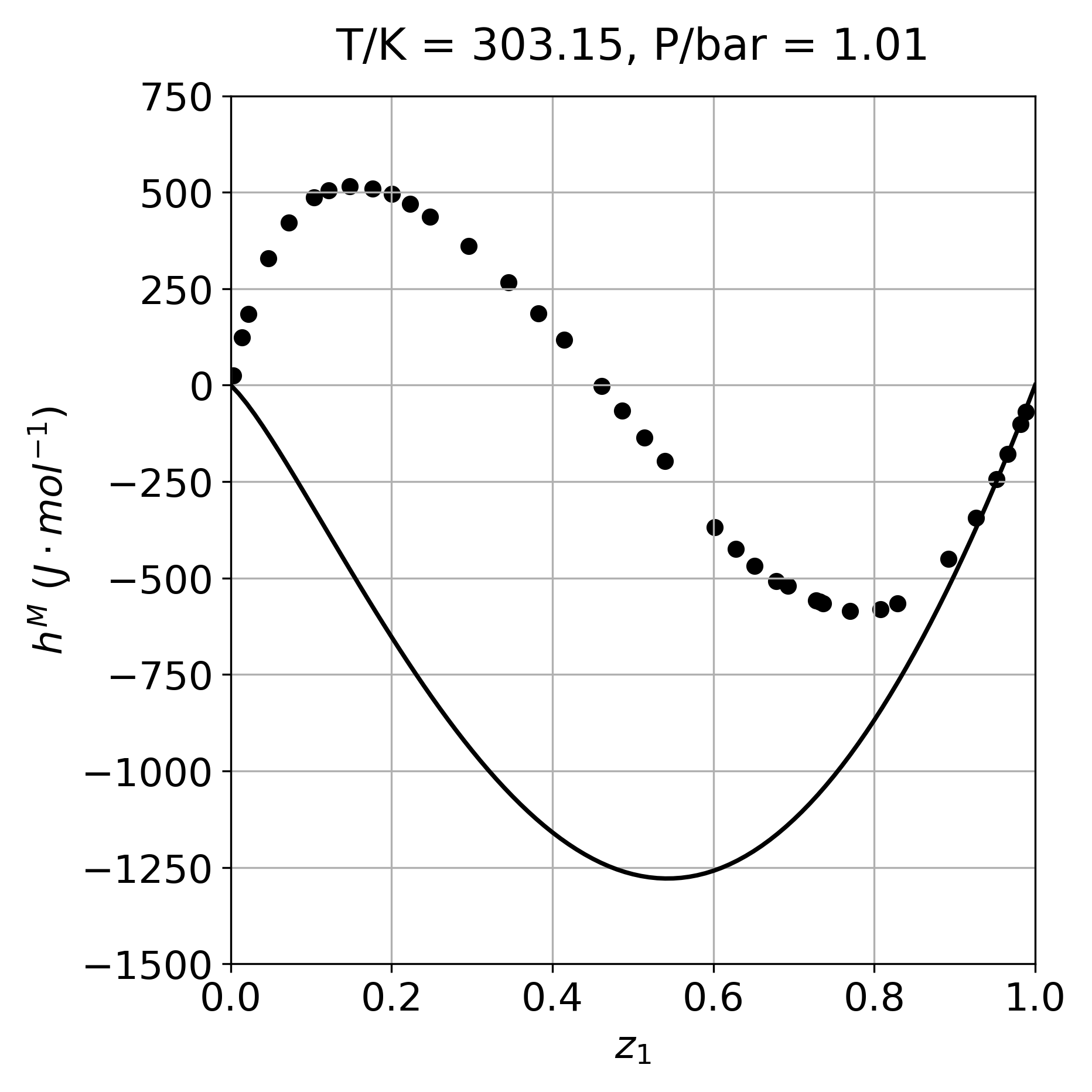}%
  }%
  \hfill
  \subcaptionbox*{\centering (h)}[.3\linewidth]{%
    \includegraphics[width=\linewidth]{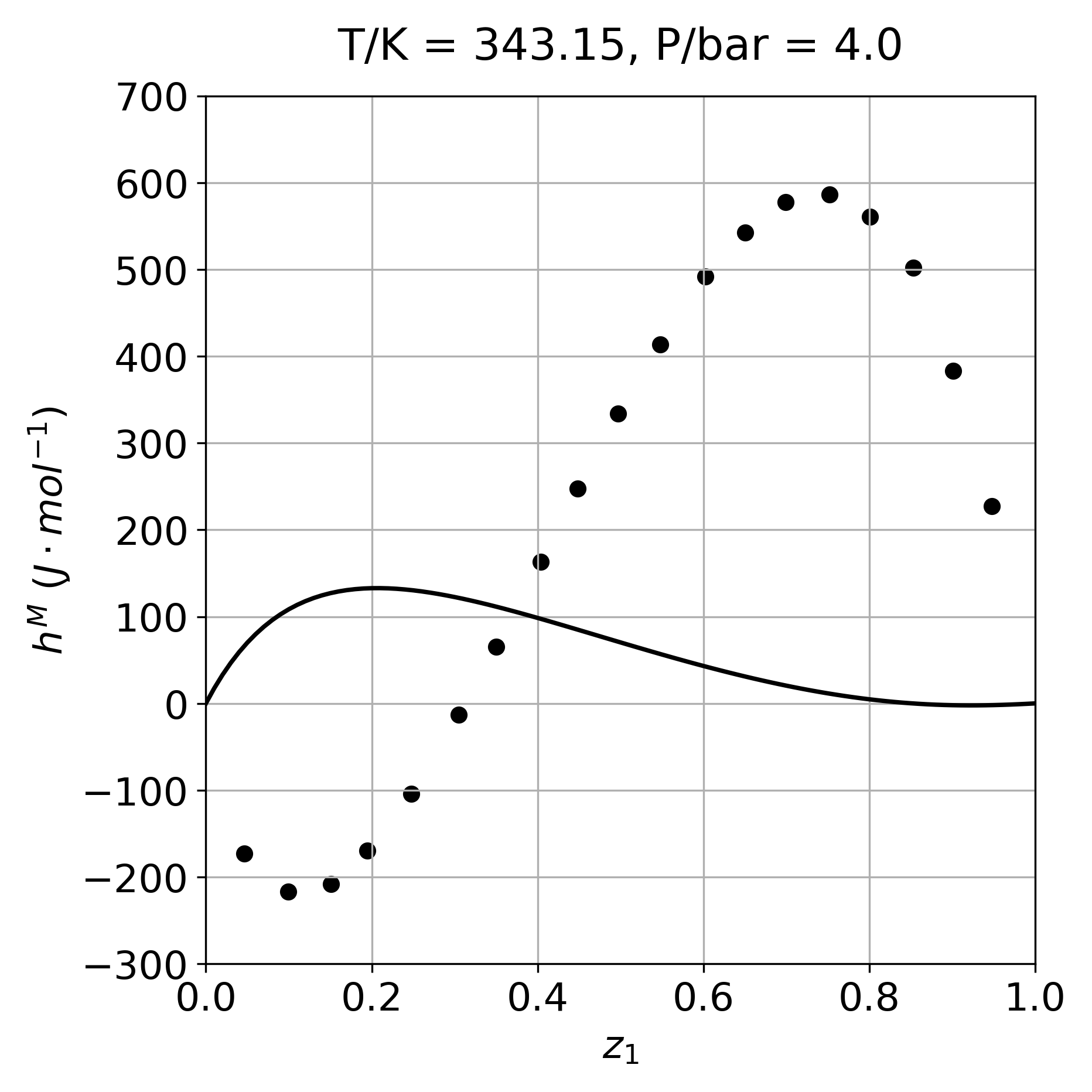}%
  }%
  \hfill
  \subcaptionbox*{\centering (i)}[.3\linewidth]{%
    \includegraphics[width=\linewidth]{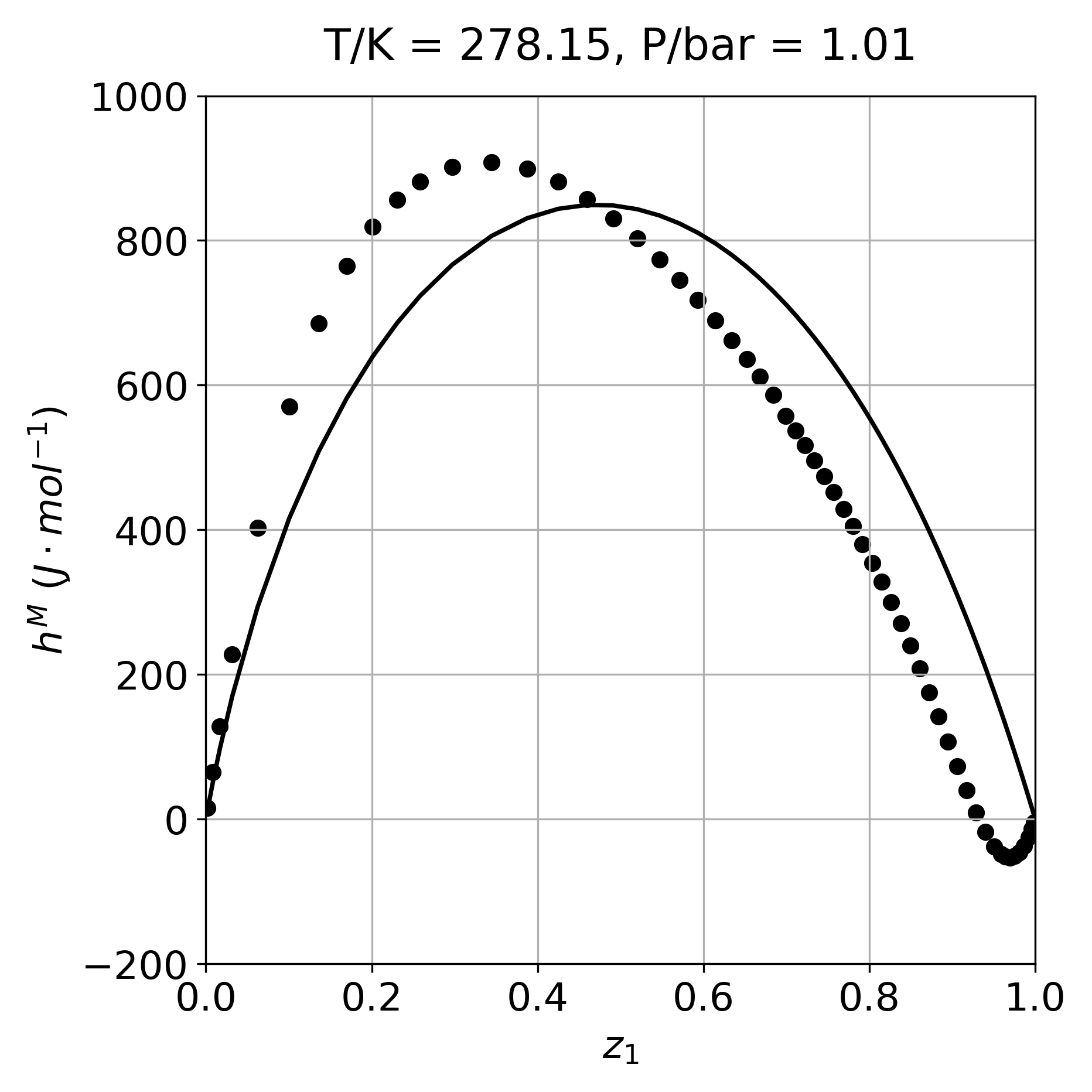}%
  }
\caption{Enthalpy changes of mixing for nine binary systems calculated with openCOSMO-RS-Phi: {\color{black} $\bullet$}: \(h^\text{M,exp}\), \; {\color{black} \rule[0.5ex]{1.5em}{0.6pt}} \(h^\text{M,calc}\);  (a) 1,2-dichloroethane (1) - carbon tetrachloride (\(\text{BAC}_1\)), (b) acetone (1) - n-hexane (\(\text{BAC}_2\)), (c) chloroform (1) - benzene (\(\text{BAC}_3\)), (d) carbon dioxide (1) - dimethyl carbonate (\(\text{BAC}_4\)), (e) ethane (1) - 1-propanol (\(\text{BAC}_5\)), (f) ethyl acetate (1) - 1,1,2,2-tetrachloroethane (\(\text{BAC}_6\)), (g) ethanol (1) - chloroform (\(\text{BAC}_7\)), (h) acetone (1) - water (\(\text{BAC}_8\)) and (i) water (1) - acetonitrile (\(\text{BAC}_9\)).}
\label{Fig:hM plots}
\end{figure}

\begin{figure}
  \subcaptionbox*{\centering (a)}[.3\linewidth]{%
    \includegraphics[width=\linewidth]{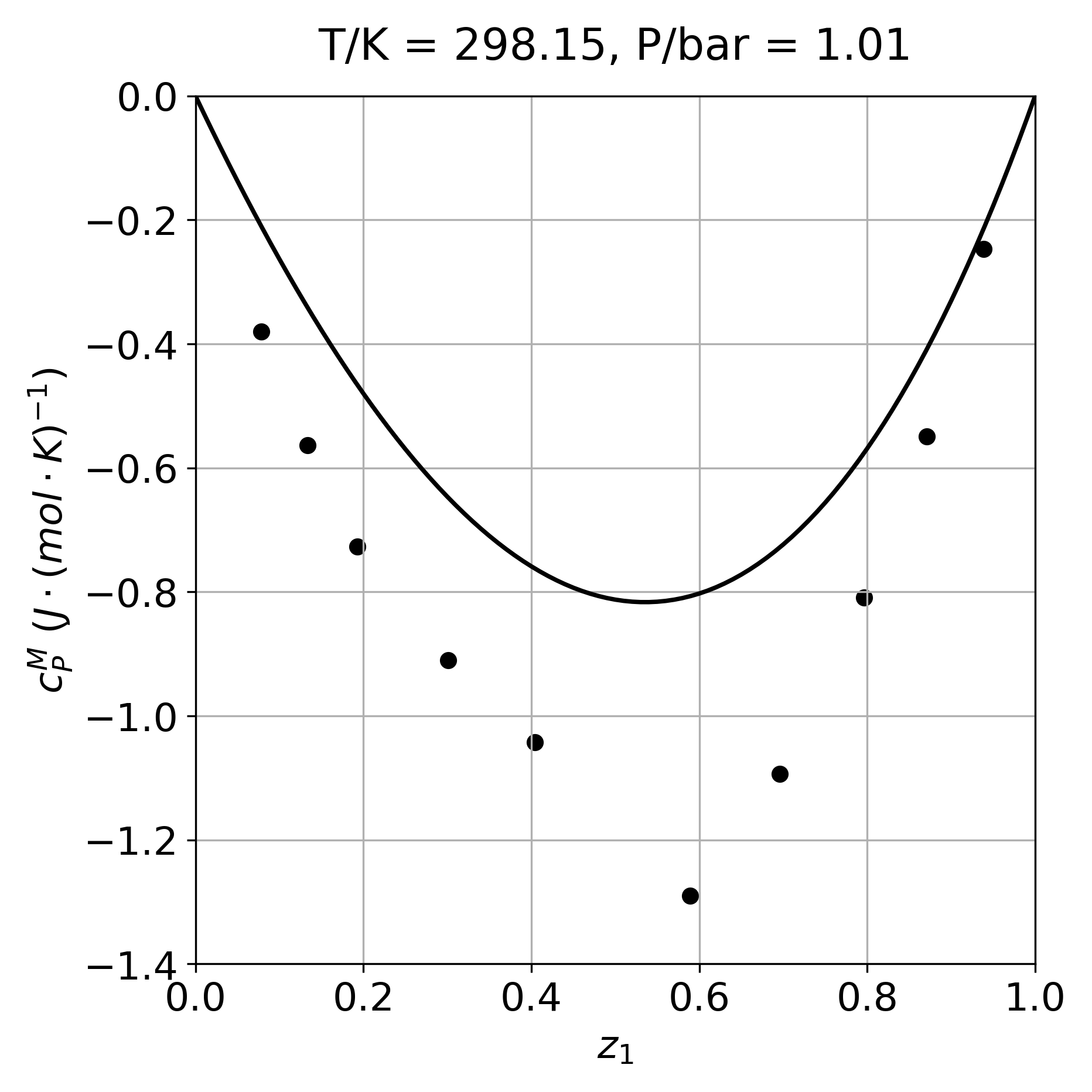}%
  }%
  \hfill
  \subcaptionbox*{\centering (b)}[.3\linewidth]{%
    \includegraphics[width=\linewidth]{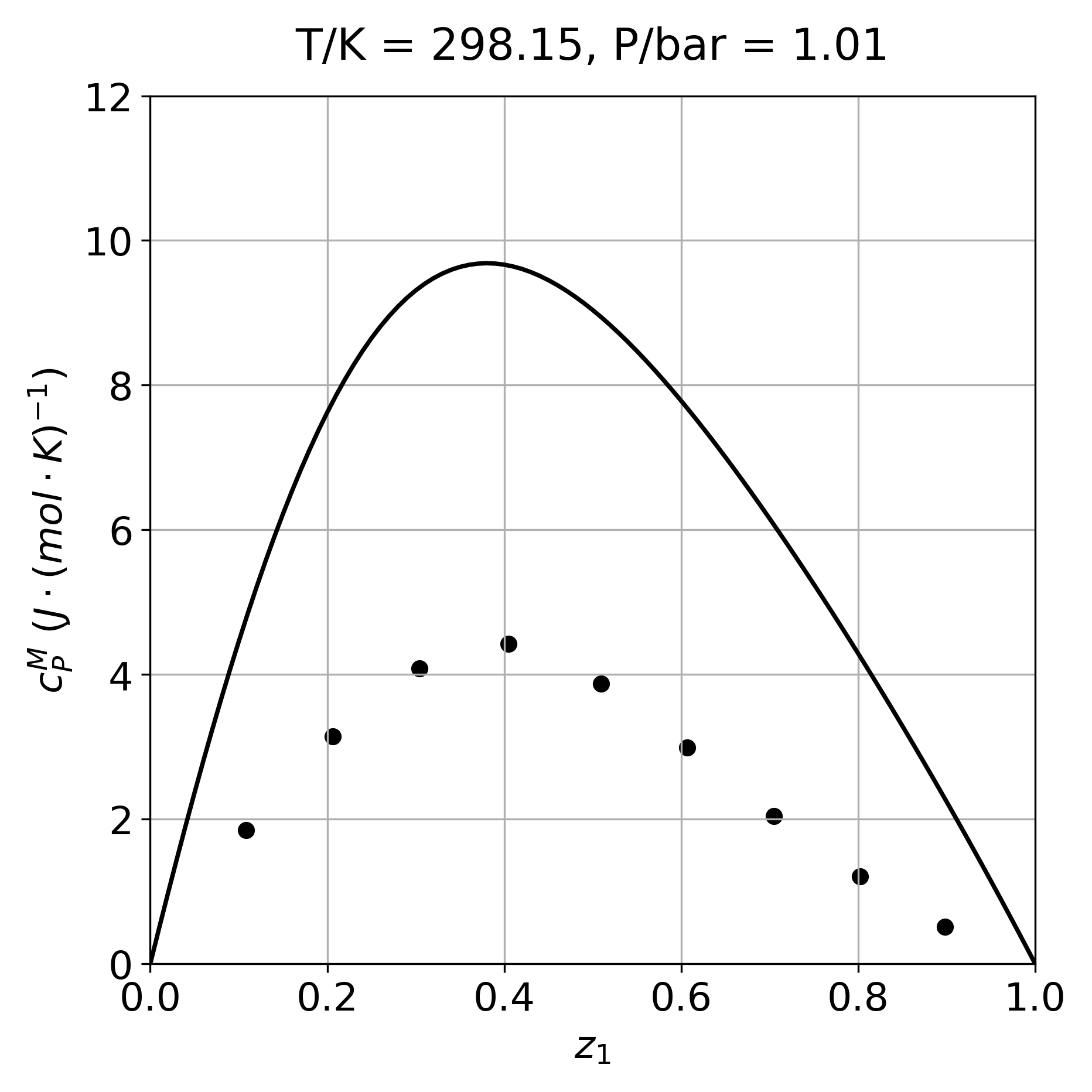}%
  }%
  \hfill
  \subcaptionbox*{\centering (c)}[.3\linewidth]{%
    \includegraphics[width=\linewidth]{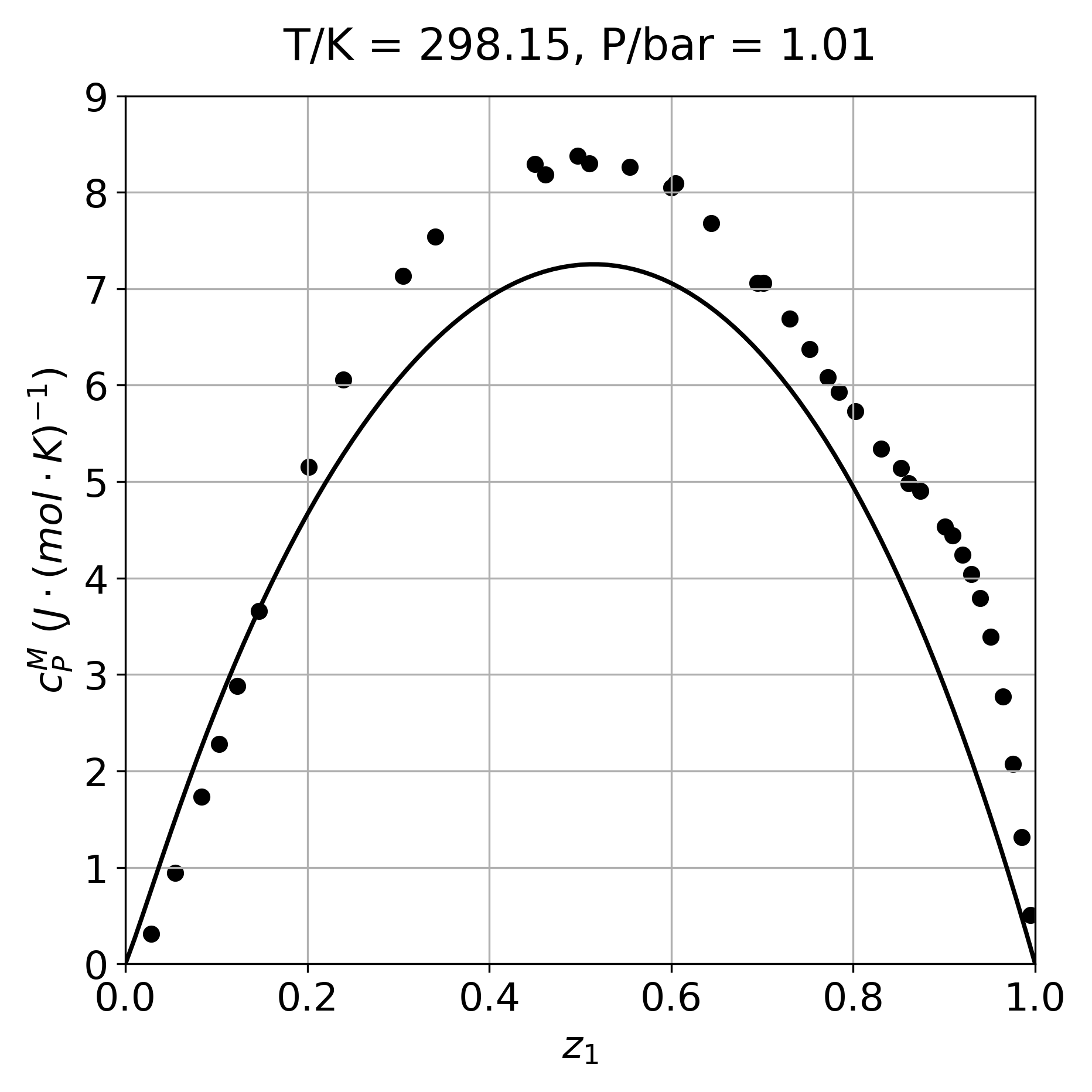}%
  }
\caption{Isobaric heat capacities of mixing for three binary mixtures calculated with openCOSMO-RS-Phi: {\color{black} $\bullet$}: \(c_\text{P}^\text{M,exp}\), \; {\color{black} \rule[0.5ex]{1.5em}{0.6pt}} \(c_\text{P}^\text{M,calc}\); (a) cyclohexane (1) - monochlorobenzene (\(\text{BAC}_1\)), (b) methyl acetate (1) - chloroform (\(\text{BAC}_6\)), (c) water (1) - acetonitrile (\(\text{BAC}_9\)).}
\label{Fig:cpM plots}
\end{figure}

Derivative properties reflect the temperature dependence of a thermodynamic model. For the enthalpy of mixing (\(h^\text{M}\)) and heat capacity of mixing (\(c_\text{P}^\text{M}\)), openCOSMO-RS-Phi shows notably strong performance. In particular, it outperforms all other predictive EoS for \(h^\text{M}\) and achieves the best overall score for \(c_\text{P}^\text{M}\) (10.7), even exceeding all predictive and correlative models, although some room for improvement remains. In addition, it should be noted that the marks reported in \autoref{tab:05} are based on the asymmetric MAPE according to \autoref{eq:44.2} which might differ from other benchmark studies. For a more detailed explanation of this topic, the reader is referred to the SI. Even with the alternative use of a symmetrical MAPE, openCOSMO-RS-Phi still shows better results than other predictive EoS models.
\\
This result is noteworthy, since COSMO-based models are often considered less reliable for temperature-dependent and energetic properties, and such data were not used in the parametrization of openCOSMO-RS. In the present implementation, temperature dependence is included only in the dispersion contribution to the free interaction energy, while the misfit and hydrogen-bonding terms are assumed to be temperature independent. Earlier versions of openCOSMO-RS included a temperature-dependent hydrogen-bonding term; however, its omission was found to improve the correlation of water vapor pressures. \\
One might therefore expect reduced accuracy for systems with significant association effects, such as \(\text{BAC}_5\) and \(\text{BAC}_7\)-\(\text{BAC}_9\), where hydrogen bonding plays a dominant role. However, no consistent deterioration is observed. While \(h^\text{M}\) predictions for \(\text{BAC}_7\)-\(\text{BAC}_9\) are slightly below average, the corresponding \(c_\text{P}^\text{M}\) results for these groups remain above average. \\
\autoref{Fig:hM plots} illustrates the performance for enthalpy-of-mixing data using nine representative binary mixtures, while three examples for the calculation of heat capacities of mixing are presented in \autoref{Fig:cpM plots}.

\hypertarget{conclusions}{
\section{Conclusions}\label{Conclusions}}
In the present work, openCOSMO-RS-Phi was introduced the first open-source implementation of the full COSMO-RS-based equation of state COSMO-SAC-Phi and its performance was systematically investigated for pure components and binary mixtures. To this end, the model was compared with other equations of state with respect to the reproduction of vapor pressures and molar volumes for up to 1800 components. In addition, its performance was evaluated using the binary-mixture database proposed by \textsc{Jaubert et al.}. \\
For pure components, openCOSMO-RS-Phi shows vapor-pressure predictions that are competitive with those of established EoS. For molar volumes, it performs better than several classical equations of state, including SRK, RK, PR, and especially tc-PR. Beyond the numerical results, a central outcome of this work is the availability of an extensive open-source parameter set and a transparent implementation that makes this type of predictive EoS accessible to the wider scientific community. \\

Based on the systematic analysis of phase-equilibrium and energetic data for binary mixtures, the fully predictive model achieves an overall score of 7.3/20. While quantitative improvements remain possible, openCOSMO-RS-Phi is already capable of qualitatively reproducing a broad range of thermodynamic behavior, including vapor-liquid equilibria, liquid-liquid equilibria, three-phase behavior, azeotropes, and derivative properties such as enthalpies and heat capacities of mixing, without the use of binary interaction parameters. Particularly encouraging is its strong performance for energetic properties and for several classes of associating mixtures. \\
Overall, this work establishes openCOSMO-RS-Phi as a fully open-source and scientifically useful predictive equation of state based on COSMO-RS theory. Future improvements may include a better repulsive term, refined treatments of misfit, hydrogen-bonding, and dispersion contributions and, where desired, the introduction of binary interaction parameters for more correlative applications.

%\Contributions{% Author Contribution statement
%Jan Markgraf: Methodology, software, formal analysis, investigation, data curation, visualization, writing - original draft.
%Davi Gustavo Lisboa Girardi: Methodology, formal analysis, validation, writing - original draft.
%Irina Smirnova: Conceptualization, supervision, writing - review and editing.
%Simon M\"uller: Conceptualization, supervision, methodology, validation, resources, writing - review and editing.
%}
%
\Acknowledgments{
Funded by the Deutsche Forschungsgemeinschaft (DFG) - Project number 503327656.
}
\DataAvailability{% Data Availability statement
The model implementation, parameter sets, and data generated or analyzed during this study are made openly available in the project repository \url{https://github.com/TUHH-TVT/openCOSMO-RS-Phi_cpp} and/or Supporting Information associated with this article.
}
\EndMatter
\end{document}